\documentclass[11pt,tightenlines,eqsecnum,floats,aps,amsmath,amssymb,nofootinbib,prd,shownopacs,floatfix,superscriptaddress]{revtex4-2}
\usepackage{graphicx}
\usepackage{epstopdf}
\usepackage{latexsym}
\usepackage{amssymb}
\usepackage{amsmath}
\usepackage{color}  
\usepackage{mathrsfs}
\usepackage{xparse}
\usepackage{float}
\usepackage{dsfont}
\usepackage{mathtools}
\usepackage[toc,title,page]{appendix}
\usepackage{xargs}                      

\makeatletter
\def\l@subsubsection#1#2{}
\makeatother
\DeclareRobustCommand{\gobblefive}[5]{}
\newcommand*{\SkipTocEntry}{\addtocontents{toc}{\gobblefive}}

\DeclareMathOperator\arctanh{arctanh}
\usepackage[sort&compress]{natbib}
\usepackage{hyperref}
\usepackage{url}                    %Will make the url look like a url

\usepackage{tensor}
\usepackage[compat=1.0.0]{tikz-feynman}
\usepackage{braket}
\renewcommand{\t}{\tensor} %Short notation for tensors
\newcommand{\mbeq}{\overset{!}{=}}
\begin{document}
\begin{abstract}
We investigate the one-particle sector for the field-theoretical model of gravitationally induced decoherence for a scalar field in \cite{Fahn:2022zql} with a special focus on the renormalisation of the one-particle master equation. In contrast to existing models in the literature, where the renormalisation is usually performed after the Markov and rotating wave approximation and often only for certain limits such as the non- or ultra-relativistic limit, here we apply the renormalisation directly after the one-particle projection. With this strategy, we show that UV-divergent contributions in the one-particle master equation can be identified with the vacuum contributions in the self-energy of the scalar field in the effective quantum field theory and depending on the chosen one-particle projection method, its vacuum bubbles, while the additional thermal contributions in the self-energy are all UV-finite. To obtain the renormalised one-particle master equation, we use an on-shell renormalisation procedure of the underlying effective QFT. We then apply the Markov and rotating wave approximation, specifying a condition under which the Markov approximation can be applied in the case of the ultra-relativistic limit. We compare our results with those available in the literature. This includes an analysis of two different kinds of one-particle projections, a comparison of the application and effects of renormalisation of quantum mechanical and field theoretical models, the non-relativistic and ultra-relativistic limits of the renormalised one-particle master equations, and a comparison with a quantum mechanical toy model for gravitationally induced decoherence in the context of neutrino oscillations.
\end{abstract}

\title{Gravitationally induced decoherence of a scalar field: investigating the one-particle sector and its interplay with renormalisation}
\author{Max Joseph Fahn}
\email{max.j.fahn@fau.de}
\affiliation{Institute for Quantum Gravity, Theoretical Physics III, Department of Physics, FAU Erlangen-N\"urnberg, Staudtstr. 7, 91058 Erlangen, Germany.}
\affiliation{Dipartimento di Fisica e Astronomia, Universita di Bologna, Via Irnerio 46, Bologna 40126, Italy}
\affiliation{INFN, Bologna, Via Irnerio 46, 40126 Bologna, Italy}
\author{Kristina Giesel}
\email{kristina.giesel@fau.de, corresponding author}
\affiliation{Institute for Quantum Gravity, Theoretical Physics III, Department of Physics, FAU Erlangen-N\"urnberg, Staudtstr. 7, 91058 Erlangen, Germany.}

\maketitle

\newpage
\tableofcontents
\newpage
\section{Introduction}\label{sec:intro}
Since we expect closed and thus isolated quantum systems to be an idealisation, there is increasing interest in various areas of physics in the investigation of open quantum systems. These are usually modelled by a total system consisting of two parts: the system under consideration and the environment as well as the interaction between the two. By tracing out the degrees of freedom of the environment, the effective dynamics of the system under consideration are obtained, which, in contrast to the closed system, is still affected by the interaction with the environment \cite{Breuer:2002pc,Hornberger,Weiss:2021uhm}. The effective dynamics of the system under consideration is often formulated in the form of a so-called master equation, of which the Lindblad equation is a special and prominent example \cite{Lindblad:1975ef,Gorini:1975nb}. An important question in the context of open quantum systems is, what to choose as the environment and how to model the interaction with the system. Since we assume that all standard matter is at least coupled to gravity, interest in open quantum systems with a gravitational environment has increased in recent years, see for instance \cite{Benatti:2000ph,Blencowe:2012mp,Anastopoulos:2013zya,Guzzo:2014jbp,Oniga:2015lro,Lagouvardos:2020laf,DEsposito:2023psn} and the reviews in \cite{Bassi:2017szd,Anastopoulos:2021jdz} and references therein. This has been investigated in the context of gravitationally induced decoherence due to a quantum gravitational environment in the framework of linearised gravity in several works in the literature. For instance in \cite{Anastopoulos:2013zya,Blencowe:2012mp,Fahn:2022zql} a scalar field is considered as the matter field, while in \cite{Lagouvardos:2020laf,fgke2024photon} a photon field is considered and in \cite{Oniga:2015lro} a model for a generic bosonic matter field is investigated. All these works have in common that they start from the underlying field theoretical model and then derive a corresponding master equation for the matter system under consideration. 

~\\
In addition, there are also phenomenological models, see for instance \cite{Lisi:2000zt,Benatti:2000ph,Sakharov:2009rn,Guzzo:2014jbp,Coelho:2017byq,Carpio:2018gum}, that do not necessarily derive a master equation from an underlying action, but rather set up physically motivated ansätze for the form of the dissipator, which encodes the effective influence of the environment in the master equation. The latter are often formulated in the context of quantum mechanics and thus for finitely many degrees of freedom, where a given master equation is often less difficult to work with or even to solve. Furthermore, many of these models already used the Lindblad equation as a starting point, so that the physical properties of the environment are often less accessible compared to the underlying field theoretical models. An interesting question is therefore, on the one hand, how the underlying field theoretical model can be linked to a given quantum-mechanical phenomenological model and, on the other hand, what additional insights the underlying field-theoretical model can provide that are no longer accessible to us in the quantum-mechanical model. In this work we want to address both question in the context of models for gravitationally induced decoherence. 

~\\
To answer these questions, the one-particle sector of field theory needs to be investigated to establish a link to microscopic quantum mechanical models. Furthermore, in order to investigate the connection to existing phenomenological models for gravitationally induced decoherence, one needs to understand more precisely how certain methods such as renormalisation and specific approximations such as the Markov or rotating wave approximation, which are often performed to finally arrive at a Lindblad-type master equation, affect the field-theoretical model respectively its one-particle sector. Some of these questions have already been discussed and answered in the works in \cite{Anastopoulos:2013zya,Blencowe:2012mp,Lagouvardos:2020laf}. There, the one-particle sector was derived from a field-theoretical model and quantum mechanical master equations were derived for the non-relativistic and ultra-relativistic cases. The new aspect we aim at investigating in this work is the role of renormalisation in this context and its interplay with the further approximations, such as the Markov and rotation wave approximation, that one needs to apply in the derivation of the final master equation. 
~\\

To the best of the authors' knowledge, renormalisation for field-theoretical models of gravitationally induced decoherence has been performed in the existing literature after applying the Markov and rotational wave approximation, often at the level of the corresponding one-particle sector in certain limits such as the ultra-relativistic limit \cite{Lagouvardos:2020laf} or the non-relativistic limit \cite{Anastopoulos:2013zya}. In contrast in this work we will perform the renormalisation at the level of the effective field theory before we apply any of the above mentioned approximations or limits. This strategy allows us to obtain a more detailed understanding about the UV-divergent contributions in the one-particle master equations. We will follow the methods introduced in \cite{Burrage:2018pyg}, where one scalar field was coupled to a second scalar field as an environment and extend those techniques to the case of a gravitational environment. These methods allow us to identify individual contributions in the one-particle master equations with specific Feynman diagrams of the underlying effective field-theoretical model. Since the starting point of the model in this paper is the canonical formulation of a scalar field coupled to linearised gravity, in a first step we introduce non-covariant Feynman rules adapted to the canonical model, following \cite{weinberg2005quantum,tong}, where this was introduced for the case of Quantum Electrodynamics (QED). Interestingly, the connection between the covariant and non-covariant Feynman rules can be used to show that the divergent contributions in the one-particle master equation are involved in the self-energy of the scalar field in the effective field theory. The self-energy can be decomposed into a vacuum and a thermal contribution, the latter vanishing when we consider the zero temperature limit of thermal gravitational waves in the environment. The vacuum contributions can then be renormalised using a standard procedure, and the renormalised one-particle master equation can be obtained. 
~\\
~\\
Equipped with this result, we can use it and investigate what kind of effect a Markov and rotating wave approximation have and compare it to the existing literature, where these two approximations are mostly done before a renormalisation is performed. At least for the case of the ultra-relativistic limit, we are able to provide a condition under which the Markov approximation can be applied in the model considered here. For the general case, this is quite a challenging task, since the integrals involved, which have to be analysed for the environmental correlation functions, are quite complicated. In the case of the rotating wave approximation in the existing literature the pre- and post-trace application (see for instance \cite{fleming2010rotating,Lagouvardos:2020laf,Burgarth:2023ppw,Wang:2023dkf} for analyses and applications) is discussed and we apply the latter in this work and determine the final one-particle master equation where both the Markov and rotating wave approximation have been applied. This is used together with the intermediate results before the individual approximations in some applications to compare with the existing results in the literature, in particular the work in \cite{Oniga:2015lro,Anastopoulos:2013zya,Lagouvardos:2020laf} for field-theoretical models and the quantum mechanical model in \cite{Domi:2024ypm} in the context of gravitationally induced decoherence in neutrino oscillations, which is based on a quantum mechanical toy model for gravitationally induced decoherence from \cite{Xu:2020lhc}. For the comparison with the quantum mechanical model, we are especially interested in the extent to which we can relate the application and effect of renormalisation in the field-theoretical model and in the quantum mechanical model and how we can thereby gain new insights into the differences, and similarities respectively as well as the physical properties of these models.
~\\
~\\
The paper is structured as follows: after the introduction in section \ref{sec:intro}, we briefly review the field-theoretical model from \cite{Fahn:2022zql} in section \ref{sec:review} whose one-particle projection is derived in section \ref{sec:onepp}, where we consider two types of projections, a non-extended and an extended one, which differ in whether each individual operator in the master equation preserves the one-particle sector or whether only the combinations of operators that enter the final master equation must do so.
The renormalisation of the one-particle master equation is discussed in section \ref{sec:renorm}. First we identify those contributions in the one-particle master equations that are UV-divergent in subsection \ref{uvren1}. Subsection \ref{uvren2} and  \ref{uvren3} introduce the non-covariant and covariant Feynman rules and discuss their relation as well as show how the UV-divergent contributions can be identified with the vacuum part of the scalar field's self-energy. The renormalisation of the self-energy is discussed in subsection \ref{uvren4} and in subsection \ref{uvren5} we use these results to determine the final form of the renormalised one-particle master equation. Afterwards in section \ref{sec:approx} we apply the Markov and rotating wave approximation that are separately discussed in subsection \ref{sec_Map} and \ref{sec_RWA} respectively. As possible applications  in section \ref{sec:appl} we discuss in subsection \ref{sec_Pop} the evolution of the populations of the one-particle master equation before and after renormalisation as well as after in addition the Markov approximation has been applied, see subsections \ref{sec:beforeRenorm}, \ref{sec:afterRenorm} and \ref{sec:afterMap} and compare our results with the ones in \cite{Oniga:2015lro}. To compare more in detail to the existing results in \cite{Anastopoulos:2013zya,Blencowe:2012mp,Lagouvardos:2020laf} we consider the non-relativistic and ultra-relativistic limit in subsections \ref{ap:Nonrandurel} and \ref{sec_apurel}. The comparison with quantum mechanical model from \cite{Domi:2024ypm} that considers gravitationally induced decoherence in the context of neutrino oscillations can be found in subsection \ref{sec:AppURel}. Finally we summarise and conclude in section \ref{sec:concl}. In addition, details of the calculations required to obtain the results of each section are provided in the appendix to make the article self-contained.

\section{Review of the underlying field-theoretical master equation}\label{sec:review}
As we aim at investigating the  one-particle sector of model considered in \cite{Fahn:2022zql} in this work and compare the results obtained here to results in the existing literature such as in \cite{Anastopoulos:2013zya,Blencowe:2012mp,Lagouvardos:2020laf}, we briefly review the main results from \cite{Fahn:2022zql} that are taken here as a starting point for the further analysis. In \cite{Fahn:2022zql} a step-by-step derivation of a second order, time-convolutionless master equation for gravitationally induced decoherence of a scalar field is presented with the aim to compute the effective dynamics of the scalar field evolving in an environment consisting of thermal gravitational waves. 
The starting point in \cite{Fahn:2022zql} is the classical action of general relativity coupled to a scalar field, where the mostly plus signature is used for the metric. To be able to apply canonical quantisation later on, general relativity is then formulated in the Hamiltonian (ADM) framework (\cite{Deser:1960zzc}) formulated in terms of Ashtekar-Barbero variables (\cite{Ashtekar:1986yd, BarberoG:1994eia, Immirzi:1996di,Ashtekar:1989ju}), which encodes the gravitational degrees of freedom in terms of an SU(2)-connection $\tensor{A}{_a^i}(\vec x,t)$ and its canonically conjugate momenta denoted as densitised triads $\tensor{E}{^a_i}(\vec x,t)$. Here, the first index $a$ is a spatial one and the second one $i$ an SU(2)-Lie algebra index. The metric can be determined by the densitised triads only up to a rotation and this yields to an additional  so-called Gauß constraint next to the Hamiltonian and spatial diffeomorphism constraint already present in the ADM formulation.
~\\
The reason that \cite{Fahn:2022zql} works with these elementary variables is that they form the elementary canonical variables in the context Loop Quantum Gravity (LQG, see \cite{rovelli_2004,Thiemann:2007pyv} for an introduction) and hence formulating the classical model in terms of them makes the application of a quantisation using LQG techniques possible, as discussed in \cite{Fahn:2022zql}. After the inclusion of a boundary term for an asymptotic flat spacetime (\cite{Thiemann:1993zq}), the gravitational sector of the system is linearised around a Minkowski spacetime, where $\kappa=\frac{8\pi G_N}{c^4}$ plays the role of the perturbation parameter and $G_N$ denotes Newton's constant. The matter contribution is included in terms of a post-Minkowski approximation \cite{PoissonBook:2014}. As the linearised model still includes gauge symmetries, suitable Dirac observables are then constructed in perturbation theory using the relational formalism (\cite{Rovelli:1990ph,Rovelli:1990pi,Rovelli:2001bz,Vytheeswaran:1994np,Dittrich:2004cb,Dittrich:2005kc,Dittrich:2006ee}) to extract the physical degrees of freedom. For gravity they correspond to the linearised symmetric transverse traceless components of the connection and the triad fields. These constitute of two degrees of freedom each, equivalent to the two polarisations of gravitational waves and their conjugate momenta which are the linearised excitations one obtains when using the metric and its conjugate momenta as the elementary variable. In addition one can construct Dirac observables for the scalar field and its canonically conjugate momentum for the matter sector. 
~\\
After a Fock quantisation, a time-convolutionless master equation (\cite{Breuer:2002pc}) is derived by tracing out the gravitational environment, which is described by a Gibbs state with temperature\footnote{See the discussion in \cite{Anastopoulos:2013zya} on the interpretation of this parameter.} parameter $\Theta$, and in this way treating the scalar field as an open quantum system. For this, the projection operator techniques (see \cite{Zwanzig:1960gvu,Nakajima:1958pnl,Breuer:2002pc}) are employed, for which one splits the Hilbert space into a relevant and an irrelevant part. In the model considered here the scalar field is the relevant and the gravitational environment the irrelevant part. Then the strategy is to solve the Liouville-von Neumann equation of the relevant part perturbatively using an approximated, truncated solution for the irrelevant part. The final master equation in \cite{Fahn:2022zql} is time-convolutionless and truncated after second order in $\sqrt{\kappa}$. It reads
\begin{equation}\label{eq:meqSPPbegin}
    \frac{\partial}{\partial t} \rho_S(t) = -i [H_S + \kappa\: U + \kappa \: H_{LS},\rho_S(t)] + \mathcal{D}[\rho_S(t)]\,.
\end{equation}
Here, $\rho_S(t)$ denotes the density matrix describing the state of the scalar field at physical\footnote{See \cite{Fahn:2022zql} for the definition of the physical time in terms of the relational formalism in this model.} time $t$. Furthermore, $H_S$ is the Hamiltonian of the free scalar field, 
\begin{equation}
    H_S := \int_{\mathbb{R}^3} d^3k \; \omega_k \, a_k^\dagger \, a_k\,,
\end{equation}
where $a^{(\dagger)}_k$ denote the annihilation/creation operator valued distributions for momentum $\vec k$ corresponding to the scalar field, that obey the standard commutation relations $[a_k,a_l^\dagger]=\delta^3(\vec k,\vec l)$, and $\omega_k=\sqrt{\vec{k}^2+m^2}$ its dispersion relation. The other terms in \eqref{eq:meqSPPbegin} arise due to the coupling to linearised gravity and represent a self-interaction term $U$ and a Lamb-shift-like term $H_{LS}$ (given in \eqref{lsham}) as well as a dissipator term $\mathcal{D}$ (given in \eqref{Dfsf}). The first one arises when the Hamiltonian is expressed in terms of the independent Dirac observables mentioned above and the Lamb-shift-like term $H_{LS}$ contributes to the unitary evolution of the scalar field, whereas the dissipator leads to to non-unitary evolution encoding effects like decoherence. A similar form for a master equation can be obtained by using the Born approximation instead of the projection operator technique, see section 4 in \cite{Fahn:2022zql} for more details.
~\\ 
While this master equation predicts the effective dynamics of the scalar field under the influence of a gravitational environment, we still need to deal with infinitely many degrees of freedom which makes investigating the solution of the master equation very challenging. 
For this reason in this work the master equation for a single scalar particle that follows from the master equation in \eqref{eq:meqSPPbegin} is derived. 
~\\
~\\
In general, this master equation, as well as the full field-theoretical equation from \cite{Fahn:2022zql}, may not be completely positive, and checking this is usually a difficult task for the field-theoretical models. This is the usual situation one is confronted with when deriving master equations from underlying (field-theoretical) microscopic models for which complete positivity has to be investigated on a model-by-model basis. While we know that Lindbladian dynamics is completely positive by construction, for non-Lindbladian dynamics this needs to be analysed on a case-by-case basis. A possible first step towards such an analysis can be to investigate whether the commonly applied approximations such as the Born and Markov approximations as well as the rotating wave approximation can also be applied in a more general context to simplify the form of the master equation and eventually write it in the form of a Lindblad equation for which full positivity is guaranteed. However, this does not allow us to draw general conclusions about the general dynamics of open microscopic models and their master equation, since models with non-Lindblad dynamics are more complicated and in general have a time-dependent Liouvillian. Often this is discussed in the context of master equations including non-Markovian dynamics, see for instance \cite{breuer2007non}, which in our case would be encoded in higher order truncations of the TCL equation of the microscopic model, corresponding to higher order contributions in $\sqrt{\kappa}$ in the model under consideration. Furthermore, to our knowledge, no necessary condition has yet been established for the generalisation of a time-dependent Liouvillian which guarantees complete positivity \cite{DAbbruzzo:2022hgp}. This differs from the situation of a time-independent Liouvillian, which is relevant for the Lindblad equation.

\section{Projection of the master equation to the one-particle case}\label{sec:onepp}
In this section, we discuss how such a master equation for a single scalar particle can be obtained starting with equation \eqref{eq:meqSPPbegin} in the field theory context. When working with field theoretical master equations such a one-particle projection is commonly applied to investigate some features of the master equation, see for instance the works in \cite{Anastopoulos:2013zya,Lagouvardos:2020laf,Oniga:2015lro,Burrage:2018pyg}. There exist however different methods on how to perform this projection in detail. In \cite{Anastopoulos:2013zya,Oniga:2015lro} the procedure is carried out such that the final one-particle master equation is probability conserving, which requires to neglect some terms that would otherwise be present in a direct projection. In contrast, in \cite{Burrage:2018pyg} a different strategy based on Thermo Field Dynamics (TFD), which is a formulation of Quantum field theory at finite temperature (see \cite{Arimitsu:1985ez,Takahashi:1996zn} and for an introduction \cite{Khanna:2009zz}), is employed, in which these terms still contribute to the one-particle master equation. Here we follow the method used in \cite{Anastopoulos:2013zya,Oniga:2015lro}, but keep all possible terms and investigate their influence on the one-particle master equation. It will turn out that after applying an on-shell renormalisation and Markov approximation, they will not play any role for decoherence, but will remove the remaining contribution of the Lamb-shift-like term to the unitary evolution after the rotating wave approximation has been applied.
~\\

To obtain the one-particle projection of the master equation in (2.1), we replace the density matrix with the corresponding density matrix for a single particle
\begin{equation}\label{eq:1pdm}
                    \rho_1(t) = \int_{\mathbb{R}^3} d^3u \int_{\mathbb{R}^3} d^3v \; \rho(\vec{u},\vec{v},t) \: a_u^\dagger \ket{0}\bra{0} a_v\,,
                \end{equation}
                in the master equation and neglect all contributions that project out of the single particle space. In this formulation, $\rho(\vec{u},\vec{v},t)$ is the (quantum mechanical) density matrix of a single particle in momentum representation. For a pure state, this density matrix in momentum representation can be expressed as the product of the particle's wave function in momentum representation $\psi(\vec{u},t)$ in the following way: 
                \begin{equation*}
                    \rho(\vec{u},\vec{v},t)=\psi(\vec{u},t)\; \psi^*(\vec{v},t) = \braket{\vec{u}|\psi(t)} \braket{\psi(t)|\vec{v}} = \bra{\vec{u}} \rho_1(t) \ket{\vec{v}}\,,
                \end{equation*}
                where a star denotes complex conjugation and we defined (normalised) single particle (generalised) momentum states as $\ket{\vec{u}} := a_u^\dagger \ket{0}$ fulfilling $\braket{\vec{u}|\vec{v}}=\delta^{(3)}(\vec{u}-\vec{v})$ with the three-dimensional Dirac delta distribution $\delta^{3}(\vec{k})$.
~\\
The projection we use in order to obtain the one-particle master equation can be understood then as
                \begin{equation*}
                    \Pi_1 := \int_{\mathbb{R}^3} d^3u \; \ket{\vec{u}}\bra{\vec{u}}\,.
                \end{equation*}
                This represents mathematically a projection given the orthogonality of Fock states with different particle numbers and given that $(\Pi_1)^2 = \Pi_1$ using the normalisation of the $\ket{\vec{u}}$. As discussed in the remainder of this section, the location of application of this projector is different in the literature, in particular whether it is applied to the master equation as a whole or to individual operators appearing in products in the master equation. In this work, we use both ways in order to be able to discuss the effect of either choice on the physical processes taken into account as well as on the final one-particle master equation in the end.
                ~\\
In the following we will discuss the corresponding individual contributions in \eqref{eq:meqSPPbegin} separately and further will discuss the assumptions used in the model considered here as well as compare them to the existing literature:
~\\
The first term of the master equation, representing the evolution of a free scalar particle, can be computed immediately to yield
\begin{equation}
    -i [H_S,\rho_1(t)] = -i\int_{\mathbb{R}^3} d^3u \int_{\mathbb{R}^3} d^3v \; (\omega_u-\omega_{v}) \rho(\vec{u},\vec{v},t) \,a_u^\dagger \, \ket{0} \bra{0} a_{v}\,.
\end{equation}
The contribution of the second term depends on the structure of the form of the operator $U$. The detailed expression is given in \cite{Fahn:2022zql} in equation (3.12), for the analysis here it is however enough to know that $U$ consists of different combinations of always four creation and/or annihilation operators for the scalar field, i.e.
\begin{align}
    U =\int d^2k_1 \int d^2k_2 \int d^2k_3 \int d^2k_4 \Big[& \zeta_1 \:a_{k_1} a_{k_2} a_{k_3} a_{k_4} + \zeta_2\: a^\dagger_{k_1} a_{k_2} a_{k_3} a_{k_4} \nonumber\\ &+ \zeta_3\: a_{k_1} a^\dagger_{k_2} a_{k_3} a_{k_4} + ... + \zeta_N\: a^\dagger_{k_1} a^\dagger_{k_2} a^\dagger_{k_3} a^\dagger_{k_4} \Big]\,, 
\end{align}
with the coefficient distributions $\zeta_i= \zeta_i(\vec{k}_1,\vec{k}_2,\vec{k}_3,\vec{k}_4)$ that contain delta distributions that relate some of the momenta. For more details, we refer to the definition in equation (3.12) in \cite{Fahn:2022zql}. When applying normal ordering to this operator, as it is done in \cite{Fahn:2022zql}, then it will not contribute after the one-particle projection: in the summands of $U$, where the number of creation operators is not equal to the number of annihilation operators, the resulting terms would project out of the one-particle space. In the other summands, there are exactly two creation and two annihilation operators which, when normal ordered, annihilate any one-particle state. 
In \cite{Anastopoulos:2013zya,Lagouvardos:2020laf} the normal ordering of $U$ is applied differently: in their work the four annihilation and/or creation operators are normal ordered pairwise\footnote{The reasoning for this is that $U$ arises as a combination of two operators that each contains two creation and/or annihilation operators.}. In that case contributions of the form $:a^\dagger_{k_1} a_{k_2}:\; :a^\dagger_{k_3} a_{k_4}:$ preserve the one-particle space and thus still contribute after the one-particle projection. To distinguish these two types of operator orderings, we denote the first one, where $U$ is normal ordered, total normal ordering, and the second one partial normal ordering. In this work we consider a totally normal ordered Hamiltonian as in \cite{Fahn:2022zql}.
~\\

The third term, the Lamb-shift-like Hamiltonian, as well as the fourth contribution, the dissipator, both contain the same building blocks. To evaluate them, it is sufficient to consider the following three combinations:
\begin{equation}
      \underbrace{j_r^A(\vec{k},\vec{p})^\dagger \, j_r^B(\vec{k},\vec{l})\, \rho_1(t)}_{(I)} \hspace{0.3in} \text{and} \hspace{0.3in} \underbrace{\rho_1(t)\,j_r^A(\vec{k},\vec{p})^\dagger \, j_r^B(\vec{k},\vec{l})}_{(II)}  \hspace{0.3in} \text{and} \hspace{0.3in} \underbrace{ j_r^B(\vec{k},\vec{l})\, \rho_1(t) \, j_r^A(\vec{k},\vec{p})^\dagger}_{(III)}\,,
\end{equation}
where the $j_r^A(\vec{k},\vec{p})$ denote individual and different normal-ordered current operators labeled by $A\in\{1,2,3,4\}$ that carry a polarisation label $r\in\{\pm\}$ and two momentum arguments. These current operators are defined in detail starting in equation \eqref{eq:defjundw} and are of the form
\begin{align}
    j_r^1(\vec{k},\vec{p}) &\propto a_p^\dagger a_{k+p}  & j_r^2(\vec{k},\vec{p}) &\propto  a_{-p-k}^\dagger a_{-p} \\
    j_r^3(\vec{k},\vec{p}) &\propto a_{-p} a_{k+p} & j_r^4(\vec{k},\vec{p}) &\propto a_p^\dagger a_{-k-p}^\dagger\,.
\end{align}
Hence they consist of two creation and/or annihilation operators with different momentum labels.  
~\\
At this point arises the question whether we want to enforce trace preservation in the one-particle master equation, which corresponds to probability conservation. In \cite{Anastopoulos:2013zya,Oniga:2015lro} this is done, which results in the exclusion of specific terms from the one-particle master equation. These terms can be identified from the general form of the master equation in \eqref{eq:meqSPPbegin} as we will discuss now: it is evident that when applying the trace the commutator vanishes and one is left with 
\begin{align}
    \frac{\partial}{\partial t} tr \left\{ \rho_S(t)\right\} &= tr \left\{ \mathcal{D}[\rho_S(t)]\right\}\,.
\end{align} 
Inserting the definition of the dissipator given in \cite{Fahn:2022zql} in equation (4.74) then yields
\begin{align}\label{eq:probcons}
    \frac{\partial}{\partial t} tr \left\{ \rho_S(t)\right\} = \frac{\kappa}{2} \int \frac{d^3k\, d^3p\, d^3l}{(2\pi)^3} &\sum_{r\in\{\pm\}} \sum_{A,B=1}^4 \; {R}_{AB}(\vec{p},\vec{l};\vec{k},t)\nonumber\\& \cdot tr\left\{ j_r^B(\vec{k},\vec{l}) \rho_S(t)j^A_r(\vec{k},\vec{p})^\dagger-\frac{1}{2} \left\{ j^A_r(\vec{k},\vec{p})^\dagger j_r^B(\vec{k},\vec{l}), \rho_S(t)\right\} \right\}\,,
\end{align}
where the $R_{AB}$ are time-dependent coefficients. 
When the current operators $j_r^A$ are individually projected onto the one-particle space, due to the cyclicity of the trace all terms in the difference of the two traces are exactly canceled and one obtains a preserved trace of the density matrix, hence probability conservation. This is the approach used for instance in \cite{Anastopoulos:2013zya,Oniga:2015lro}. Another option is to apply the one-particle projection in such a way that each entire term in the master equation has to preserve the one-particle space. This is for instance done in \cite{Burrage:2018pyg}, where two scalar fields are considered, one as the system and the other one as the environment. In this case there will remain terms in the one-particle projection of the product of two current operators in the last term of \eqref{eq:probcons} that have no counterpart in the first term of \eqref{eq:probcons} and thus will not cancel in the difference of the two  traces. 
~\\
To keep our analysis as generic as possible, we will include these terms in this work and investigate their effect in the one-particle master equation and denote this one-particle projection the extended one-particle projection. To take them into account in our further calculations, we will introduce a factor $\delta_P$ in these contributions to be able to switch between the extended one-particle projection $(\delta_P=1)$ and the non-extended one ($\delta_P=0$). 
~\\
~\\
The detailed derivation of the one-particle projection of the master equation following these methods introduced here can be found in appendix \ref{app:OPP}. The additional terms that are present in the extended one-particle projection correspond to physical situations in Quantum Field
Theory (QFT) in which in the intermediate steps two particles are created and annihilated afterwards. This also includes the case where the original particle is left invariant and a vacuum bubble is created. The latter case thus requires a renormalisation, which is also carried out in appendix \ref{app:OPP}. 
Following these projection methods, the one-particle master equation for the density matrix in momentum representation is then given in \eqref{eq:meqopff} and reads
\begin{align}\label{eq:meqopf}
\frac{\partial}{\partial t} \rho(\vec{u},\vec{v},t) = &-i \rho(\vec{u},\vec{v},t) \; (\omega_u-\omega_v) \nonumber\\
&-\frac{\kappa}{2} \int \frac{d^3k}{(2\pi)^3} \bigg\{ \frac{ P_u(\vec k)}{\omega_{u-k}\omega_u}\; \left[C(\vec{u},\vec{k},t)+\delta_P C_P(\vec{u},\vec{k},t)\right] \nonumber\\ &\hspace{1.1in} +\frac{P_v(\vec k)}{\omega_{v-k}\omega_v}\; \left[C^*(\vec{v},\vec{k},t)+\delta_P C_P^*(\vec{v},\vec{k},t)\right]\bigg\}\;\rho(\vec{u},\vec{v},t)\nonumber\\
&+\frac{\kappa}{2} \int \frac{d^3k}{(2\pi)^3} \frac{P^{ijln}(\vec{k})\: u_i u_j  v_l v_n}{\sqrt{\omega_{u+k}\omega_u\omega_{v+k}\omega_v}}\; \left\{C(\vec{u}+\vec{k},\vec{k},t)+ C^*(\vec{v}+\vec{k},\vec{k},t)\right\}\: \rho(\vec{u}+\vec{k},\vec{v}+\vec{k},t) 
\end{align}
where a star $^*$ denotes complex conjugation, and with $P_u(\vec k) := {P}\t{}{^i^j^l^n}(\vec{k}) u_i u_j u_l u_n$ which contains the symmetric transverse traceless (STT-)projector
\begin{equation}\label{eq:ttproj}
{P}\t{}{^i^j^l^n}(\vec{k}) = \frac{1}{2} [{P}^{il}(\vec{k}) {P}^{jn}(\vec{k})+{P}^{in}(\vec{k}) {P}^{jl}(\vec{k})-{P}^{ij}(\vec{k}) {P}^{ln}(\vec{k})]\,,
\end{equation}
that in turn consists of combinations of the transverse projectors
\begin{equation}
{P}^{ij}(\vec{k}) = \delta^{ij}-\frac{k^i k^j}{\vec{k}^2}\,.
\end{equation}
The presence of this projector is a consequence of the chosen Dirac observables, and thus the physical degrees of freedom of the linearised gravitational field. The coefficients in \eqref{eq:meqopf} are defined as
\begin{align}
C(\vec{u},\vec{k},t) &= \int_{0}^{t-t_0} \frac{d\tau}{\Omega_k} \Big\{ [N(k)+1]\: e^{- i(\Omega_k+\omega_{u-k}-\omega_u)\tau} +N(k)\;e^{ i(\Omega_k-\omega_{u-k}+\omega_u)\tau}\Big\}\\
C_P(\vec{u},\vec{k},t) &= \int_{0}^{t-t_0} \frac{d\tau}{\Omega_k} \Big\{ [N(k)+1]\: e^{- i(\Omega_k+\omega_{u-k}+\omega_u)\tau} +N(k)\;e^{ i(\Omega_k-\omega_{u-k}-\omega_u)\tau}\Big\}\,,
\end{align}
where $N(k) = \frac{1}{e^{\beta \Omega_k}-1}$ is the Bose-Einstein distribution of the gravitational waves in the environment with frequencies $\Omega_k := \sqrt{\vec{k}^2}$ and $\beta=(k_B \Theta)^{-1}$, where $k_B$ is the Boltzmann constant and $\Theta$ the temperature parameter of the Gibbs state that characterises the environment of thermal gravitational waves. The term in the first line of the master equation in \eqref{eq:meqopf} represents the standard unitary evolution of the free scalar particle. The remaining terms describe the influence of the environment and encode in general different physical processes like energy shifts, dissipation and decoherence. While the expressions in lines two and three only depend on the state $\rho(\vec u,\vec v,t)$ considered at time $t$, the last line links this state to other states $\rho(\vec u+\vec k,\vec v+\vec k,t)$ at time $t$. 
This master equation however still has some problematic contributions that possess UV-divergences and hence needs to be UV-renormalised, as will be discussed in the next section.

\section{Renormalisation of the TCL one-particle master equation}\label{sec:renorm}
Upon investigation of the individual contributions in \eqref{eq:meqopf} it becomes evident that some terms exhibit divergences as will be discussed in detail below.  This raises the question of at what stage of the derivation of the master equation the renormalisation procedure should be carried out. In the literature, there are different strategies how to deal with these divergent contributions. For gravitationally induced decoherence, they have often not been computed in detail due to the reason that they are expected to not modify decoherence but only influence the unitary evolution, see for instance the discussion in \cite{Blencowe:2012mp,Oniga:2015lro}. In \cite{Anastopoulos:2013zya,Lagouvardos:2020laf} the renormalisation of these contributions has been performed in the end after a Markov and rotating wave approximation have already been applied.
~\\
~\\
 In this work we choose the strategy to renormalise the master equation first before applying further approximations or deriving physical implications. It will turn out that effects predicted with a non-renormalised master equation might get modified or even vanish when working with the renormalised version instead. An example of this kind is also discussed in \cite{Domi:2024ypm}, where a quantum mechanical toy model for gravitationally induced  decoherence based on the model in \cite{Xu:2020lhc} is applied in the context of neutrino oscillations. In that case the necessary renormalisation is very trivial compared to the model considered in this work. There, the renormalisation causes the contributions of the Lamb-shift Hamiltonian to cancel exactly. Consequently, all physical implications involving contributions of the Lamb-shift Hamiltonian, as discussed for example in \cite{Benatti:2001fa}, would be absent when working with the renormalised model presented in \cite{Domi:2024ypm}.
 ~\\
~\\ 
  In order to carry out the renormalisation, we will first identify the diverging terms. As we will discuss below in more detail, these are in particular the terms in the second and third line of the master equation \eqref{eq:meqopf} that will also be present in the case where the temperature parameter vanishes, that is for $\Theta=0$, in which the thermal state merges into a vacuum state. They are of the form $\int d^3k \frac{1}{|\vec k|^3}$ and thus yield a logarithmic UV-divergence. Once these contributions are identified, we express them in the form of Feynman diagrams of the underlying effective QFT. For this purpose, we follow the strategy in \cite{Burrage:2018pyg}, where a master equation for a scalar field with an environment consisting of another scalar field is presented. Here the treatment is extended so that the linearised gravitational field can be included as an environment. We will proceed in five steps: first in subsection \ref{uvren1} we will identify the divergent contributions in the master equation and then present the corresponding Feynman rules following from the underlying effective QFT based on the non-covariant formulation in subsection \ref{uvren2}. Afterwards in subsection \ref{uvren3} we provide a set of equivalent, covariant Feynman rules in terms of which we perform the renormalisation of the divergent contributions in subsection \ref{uvren4}. Finally we discuss the resulting renormalised one-particle master equation in subsection \ref{uvren5}.\\

\subsection{Identification of the UV divergences in the one-particle master equation}\label{uvren1}
Starting from the master equation in \eqref{eq:meqopf}, we want to investigate which terms on the right-hand side are UV-divergent with respect to the $\int_{\mathbb{R}^3} d^3k$ integration. As the projector $P_{ijln}(\vec{k})$ is independent of the absolute value of $\vec k$, it does not influence the UV behaviour. Then, one can identify four different types of contributions in the integrands after performing the $\tau$-integration and introducing the following sign factors $\sigma,\sigma_1,\sigma_2\in\{\pm 1\}$:
\begin{itemize}
\item[(a)] $\frac{1}{\omega_{u-k}\Omega_k} \frac{1}{\Omega_k+\omega_{u-k}+\sigma\omega_u}$: for large $|\vec{k}|$, i.e. for $|\vec{k}| >> |\vec{u}|,m$ this term becomes $\frac{1}{|\vec{k}|^3}$ and thus leads to a logarithmic UV-divergence under the integral.
\item[(b)] $\frac{1}{\omega_{u-k}\Omega_k} \frac{1}{\Omega_k+\omega_{u-k}+\sigma\omega_u} \rho(\vec{u}+\vec{k},\vec{v}+\vec{k},t)$: assuming that $\rho(\vec{x},\vec{y},t)$ is a proper, normalisable density matrix in position space for which the Fourier transform exists leads to the requirement that $\rho(\vec{u},\vec{v},t)$ has to decrease rapidly for large $\vec{u},\vec{v}$. Therefore this expression is UV-finite.
\item[(c)] $N(k)\frac{1}{\omega_{u-k}\Omega_k} \frac{1}{\Omega_k+\sigma_1\omega_{u-k}+\sigma_2\omega_u}$: a series expansion of the denominator of $N(k)$ yields $N(k) = \frac{1}{\beta |\vec{k}| + O(|\vec{k}|^2)}$. Hence, this term tends to zero for large $|\vec{k}|$ and also the combination $\frac{x^n}{e^x-1}$ decreases rapidly for $x\rightarrow\infty$, thus this kind of contribution is UV-finite.
\item[(d)] $N(k)\frac{1}{\omega_{u-k}\Omega_k} \frac{1}{\Omega_k+\sigma_1\omega_{u-k}+\sigma_2\omega_u} \rho(\vec{u}+\vec{k},\vec{v}+\vec{k},t)$: this contribution is a combination of cases (b) and (c) and also UV-finite.
\end{itemize}
From this analysis follows that the expressions involving $N(k)$, that would be absent in the vacuum case and are thus denoted as thermal contributions in the following,  are all UV-finite. Some of the vacuum contributions, these are the ones that do not involve $N(k)$, lead to  UV-divergences which we want to cure by a renormalisation. To achieve this, in the next section we show in a first step that these terms correspond exactly to the self-energy diagrams for the scalar particle in the form of Feynman diagrams.

\subsection{Non-covariant Feynman rules and self-energy}\label{uvren2}
In this section we present the Feynman rules in non-covariant form corresponding to the effective quantum field theory containing a scalar field coupled to linearised gravity in \cite{Fahn:2022zql}, where the latter is considered as the environment, which is the basis for the master equation in \eqref{eq:meqopf}. Then, we rewrite these rules in the next section in a covariant form to be able to follow the strategy of \cite{Burrage:2018pyg}, where a suitable renormalisation for a master equation for a scalar fields with a second scalar field as the environment is discussed. Here, we slightly extend these methods in order to apply them to the case where the linearised gravitational field is treated as the environment. The Feynman rules can be constructed from \cite{Fahn:2022zql}:
\begin{itemize}
\item The \textbf{scalar field} has the standard \textbf{propagator}, which follows from its quantised mode expansion in (3.4) and (3.5) in \cite{Fahn:2022zql}, which we denote by a solid line and which reads in the mostly plus signature convention:
\begin{equation}
\feynmandiagram [horizontal=a to b] {
  a -- [plain] b,
}; = \frac{-i}{k^2+m^2-i\epsilon}\,.
\end{equation}
Here, $k^2=-\left(k^0\right)^2 + \vec{k}^2$.

\item The \textbf{propagator of the triad field} was derived in \cite{Fahn:2022zql} in equation (4.46) and is denoted in terms of a curved line:
\begin{equation}
\feynmandiagram [horizontal=a to b] {
  a -- [boson] b,
}; = \frac{1}{\kappa} {P}\t{}{^a^b^c^d}(\vec{k}) \left[ \frac{-i}{k^2-i\epsilon} +2\pi N(k) \delta(k^2)\right] \,,
\end{equation}
where the first summand is the vacuum and the second one the thermal part.
The existing tensor structure manifests in the form of the tensor structure of the STT-projector defined above in \eqref{eq:ttproj}. When contracted with a quantity that is symmetric in $(ab)$ as well as  in $(cd)$, like it is the case for the interaction vertex introduced below, the STT-projector reduces to
\begin{align}
{P}\t{}{^a^b^c^d}(\vec{k}) = \delta^{ac} \delta^{bd}-\frac{1}{2}\delta^{ab}\delta^{cd} +\frac{1}{2\vec{k}^4} k^ak^bk^ck^d -\frac{1}{\vec{k}^2}\Bigg(2\delta^{ac} k^bk^d -\frac{1}{2}\delta^{ab} k^c k^d -\frac{1}{2}\delta^{cd} k^a k^b \Bigg) \,.
\end{align}

\item The coupling between the scalar field and linearised gravity is encoded in the interaction part of the total action in \cite{Fahn:2022zql} that is given as a reformulation of equation (3.17) in that work by:
\begin{align}
S_{int} &=- \int dt \int d^3x\; \mathcal{H}_{int}(\vec x,t) \nonumber\\
&= -\int dt \int \frac{d^3k}{\vec{k}^2}\; \kappa \bigg\{\vec{k}^2\: {\delta \mathcal{E}}{}_{ab}(\vec k,t)\: T^{ab}(-\vec k,t) - \frac{1}{2} T_{00}(\vec{k},t) T_a^a(-\vec{k},t)\nonumber\\ &\hspace{1.3in} -\frac{1}{4} T_{00}(\vec{k},t) T_{00}(-\vec{k},t)+T_{0a}(\vec{k},t) T_{0b}(-\vec{k},t) \left( \delta^{ab} -\frac{k^a k^b}{4\vec{k}^2}\right) \bigg\}\,,\label{eq:intact}
\end{align}
where $\delta \mathcal{E}_{ab}(\vec{k},t)$ denotes the Fourier transform of the observable for the linearised densitised triad $\delta \tensor{E}{^a_i}(\vec{x},t)$, that is in this case the physical degrees of freedom of the linearised gravitational field expressed in Ashtekar variables. Furthermore, $T^{\mu\nu}(\vec k,t)$ denotes the Fourier transform of the scalar field's energy momentum tensor\footnote{In position space, its components are given by \begin{align*}
T^{00} &= T_{00} =  \frac{1}{2} \left[ \pi^2 + (\t{\partial}{_a} \phi)\, (\partial^a \phi) + m^2\phi^2\right] = \epsilon(\phi,\pi)\,,\\
T^{0a}&=T^{a0} = - \delta^{ab} \,T_{0b}=-\delta^{ab}\, \pi\, \partial_b \phi =- \delta^{ab} \,p_b(\phi,\pi)\,,\\
T^{ab} &= \delta^{ac}\, \delta^{bd}\, T_{cd} = \frac{\delta^{ab}}{2} \left[\pi^2 - (\partial_c\phi)\, (\partial^c \phi) -m^2 \phi^2 \right] + (\partial^a \phi) \, (\partial^b \phi)\,.
\end{align*}
}.
The interaction vertex between the scalar and the triad field can then be read off and is related to $T_{ab}$. 
Due to the fact that $T_{ab}$ depends on derivatives of the scalar field, the expression for the triad-scalar-field vertex is different depending on the direction of the momenta involved in the diagrams. Considering the Fourier transform of $T_{ab}(\vec x,t)$, where the the scalar fields can be factorised, we find the expression $\widetilde{{T}}{}_{ab}(p,q) \phi(p) \phi(q)$ with 
\begin{equation}
\label{eq:tildeTab}
\widetilde{{T}}{}_{ab}(p,q) := \frac{1}{2} \delta_{ab} [-p_0 q_0 + \vec{p}\cdot \vec{q} -m^2] -\frac{1}{2} [p_a q_b + p_b q_a]\,.
\end{equation}
Hence, the \textbf{triad-scalar-field-vertex} is given by\\
\begin{tikzpicture}
  \begin{feynman}[small]
    \vertex (a);
    \vertex [left=of a] (b);
    \vertex [above left=of b] (d);
    \vertex [below left=of b] (c);   
    \diagram* {
      (a) -- [boson] (b) -- [plain, reversed momentum'=\(p\)] (d),
      (b) -- [plain, reversed momentum=\(q\)] (c),
    };
  \end{feynman}
  \hspace{1em}$=$\hspace{1em}
  \begin{feynman}[small]
    \vertex (a);
    \vertex [right=of a] (b);
    \vertex [above right=of b] (d);
    \vertex [below right=of b] (c);   
    \diagram* {
      (a) -- [boson] (b) -- [plain, momentum=\(p\)] (d),
      (b) -- [plain, momentum'=\(q\)] (c),
    };
  \end{feynman}
  \hspace{5em}$= i\kappa {\tilde{{T}}}{}_{ab}(\sigma_p p,\sigma_q q)\,,$
\end{tikzpicture}
\\
where $\sigma_p$ is $+$ if particle $p$ is incoming and $-$ if it is outgoing.

\item The remaining terms in the interaction part of the action \eqref{eq:intact} give rise to an additional \textbf{second order vertex} which cannot be split into first order vertices due to the lack of a suitable intermediate particle in this effective field theory. They have the form
\begin{align}
-&\kappa \int d^3x \int d^3y \frac{-\frac{1}{4}T_{00}(\vec{x})T_{00}(\vec{y}) + T_{0a}(\vec{x})T_{0b}(\vec{y})\delta^{ab} -\frac{1}{4} \left( T_{00}(\vec{x}) T_a^a(\vec{y}) + T_a^a(\vec{x}) T_{00}(\vec{y})  \right)}{4\pi ||\vec{x}-\vec{y}||} \nonumber\\  -&\kappa  \int d^3x \int d^3y \int d^3z \frac{-\frac{1}{4} \partial^a_{\vec{x}} T_{0a}(\vec{x}) \;\partial^b_{\vec{y}} T_{0b}(\vec{y}) }{(4\pi)^2 ||\vec{x}-\vec{z}|| ||\vec{y}-\vec{z}||}\,.\label{eq:nonlocintX}
\end{align}
The corresponding symmetrised Feynman rule reads: \\
\begin{tikzpicture}
  \begin{feynman}[small]
    \vertex (a);
    \vertex [left=of a] (b);
    \vertex [above left=of b] (g);
    \vertex [below left=of b] (c); 
    \vertex [above right=of a] (h);
    \vertex [below right=of a] (f);  
    \diagram* {
      (a) -- [scalar, edge label=\(k\)] (b) -- [plain, reversed momentum'=\(p\)] (g),
      (b) -- [plain, reversed momentum=\(q\)] (c),
      (a) -- [plain, momentum=\(u\)] (h), (a) -- [plain, momentum'=\(v\)] (f)
    };
  \end{feynman}
 \hspace{2em}$=-\frac{i \kappa}{\vec{k}^2} NI(p,q,u,v)$ 
\end{tikzpicture}\\
with \begin{align}
    NI(p,q,u,v):=\Bigg[-\frac{1}{4} \tilde{T}_{00}(p,q) \tilde{T}_{00}(-u,-v) + \tilde{T}_{0a}(p,q) \tilde{T}_{0b}(-u,-v) \left( \delta^{ab} -\frac{k^a k^b}{4\vec{k}^2} \right) \nonumber\\
  -\frac{1}{4} \widetilde{T}_{00}(p,q) \widetilde{T}_a^a(-u, -v) -\frac{1}{4} \widetilde{T}_a^a(p,q) \widetilde{T}_{00}(-u,-v) \Bigg] \,,
\end{align}
with $\vec k$ defined using momentum conservation as $\vec k=\vec p +\vec q = \vec u + \vec v$. Note that a similar vertex does also appear in QED when quantised in Coulomb gauge and there it represents the Coulomb interaction, see  for instance \cite{weinberg2005quantum,tong}.
\item As \textbf{external lines} we only have the scalar field in the cases we are interested in here. This follows the standard case for a quantised scalar field, the detailed expressions are however not required for the following discussions.
\end{itemize}

Equipped with these Feynman rules, we will now show in the subsequent section that the divergent contributions in the one-particle master equation can be identified exactly with the contribution of the self-energy diagram constructed with the above Feynman rules. In the model considered here this corresponds to the following Feynman diagram: 
\begin{equation}
\begin{tikzpicture}
  \begin{feynman}
    \vertex (a);
    \vertex [right=of a] (b);
    \vertex [right=of b] (c);
    \vertex [right=of c] (d);
    \diagram* {
      (a) -- [plain, edge label=\(u\)] (b) -- [plain, edge label'=\(u-k\)] (c) -- [plain, edge label=\(u\)] (d),
      (b) -- [boson, quarter left, edge label = \(k\)] (c)
    };
  \end{feynman}
\end{tikzpicture}
\end{equation}
With the Feynman rules introduced above, the amplitude represented by this diagram and denoted by $\Pi(u^2)$ has the form
\begin{align}
\Pi(u^2) :=& \int_{\mathbb{R}^4} \frac{d^4k}{(2\pi)^4}\frac{1}{\kappa} P^{abcd}(\vec{k}) \left[\frac{-i}{k^2-i\epsilon}+2\pi N(k)\delta(k^2) \right]\left[i\kappa\widetilde{T}_{ab}(u,-(u-k))\right] \nonumber\\ &\hspace{1in}\cdot \frac{-i}{(u-k)^2+m^2-i\epsilon}\left[i\kappa\widetilde{T}_{cd}(u-k,-u)\right]\nonumber\\
=&\kappa u_au_bu_cu_d\int_{\mathbb{R}^4} \frac{d^4k}{(2\pi)^4} P^{abcd}(\vec k)\left[ \frac{-i}{k^2-i\epsilon }+2\pi N(k) \delta(k^2)\right] \frac{1}{(u-k)^2+m^2-i\epsilon} \nonumber\\ =& \kappa \int_{\mathbb{R}^3} \frac{d^3k}{2(2\pi)^3} P_u(\vec{k}) \Bigg\{ \frac{(\Omega_k+\omega_{u-k})}{\Omega_k\omega_{u-k}(\Omega_k+\omega_{u-k}+u^0-i\epsilon)(\Omega_k+\omega_{u-k}-u^0-i\epsilon)}\nonumber\\
&\hspace{1.4in} - N(\Omega_k) \Bigg[ \frac{1}{\Omega_k(u^0-\Omega_k+\omega_{u-k}-i\epsilon)(u^0-\Omega_k-\omega_{u-k}+i\epsilon)} \nonumber\\
&\hspace{2in} +\frac{1}{\Omega_k(u^0+\Omega_k+\omega_{u-k}-i\epsilon)(u^0+\Omega_k-\omega_{u-k}+i\epsilon)} \Bigg]\Bigg\}\nonumber\\
=&\Pi_{vac}(u^2)+\Pi_\Theta(u^2)\,
\end{align}
In the first step the definition of $\widetilde{T}_{ab}$ in \eqref{eq:tildeTab} was used and in the second step the $k^0$-integration was performed, where for the vacuum part the residue theorem was applied. In the last step, we have defined the vacuum and thermal contribution to the self-energy as
\begin{align}
    \Pi_{vac}(u^2) :=& \kappa \int_{\mathbb{R}^3} \frac{d^3k}{2(2\pi)^3} P_u(\vec{k}) \Bigg\{ \frac{(\Omega_k+\omega_{u-k})}{\Omega_k\omega_{u-k}(\Omega_k+\omega_{u-k}+u^0-i\epsilon)(\Omega_k+\omega_{u-k}-u^0-i\epsilon)}\label{eq:sendiagvacpropdef} \\
    \Pi_\Theta(u^2) :=& -\kappa \int_{\mathbb{R}^3} \frac{d^3k}{2(2\pi)^3} P_u(\vec{k}) N(\Omega_k) \Bigg[ \frac{1}{\Omega_k(u^0-\Omega_k+\omega_{u-k}-i\epsilon)(u^0-\Omega_k-\omega_{u-k}+i\epsilon)} \nonumber\\
&\hspace{2in} +\frac{1}{\Omega_k(u^0+\Omega_k+\omega_{u-k}-i\epsilon)(u^0+\Omega_k-\omega_{u-k}+i\epsilon)} \Bigg]\Bigg\}\,.\label{eq:sendiagthermpropdef}
\end{align}
~\\
~\\
If we now want to identify contributions in the one-particle master equation with the self-energy, the following subtlety results: a key difference between the master equation in \eqref{eq:meqopf} and standard quantum field theory is that the latter is constructed for the limit $t_0\rightarrow -\infty$, $t\rightarrow\infty$ when evaluating scattering amplitudes. To take this into account, we apply the method presented in \cite{Burrage:2018pyg} for situations where there is a finite temporal interval. In this way, we can transform the self-energy diagram into the second line of the right-hand side of the master equation in \eqref{eq:meqopf}:
\begin{equation}\label{eq:cosconv}
    \Xi(\omega_u,\vec u,t_0,t):=\int_{t_0}^t d\tau \int_\mathbb{R} du^0\;\Pi(u^2) \cos[(u^0-\omega_u)(t-\tau)] = \int_{0}^{t-t_0} d\tau \int_\mathbb{R} du^0\;\Pi(u^2) \cos[(u^0-\omega_u)\tau]\,.
\end{equation}
The standard QFT-limit can be recovered, in which $t\rightarrow \infty$ and $t_0 \rightarrow -\infty$, and using this the integral over the temporal interval can be rewritten as a $\delta$-distribution as $\int_{0}^\infty d\tau \cos[(u^0-\omega_u)\tau] = \frac{1}{2} \int_{-\infty}^\infty d\tau e^{-i(u^0-\omega_u)\tau} = \pi \delta(u^0-\omega_u)$, that will set the external momentum $u$ on-shell. \\
After evaluating the $u^0$ integration we obtain for finite times $t$ and $t_0$:
\begin{align}
    \Xi(\omega_u,\vec u,t_0,t)=\frac{\kappa}{2} \int_{\mathbb{R}^3} \frac{d^3k}{(2\pi)^3} P_u(\vec{k})\frac{\pi i}{2\omega_{u-k}} \left[C(\vec u,\vec k,t-t_0) + C_P(\vec u,\vec k,t-t_0)\right]\,,
\end{align}
which indeed, multiplied by a factor $\frac{2i}{\pi\omega_u}$, can be identified with the first term on the right hand side of the second line of the one-particle master equation in \eqref{eq:meqopf} in the extended one-particle projection. Given this results, it is now also easy to discuss the case of the non-extended one-particle projection: the master equation for this case, i.e. for $\delta_P=0$, can just be obtained by replacing the cosine in \eqref{eq:cosconv} by $\frac{1}{2} e^{-i(u^0-\omega_u)\tau}$. We find that in the QFT-limit the difference between the extended and non-extended one-particle projection manifests itself in a factor of $2$. 
~\\
To obtain the second term in the second line of \eqref{eq:meqopf}, we can follow the same steps and just have to replace $\vec u$ by $\vec v$ and take the complex conjugate.\\
~\\
With the results in this section we have shown that the UV-divergent terms in the one-particle master equation correspond to the self-energies of the scalar particle. What remains is to discuss the renormalisation of this self-energy. In order to be able to apply the standard procedure for renormalisation in this case, however, the corresponding covariant Feynman rules of the model considered here must first be derived.

\subsection{Covariant Feynman rules}\label{uvren3}
To be able to employ the standard renormalisation technique for the loop associated with the scalar particle's self-energy, we introduce in this section the covariant Feynman rules corresponding to the effective QFT under consideration here. For this, we follow \cite{weinberg2005quantum,tong}, where the procedure is outlined for QED. 
~\\
~\\
In a first step we will demonstrate that specific sums of non-covariant Feynman diagrams add up to the corresponding covariant Feynman diagram. For this purpose, we consider the sum of the second order vertex in \eqref{eq:nonlocintX} with a second order combination of the non-covariant scalar field-triad vertex, shown below \eqref{eq:tildeTab}. As will be derived below, the second order vertex in \eqref{eq:nonlocintX} is precisely that term which restores covariance if we work with a fully covariant triad propagator and a covariant vertex. 

We will restrict our discussions mainly to a Coulomb-scattering type of diagram here which is sufficient for our later applications. At the end of this section we will also briefly discuss the diagram associated with the scalar particle's self-energy.
In the case of the Coulomb-scattering type diagram, the above mentioned equivalence in terms of Feynman diagrams reads
\\
\\
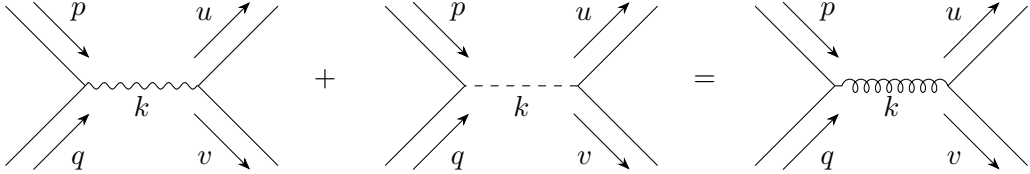
\begin{tikzpicture}
  \begin{feynman}
    \vertex (a);
    \vertex [left=of a] (b);
    \vertex [above left=of b] (g);
    \vertex [below left=of b] (c); 
    \vertex [above right=of a] (h);
    \vertex [below right=of a] (f);  
    \diagram* {
      (a) -- [boson, edge label=\(k\)] (b) -- [plain, reversed momentum'=\(p\)] (g),
      (b) -- [plain, reversed momentum=\(q\)] (c),
      (a) -- [plain, momentum=\(u\)] (h), (a) -- [plain, momentum'=\(v\)] (f)
    };
  \end{feynman} \hspace{4em} $ + $\end{tikzpicture} \hspace{3em}
  \begin{tikzpicture}
  \begin{feynman}
    \vertex (a);
    \vertex [left=of a] (b);
    \vertex [above left=of b] (g);
    \vertex [below left=of b] (c); 
    \vertex [above right=of a] (h);
    \vertex [below right=of a] (f);  
    \diagram* {
      (a) -- [scalar, edge label=\(k\)] (b) -- [plain, reversed momentum'=\(p\)] (g),
      (b) -- [plain, reversed momentum=\(q\)] (c),
      (a) -- [plain, momentum=\(u\)] (h), (a) -- [plain, momentum'=\(v\)] (f)
    };
  \end{feynman} \hspace{4em} $ = $
\end{tikzpicture}\hspace{3em}
\begin{tikzpicture}
  \begin{feynman}
    \vertex (a);
    \vertex [left=of a] (b);
    \vertex [above left=of b] (g);
    \vertex [below left=of b] (c); 
    \vertex [above right=of a] (h);
    \vertex [below right=of a] (f);  
    \diagram* {
      (a) -- [gluon, edge label=\(k\)] (b) -- [plain, reversed momentum'=\(p\)] (g),
      (b) -- [plain, reversed momentum=\(q\)] (c),
      (a) -- [plain, momentum=\(u\)] (h), (a) -- [plain, momentum'=\(v\)] (f)
    };
  \end{feynman}
\end{tikzpicture}\\
\\
where the curly line corresponds to the covariant triad propagator. 

Next, we will present the covariant Feynman rules, then specialise them to the case of the Coulomb-scattering type diagram to show the above equivalence. The corresponding covariant Feynman rules for the propagators and vertices discussed in the last section are as follows:
\begin{itemize}
\item The \textbf{scalar propagator} remains unchanged
\begin{equation}
\feynmandiagram [horizontal=a to b] {
  a -- [plain] b,
};=\frac{-i}{k^2+m^2-i\epsilon}\,.
\end{equation}

\item The \textbf{covariant triad propagator} becomes
\begin{equation}
\feynmandiagram [horizontal=a to b] {
  a -- [gluon] b,
};=\frac{1}{\kappa} \frac{-i}{k^2-i\epsilon} P_{\mu\nu\rho\sigma}
\end{equation}
with
\begin{equation}
P_{\mu\nu\rho\sigma}:= \frac{1}{2}[\eta_{\mu\rho}\eta_{\nu\sigma} + \eta_{\mu\sigma}\eta_{\nu\rho} - \eta_{\mu\nu} \eta_{\rho\sigma}]\,.
\end{equation}
In the context of a linearised gravitational environment there is no multi-triad vertex and therefore in the effective QFT considered in this work the triad propagator always couples only to the scalar field-triad-vertex, see also below. The latter is symmetric in $(\mu\nu)$ as well as  in $(\rho\sigma)$. This allows us to slightly simplify the projector $P_{\mu\nu\rho\sigma}$ whenever it occurs in combination with scalar field-triad-vertices and express it as
\begin{equation}
P_{\mu\nu\rho\sigma}:= \eta_{\mu\rho}\eta_{\nu\sigma} -\frac{1}{2} \eta_{\mu\nu} \eta_{\rho\sigma}\,.
\end{equation}

\item The \textbf{covariant vertex} is given by\\
\begin{tikzpicture}
  \begin{feynman}[small]
    \vertex (a);
    \vertex [left=of a] (b);
    \vertex [above left=of b] (d);
    \vertex [below left=of b] (c);   
    \diagram* {
      (a) -- [gluon] (b) -- [plain, reversed momentum'=\(p\)] (d),
      (b) -- [plain, reversed momentum=\(q\)] (c),
    };
  \end{feynman} \hspace{1em}$ = $
\end{tikzpicture} \hspace{2em}
\begin{tikzpicture}
  \begin{feynman}[small]
    \vertex (a);
    \vertex [right=of a] (b);
    \vertex [above right=of b] (d);
    \vertex [below right=of b] (c);   
    \diagram* {
      (a) -- [gluon] (b) -- [plain, momentum=\(p\)] (d),
      (b) -- [plain, momentum'=\(q\)] (c),
    };
  \end{feynman} \hspace{5em} $=\frac{-i\kappa}{2} (-2 \tilde{T}{}^{\mu\nu}(\sigma_p p,\sigma_q q) + \eta^{\mu\nu} \tilde{T}{}^\rho_\rho(\sigma_p p, \sigma_q q))\,,$
\end{tikzpicture}
\\ where
\begin{equation}
\widetilde{T}^{\mu\nu}(p,q)= \frac{1}{2}\eta^{\mu\nu} \left(p_\rho q^\rho - m^2\right) -\frac{1}{2} (p^\mu q^\nu + p^\nu q^\mu)\,.
\end{equation}
Whenever this is combined with a triad propagator, the second term of the vertex contribution vanishes because we have
\begin{equation}
\eta^{\mu\nu} \widetilde{T}^\alpha_\alpha [\eta_{\mu\rho}\eta_{\nu\sigma} + \eta_{\mu\sigma}\eta_{\nu\rho} - \eta_{\mu\nu} \eta_{\rho\sigma}] = 2\eta_{\rho\sigma} \widetilde{T}^\alpha_\alpha -2\eta_{\rho\sigma} \widetilde{T}^\alpha_\alpha=0\,,
\end{equation}
where we used that $\eta^{\mu\nu} \eta_{\mu\nu}=4$. Hence, for processes like Coulomb scattering we can replace the expression for the vertex with
\begin{equation}
i\kappa \widetilde{T}{}^{\mu\nu}(\sigma_p p,\sigma_q q)\,.
\end{equation}
\item Since the second order vertex was used to obtain a covariant propagator and vertex, there is no analogue of the second order vertex in the covariant case.
\item The \textbf{external lines} for the scalar field remain unmodified and the ones for the triad field are not important for this work.
\end{itemize}
These covariant Feynman rules are in accordance with the ones\footnote{Their vertex has one incoming and one outgoing scalar particle, hence is equivalent to $i\kappa\widetilde{T}^{\mu\nu}(p,-q)$ in the notation used here.} presented in \cite{Donoghue:1994dn,Arteaga:2003we}, where also a scalar field is coupled to a linearised gravitational field, except for the usual differences caused by the choice of different signatures for the metric, as they use the mostly minus signature. Using momentum conservation, i.e. $k=p+q=u+v$, yields
\begin{align}\label{eq:momcons1}
k^\mu \widetilde{T}_{\mu 0}(p,q) &= k^0 \widetilde{T}_{00}(p,q) + k^a \widetilde{T}_{a0}(p,q) = \frac{1}{2} [q_0 (p_0^2 -\vec{p}^2 -m^2) + p_0 (q_0^2 -\vec{q}^2 -m^2)]\,,\\\label{eq:momcons2}
k^\mu \widetilde{T}_{\mu a}(p,q) &= k^0 \widetilde{T}_{0a}(p,q) + k^b \widetilde{T}_{ba}(p,q) = \frac{1}{2} [q_a (p_0^2 -\vec{p}^2 -m^2) + p_a (q_0^2 -\vec{q}^2 -m^2)]\,.
\end{align}
If the scalar field is on-shell, which we assume for a moment for the Coulomb-scattering diagram, then the right hand side of both expressions vanishes. With this, one can directly show the equivalence of using the covariant set of Feynman rules for the Coulomb scattering diagrams discussed above of this. We do present this in appendix \ref{ap:equivFR}.
~\\
Note that in perturbation series, the diagram containing the two vertices is of second order in the expansion and hence obtains an additional factor $\frac{1}{2}$ compared to the second order vertex diagram.\\
~\\
As a next step we would like to show a similar equivalence between the non-covariant and covariant Feynman rules for the self-energy diagram, namely the following equivalence in terms of Feynman diagrams 
\\
\begin{tikzpicture}
  \begin{feynman}
    \vertex (a);
    \vertex [right=of a] (b);
    \vertex [right=of b] (c);
    \vertex [right=of c] (d);
    \diagram* {
      (a) -- [plain] (b) -- [plain] (c) -- [plain] (d),
      (b) -- [boson, quarter left] (c)
    };
  \end{feynman}
\end{tikzpicture} + 
\begin{tikzpicture}
  \begin{feynman}
    \vertex (a);
    \vertex [right=of a] (b);
    \vertex [right=of b] (c);
    \vertex [right=of c] (d);
    \diagram* {
      (a) -- [plain] (b) -- [plain] (c) -- [plain] (d),
      (b) -- [scalar, quarter left] (c)
    };
  \end{feynman}
\end{tikzpicture}
$=$
\begin{tikzpicture}
  \begin{feynman}
    \vertex (a);
    \vertex [right=of a] (b);
    \vertex [right=of b] (c);
    \vertex [right=of c] (d);
    \diagram* {
      (a) -- [plain] (b) -- [plain] (c) -- [plain] (d),
      (b) -- [gluon, quarter left] (c)
    };
  \end{feynman}
\end{tikzpicture}
~\\
Note that the diverging term actually only contains the first of the two Feynman diagrams on the left hand side. However, the second term is the self-energy contribution which vanishes in the one-particle projection of the master equation, so we can add this diagram as in the one-particle master equation its contribution vanishes for normal ordering. \\
The equivalence for the Feynman diagrams on both sides of this equation is however much more difficult to prove compared to the Coulomb scattering diagram, which is also the case in QED, since the momentum inside the loop is not on-shell, which prevents a similar calculations as done for the Coulomb scattering tree level graph. Given that, to prove this equivalence goes beyond the scope of this work here and
we refer here to the fact that the covariant set of Feynman rules can also be derived from the same underlying action using a different approach and gauge, which are then used for instance in \cite{Donoghue:1994dn,Arteaga:2003we}. Hence, independently of the derivation, we expect that they describe the same physics. Based on this, it is now possible to specify the expression corresponding to the scalar particle self-energy diagram in covariant form and renormalise it, which will be discussed in detail in the next section. As mentioned at the beginning of this section, such a replacement of non-covariant by covariant Feynman rules along the lines presented here is also employed in QED when quantising in Coulomb gauge, as for instance in \cite{weinberg2005quantum,tong}.

\subsection{UV-renormalisation of the self-energy of the scalar particle}\label{uvren4}
In terms of the Feynman rules introduced in the previous section, the self-energy diagram for the vacuum propagator, which was defined in \eqref{eq:sendiagvacpropdef}, can be expressed as 
\begin{align}
\Pi_{vac}(u^2) = & \int \frac{d^4k}{(2\pi)^4} \left[i\kappa\widetilde{T}^{\mu\nu}(u,-(u-k)) \right] \left[i\kappa\widetilde{T}^{\rho\sigma}(u-k,-u) \right] \frac{1}{\kappa} \frac{-i}{k^2-i\epsilon}  \nonumber\\ &\hspace{2in} \cdot \frac{1}{2} \left[\eta_{\mu\rho} \eta_{\nu\sigma} + \eta_{\mu\sigma}\eta_{\nu\rho} -\eta_{\mu\nu} \eta_{\rho\sigma}\right] \frac{-i}{(u-k)^2+m^2-i\epsilon}\nonumber\\
&= \frac{\kappa}{2} \int d^4k \frac{u^2 k^2 +2m^2 uk -2m^4}{[(k+u)^2-i\epsilon] [k^2+m^2-i\epsilon]} \,.
\end{align}
As the thermal part $\Pi_\Theta(u^2)$, defined in \eqref{eq:sendiagthermpropdef}, is not divergent, as it has been discussed in subsection \ref{uvren1}, we only consider the vacuum part here. For the renormalisation we follow the strategy in \cite{Hatfield:2019sox}. Using dimensional regularisation with $d=4-\epsilon$, the STT-projector is slightly modified in $d$ dimensions and reads (see e.g. \cite{Arteaga:2003we}):
\begin{equation}
P_{\mu\nu\rho\sigma}:= \frac{1}{2}[\eta_{\mu\rho}\eta_{\nu\sigma} + \eta_{\mu\sigma}\eta_{\nu\rho} - \frac{2}{2-\epsilon} \eta_{\mu\nu} \eta_{\rho\sigma}]\,.  
\end{equation}
Due to this, the expression for the self-energy diagram slightly changes and becomes (for $\epsilon \ll 1$):
\begin{align}
\Pi_{vac}(u^2) = \frac{\kappa}{2} \mu^\epsilon \int d^dk \frac{u^2 k^2 +2m^2 uk -2m^4 \left( 1+\frac{\epsilon}{4}\right)}{[(k+u)^2-i\epsilon] [k^2+m^2-i\epsilon]} \,,
\end{align}
which coincides with the expression derived in \cite{Arteaga:2003we}. Here, we rescaled $\kappa \rightarrow \kappa\mu^\epsilon$ to keep the dimension of $\kappa$ for any value of $d$. As later we will encounter also IR-divergences, we introduce a small artificial triad mass $\lambda$ in the triad propagator that becomes
\begin{equation}
    \frac{1}{\kappa} \frac{-i}{k^2+\lambda^2-i\epsilon} P_{\mu\nu\rho\sigma}\,.
\end{equation}
With this, the self-energy diagram reads
\begin{align}\label{eq:selfendiag}
\Pi_{vac}(u^2) = \frac{\kappa}{2} \mu^\epsilon \int d^dk \frac{u^2 k^2 +2m^2 uk -2m^4 \left( 1+\frac{\epsilon}{4}\right)}{[(k+u)^2+\lambda^2-i\epsilon] [k^2+m^2-i\epsilon]} \,,
\end{align}

This can then be evaluated using the standard methods for dimensional regularisation (see appendix \ref{sec:apLoopInt}). The result is that the divergent part can be isolated such that one obtains
\begin{align}
    \Pi_{vac}(u^2) = -\frac{2\pi^2\kappa m^2}{\epsilon} (m^2+u^2) + \Pi_{vac}^{\text{reg}}(u^2) 
\end{align}
with the finite part $\Pi_{vac}^{\text{reg}}(u^2)$. The infinite part then has to be renormalised by introducing a suitable counter term. As the finite part of this counter term can in principle be chosen arbitrarily, $\Pi_{vac}^{\text{reg}}(u^2)$ can still change. In our case, we choose the finite part of the counter term according to the on-shell renormalisation procedure. This then yields for the final renormalised loop $\Pi^R_{vac}(u^2)$: 
\begin{equation}\label{eq:rloopdefreg}
\Pi^R_{vac}(u^2) := \Pi_{vac}^{\text{reg}}(u^2) - \Pi_{vac}^{\text{reg}}(-m^2) - (u^2+m^2) \frac{\partial}{\partial u^2} \Pi_{vac}^{\text{reg}}(-m^2)\,.    
\end{equation}
This specific form is determined by the on-shell renormalisation scheme that sets the pole of the scalar propagator to $m^2$ and also fixes its residue according to the following two conditions:
\begin{align}
    \Pi^R_{vac}(u^2=-m^2)&\mbeq 0\\
    \frac{\partial}{\partial u^2}\Pi^R_{vac}(u^2=-m^2)&\mbeq 0\,.
\end{align}
It can readily be seen that the definition in \eqref{eq:rloopdefreg} satisfies these two conditions. Note that in \cite{Arteaga:2003we}, they apply a similar procedure without fixing the residue of the pole and therefore also not including an artificial triad mass, because for their purposes it is sufficient to fix the pole of the propagator.\\
The consideration above suggests that we have to include the following counter term:
\begin{equation}
    \delta \Pi(u^2) = \frac{2\pi^2 \kappa m^2}{\epsilon} (m^2+u^2) - \Pi_{vac}^{\text{reg}}(-m^2) - (u^2+m^2) \frac{\partial}{\partial u^2} \Pi_{vac}^{\text{reg}}(-m^2)
\end{equation}
such that
\begin{equation}
    \Pi(u^2) + \delta \Pi(u^2) = \Pi^R_{vac}(u^2) + \Pi_\Theta(u^2)\,,
\end{equation}
where $\Pi_\Theta(u^2)$ denotes the finite thermal contribution to the loop defined in \eqref{eq:sendiagthermpropdef}.
From \eqref{apploop:piregm2} we have for $\lambda\rightarrow 0$:
\begin{align}
    \Pi_{vac}^{\text{reg}}(-m^2) =0\,,
\end{align}
thus
\begin{equation}
    \delta \Pi(u^2) = \left[\frac{2\pi^2 \kappa m^2}{\epsilon} - \frac{\partial}{\partial u^2} \Pi_{vac}^{\text{reg}}(-m^2)\right] (m^2+u^2)\,,
\end{equation}
where the expression in the square brackets only depends on $m^2$. In order to implement a suitable counter term, we introduce a renormalised mass $m_R$ by $m^2 = m_R^2 + m_R^2 \delta_m$, where $m$ denotes the bare mass we have used so far and $\delta_m$ a mass counterterm, as well as a renormalised wave function $\varphi_R= \frac{1}{\sqrt{Z_2}}\varphi$. Then the renormalised scalar field propagator (a Greens function containing twice $\varphi$) reads up to the one-loop contribution:
\begin{equation}
iG^{(1)}(u^2) = \frac{1}{Z_2}\frac{-i}{u^2 + m^2} = \frac{-i}{u^2 +m_R^2 } + \frac{-i}{u^2+m_R^2}[-i(u^2 \delta_2 + m_R^2 (\delta_2 +\delta_m)) +\Pi(u^2)] \frac{-i}{u^2+m_R^2}+O(\kappa^2)\,,
\end{equation}
where we expanded $Z_2 = 1+\delta_2$. From this follows that 
\begin{equation}
    -i(u^2 \delta_2 + m_R^2 (\delta_2 +\delta_m)) \mbeq \delta \Pi(u^2)
\end{equation}
and it becomes evident that only the wave function has to be renormalised in the following manner:
\begin{align}\label{eq:renormctermsdef}
\delta_m = 0 \hspace{1in} \delta_2 = i\left[\frac{2\pi^2 \kappa m_R^2}{\epsilon} - \frac{\partial}{\partial u^2} \Pi^{\text{reg}}(-m_R^2)\right] \,.
\end{align}
Due to the counter-term, We have to replace in the old set of Feynman rules $m$ by $m_R$ and obtain the following additional Feynman rule of order $\kappa$:
\begin{equation}
\feynmandiagram [horizontal=a to b] {
  a -- c [crossed dot] -- b
}; = -i\delta_2 (u^2+m_R^2) \,.
\end{equation}
To simplify notation, we continue to use $m$, in particular as we have seen that $m_R=m$.
The result is therefore an additional counterterm in the Lagrangian which leads in renormalised perturbation theory to an additional interaction of order $\kappa$ that we have to include when evaluating the loop. Then the former diverging term $\Pi(u^2)$ becomes finite and only $\Pi^R_{vac}(u^2)$ is left.
This yields a modification of the right hand side of the master equation. As calculated in appendix \ref{secapren2}, the contribution of the renormalised vacuum loop terms to the master equation vanishes:
\begin{align}
 \Xi_R(\omega_u,\vec u,t_0,t)&=\int_{t_0}^t d\tau \int_\mathbb{R} du^0\;\Pi^R_{vac}(u^2) \cos[(u^0-\omega_u)(t-\tau)] =0  \,,\\
 \Xi_R(\omega_v,\vec v,t_0,t)&=\int_{t_0}^t d\tau \int_\mathbb{R} dv^0\;\Pi^R_{vac}(v^2) \cos[(v^0-\omega_v)(t-\tau)] =0  \,.
\end{align}
Therefore neither the dimensional constant $\mu$, nor the artificial triad mass $\lambda$ play a role in the physical predictions made with the master equation.

\subsection{Renormalised one-particle master equation}\label{uvren5}
With the renormalisation carried out in the previous subsections, a first renormalised version of the one-particle master equation \eqref{eq:meqopf}, where only the former diverging terms are modified, reads:
\begin{align}\label{eq:meqnrenY}
\frac{\partial}{\partial t} \rho(\vec{u},\vec{v},t) = &-i \rho(\vec{u},\vec{v},t) \; (\omega_u-\omega_v) \nonumber\\
&-\frac{\kappa}{2} \int \frac{d^3k}{(2\pi)^3} \bigg\{ \frac{ P_u(\vec k)}{\omega_{u-k}\omega_u}\; \left[C^R(\vec{u},\vec{k},t)+\delta_P C_P^R(\vec{u},\vec{k},t)\right] \nonumber\\ &\hspace{1.1in} +\frac{P_v(\vec k)}{\omega_{v-k}\omega_v}\; \left[\left(C^R(\vec{v},\vec{k},t)\right)^*+\delta_P \left(C_P^R(\vec{v},\vec{k},t)\right)^*\right]\bigg\}\;\rho(\vec{u},\vec{v},t)\nonumber\\
&+\frac{\kappa}{2} \int \frac{d^3k}{(2\pi)^3} \frac{P_{ijln}(\vec{k})\: u^i u^j  v^l v^n}{\sqrt{\omega_{u+k}\omega_u\omega_{v+k}\omega_v}}\; \left\{C(\vec{u}+\vec{k},\vec{k},t)+ C^*(\vec{v}+\vec{k},\vec{k},t)\right\}\: \rho(\vec{u}+\vec{k},\vec{v}+\vec{k},t) 
\end{align}
with
\begin{align}\label{eq:crenorm}
C^R(\vec{u},\vec{k},t) &= 2 \int_0^{t-t_0} \frac{d\tau}{\Omega_k} N(k)\:\cos[\Omega_k \tau] e^{ -i(\omega_{u-k}-\omega_u)\tau}\\
C_P^R(\vec{u},\vec{k},t) &= 2\int_0^{t-t_0} \frac{d\tau}{\Omega_k} N(k)\:\cos[\Omega_k \tau] e^{ -i(\omega_{u-k}+\omega_u)\tau}
\end{align}
and
\begin{align}\label{eq:Cunrenor}
C(\vec{u},\vec{k},t) &= \int_0^{t-t_0} \frac{d\tau}{\Omega_k} \Big\{ [N(k)+1]\: e^{- i(\Omega_k+\omega_{u-k}-\omega_u)\tau} +N(k)\;e^{ i(\Omega_k-\omega_{u-k}+\omega_u)\tau}\Big\}\,.
\end{align}
At the level of the operator equation, the renormalisation removed the $\Theta$-independent terms from the terms in the second and third line of \eqref{eq:meqnrenY}, hence leaving us with the following dissipator:
\begin{align}
\mathcal{D}[\rho_S]= - \frac{\kappa}{2}\sum_{r\in\{+,-\}} \sum_{a,b=1}^4 \int_{\mathbb{R}^3} \frac{d^3k\: d^3p\: d^3l}{(2\pi)^3}\frac{1}{\Omega_k} \Bigg\{ -\left( j_r^b(\vec{k},\vec{l})\rho_S(t)j^a_r(\vec{k},\vec{p})^\dagger\right) \; f(\Omega_k+\omega_b(\vec{k},\vec{l}))+h.c.\nonumber \\ + N(k) \; \left[j_r^a(\vec{k},\vec{p})^\dagger,\left[ j^b_r(\vec{k},\vec{l}),\rho_S(t)\right]\right]\; f(\Omega_k+\omega_b(\vec{k},\vec{l})) +h.c.\Bigg\}\,.
\end{align}
If working with the non-extended projection $\delta_P=0$, then there was probability conservation before the renormalisation, i.e. $\int d^3u \frac{\partial}{\partial t}\rho(\vec u,\vec u,t) =0$. Now, due to the vacuum term in \eqref{eq:Cunrenor} this probability conservation is destroyed. As the renormalisation is a purely technical procedure that should not change the physics, in particular not basic principles as probability conservation, we also replace $C(\vec u,\vec k,t)$ by $C^R(\vec u,\vec k,t)$ in the last line of the master equation. Another reason for this is that the term in the last line of the master equation is based on the same QFT as the terms in the second and third line, hence they should be renormalised in the same way\footnote{If one keeps these terms, they will drop from the dissipator part in the Markov approximation and from the Lamb-shift Hamiltonian after the rotating wave approximation, hence not form part of a Lindblad equation derived using these two approximations.}. The final renormalised one-particle master equation is thus
\begin{align}\label{eq:meqren}
\frac{\partial}{\partial t} \rho(\vec{u},\vec{v},t) = &-i \rho(\vec{u},\vec{v},t) \; (\omega_u-\omega_v) \nonumber\\
&-\frac{\kappa}{2} \int \frac{d^3k}{(2\pi)^3} \bigg\{ \frac{ P_u(\vec k)}{\omega_{u-k}\omega_u}\; \left[C^R(\vec{u},\vec{k},t)+\delta_P C_P^R(\vec{u},\vec{k},t)\right] \nonumber\\ &\hspace{1.1in} +\frac{P_v(\vec k)}{\omega_{v-k}\omega_v}\; \left[\left(C^R(\vec{v},\vec{k},t)\right)^*+\delta_P \left(C_P^R(\vec{v},\vec{k},t)\right)^*\right]\bigg\}\rho(\vec{u},\vec{v},t)\nonumber\\
&+\frac{\kappa}{2} \int \frac{d^3k}{(2\pi)^3} \frac{P_{ijln}(\vec{k})\: u^i u^j  v^l v^n}{\sqrt{\omega_{u+k}\omega_u\omega_{v+k}\omega_v}}\; \left\{C^R(\vec{u}+\vec{k},\vec{k},t)+ \left(C^R(\vec{v}+\vec{k},\vec{k},t)\right)^*\right\} \rho(\vec{u}+\vec{k},\vec{v}+\vec{k},t) 
\end{align}
and the dissipator at operator level
\begin{align}\label{eq:renodisoplvl}
\mathcal{D}[\rho_S]= - \frac{\kappa}{2}\sum_{r\in\{+,-\}} \sum_{a,b=1}^4 \int_{\mathbb{R}^3} \frac{d^3k\: d^3p\: d^3l}{(2\pi)^3}\frac{N(k) }{\Omega_k} \Bigg\{ \left[j_r^a(\vec{k},\vec{p})^\dagger,\left[ j^b_r(\vec{k},\vec{l}),\rho_S(t)\right]\right]\; f(\Omega_k+\omega_b(\vec{k},\vec{l})) +h.c.\Bigg\}\,.
\end{align}
Compared to the expression before renormalisation given in \cite{Fahn:2022zql} in equation (4.60), the vacuum contribution in the dissipator vanishes. 
~\\
In \cite{Domi:2024ypm} a quantum mechanical model based on the model in \cite{Xu:2020lhc}
is considered where a system is coupled to an environment of harmonic oscillators. The bath of harmonic oscillators mimics the thermal gravitational waves and the model serves as toy model for gravitationally induced decoherence. In \cite{Domi:2024ypm} the system was then specified to neutrinos in order to investigate gravitationally induced decoherence in the context of neutrino oscillations. A more detailed discussion on the model from \cite{Domi:2024ypm} can be found in section \ref{sec:AppURel}.
~\\
The quantum mechanical master equation in that work consists of a dissipator term and a Lamb-shift, where the latter contained divergences and is finally removed by a renormalisation in \cite{Domi:2024ypm}. We are now interested to discuss the similarities and differences of that quantum mechanical renormalisation and the renormalisation of the one-particle master equation presented in this work. In the quantum mechanical case, the interaction Hamiltonian has the following form:
\begin{equation}
    \hat{H}_I = \hat{H}_S \otimes \sum_{i=1}^N g_i \hat{q}_i
\end{equation}
with the Hamiltonian $\hat{H}_S$, describing the neutrino propagation, coupling constants $g_i$ and the position operators of the harmonic oscillators in the environment $\hat{q}_i$. This provides a toy model for the coupling of the energy momentum tensor and the metric perturbations in the field theoretical model. The form of the coupling  in the quantum mechanical model implies that the coefficients $\Lambda$ of the Lamb-shift and $\Gamma$ of the dissipator in the final Lindblad equation only depend on the environment and are defined in the following way, see \cite{Domi:2024ypm} equations (4) and (5):
\begin{align}\label{eq:deflambqmvgl}
    \Lambda (t-t_0) &:=\int_0^{t-t_0} d\tau \int_0^\infty d\omega \ J(\omega;\Omega,\eta) \sin(\omega\tau) \\\label{eq:defgamqmvgl}
    \Gamma (t-t_0) &:= 2 \int_0^{t-t_0} d\tau \int_0^\infty d\omega \ J(\omega;\Omega,\eta) \cos(\omega\tau) \coth\left( \frac{\beta\hbar\omega}{2}\right)\,,
\end{align}
where a spectral density $J(\omega;\Omega,\eta)$ was used which depends on the frequencies of the harmonic oscillators denoted by $\omega$, an effective coupling parameter $\eta$ and a UV-cutoff frequency $\Omega$. From this form it becomes evident that the Lamb-shift term is independent of the temperature parameter $\beta=\frac{1}{k_B\Theta}$, where $k_B$ denotes  Boltzmann's constant. Thus the Lamb-shift contribution only encodes vacuum effects, while the prefactor $\Gamma$ of the dissipator depends on $\Theta$ and yields a non-vanishing contribution for $\Theta=0$. The renormalisation in this model applied in \cite{Domi:2024ypm} then removes the Lamb-shift contribution completely, as it depends on the unphysical cutoff frequency $\Omega$, while the prefactor of the dissipator is not altered, as here the dependency on $\Omega$ vanishes after the Markov approximation.
\\
\\
Next, let us discuss to what extent it is possible to connect the renormalisation and its effects of the one-particle master equation presented above with the renormalisation applied in the quantum mechanical toy model. For this purpose first we discuss the two forms of the original full field theoretical master equation derived in \cite{Fahn:2022zql}. The first form is given in that work in equation (4.52):
\begin{align}\label{eq:meqpap1form1}
\frac{\partial}{\partial t} \rho_S(t) =& -i \left[H_S+\kappa \: {U}, \rho_S(t)\right]\nonumber\\ &-\frac{\kappa}{2}\int_{0}^t ds \sum\limits_{r} \int_{\mathbb{R}^3} d^3k \; \bigg\{ i D(\vec{k},t-s) \left[{J}_{r}(\vec k),\left\{ J_{r}(-\vec k,s-t), {\rho}_S(t)\right\}\right] \nonumber\\&\hspace{2.0in} + D_1(\vec{k},t-s) \left[{J}_{r}(\vec k),\left[ J_{r}(-\vec k,s-t),{\rho}_S(t)\right]\right] \bigg\}\,,
\end{align}
where $D$ and $D_1$ are two coefficients that arise from combinations of the environmental correlation functions similar to $\Lambda$ and $\Gamma$ in the quantum mechanical model and they read:
\begin{align}\label{eq:diskern}
    D(\vec{k},t-s) &:=  -\frac{\sin(\Omega_k(t-s))}{\Omega_k}\\
\label{eq:noikern} D_1(\vec{k},t-s) &:= \coth\left( \frac{\beta \Omega_k}{2}\right)\,\frac{\cos(\Omega_k(t-s))}{\Omega_k}\,.
\end{align}
The operators $J_r(\vec{k},t)$ and $J_r(\vec{k}):= J_r(\vec{k},0)$ were defined in \cite{Fahn:2022zql} in equation (3.18) and contain a combination of two creation and/or annihilation operators of the scalar field along with their time evolution. As a first difference to the master equation in \cite{Domi:2024ypm} it turns out that the term proportional to $D$ in \eqref{eq:meqpap1form1} cannot be written as a simple commutator, as it is the case with the Lamb-shift contribution in the quantum mechanical toy model. If the system's operator $J$ were to commute with the system Hamiltonian, then this would be possible, and this is the case in the quantum mechanical toy model in \cite{Domi:2024ypm}. Then this would imply that the Lamb-shift is independent of the temperature parameter $\Theta$ and therefore a pure vacuum effect.
\\ 
For the field-theoretical model, a similar form where one has a Lamb-shift contribution and a dissipator is the one given at the beginning of Appendix \ref{Apoppsu1}. Here, the coefficients of the Lamb-shift term are $S_{ab}$ and the prefactors of the dissipator are $R_{ab}$. From their definitions in \eqref{eq:SabMatrix} and \eqref{eq:RabMatrix} one can see that in general they have a different form as $\Lambda$ and $\Gamma$ in \eqref{eq:deflambqmvgl} and \eqref{eq:defgamqmvgl} above and the Lamb-shift includes vacuum as well as thermal contributions. A similar result is obtained in \cite{Breuer:2002pc}, where quantum electrodynamics is discussed from the point of view of open quantum systems with the standard interaction Hamiltonian of QED. There, the resulting Lamb-shift Hamiltonian is therefore split into a vacuum part, denoted as Lamb-shift and a thermal part, denoted as Stark-shift. 
\\
As discussed above, we would expect that if the system's operator in the interaction Hamiltonian commutes with the system Hamiltonian, that we can then recover the form of the quantum mechanical model in \cite{Domi:2024ypm}. Indeed, if we had $[J,H_S]=0$, then the phases $e^{\pm i \omega_a(\vec{k},\vec{p}) t}$ and $e^{\pm i \omega_b(\vec{k},\vec{l}) t}$ coming from the time evolution of the $J$ operators would vanish in the definitions of $S_{ab}$ and $R_{ab}$ in \eqref{eq:SabMatrix} and \eqref{eq:RabMatrix}. This allows the remaining terms to be combined into a form similar to $\Lambda$ and $\Gamma$. In particular the thermal contribution of the Lamb-shift would vanish, as it is the case in the quantum mechanical toy model. 
\\
Let us now compare the renormalisations of the two models: in the quantum mechanical model in \cite{Domi:2024ypm}, the effect of the renormalisation is to remove the Lamb-shift Hamiltonian which only consisted of a vacuum part. The renormalisation applied in the present work removes the vacuum parts in the Lamb-shift Hamiltonian and the dissipator. The thermal part of the Lamb-shift Hamiltonian however remains. From the discussion of the open QED model from \cite{Breuer:2002pc}, one would expect a similar result in a quantum mechanical model where a thermal contribution in the Lamb-shift is present. The dissipator of the quantum mechanical model \cite{Domi:2024ypm} is left unmodified by the renormalisation, in particular the vacuum contribution is present there.  This is in contrast to the procedure here, where the renormalisation removes all vacuum terms, also the ones from the dissipator. In the quantum mechanical toy model, these contributions are however removed at a later stage when the Markov approximation is applied and hence also not present in the final Lindblad equation.
~\\~\\
This concludes the discussion on the renormalisation of the one-particle master equation. In the next section, we discuss how one can apply specific physical approximations to draw physical implications from the renormalised one-article master equation.

\section{Application of the Markov and rotating wave approximations to transform the TCL master equation into Lindblad form}\label{sec:approx}
The renormalised TCL one-particle master equation \eqref{eq:meqren} describes the evolution of a single scalar particle in an environment filled with thermal gravitational waves. Since this master equation is not in Lindblad form, we cannot directly conclude that it is completely positive and provides physically meaningful implications based on positive probabilities for all chosen time intervals. For such models, one usually has to investigate case by case whether further assumptions such as the Markov and rotating wave approximation are justified that are usually used to obtain a master equation in Lindblad form. It is often possible to understand from the involved time scales in the system and environment of the open quantum model in which scenarios these approximations are a good choice, see for example \cite{Breuer:2002pc} for a discussion in quantum optics. For models with finitely many degrees of freedom, there are also results that suggest time scales which allow to judge when the Markov approximation can be applied that are completely determined by the properties of the environment, such as its spectral density as well as the coupling constant, which encodes the strength of the coupling to the system in the interaction Hamiltonian \cite{Nathan:2020}.
~\\
The derivation of master equations in the context of field-theoretical models with gravity as an environment is less well explored in the literature in comparison and has been presented in the context of gravitationally induced decoherence recently for instance in  \cite{Anastopoulos:2013zya,Blencowe:2012mp,Oniga:2015lro,Lagouvardos:2020laf,Fahn:2022zql,fgke2024photon}. While the works in \cite{Blencowe:2012mp,Oniga:2015lro,Fahn:2022zql,fgke2024photon} focus on the derivation of a TCL master equation, in \cite{Anastopoulos:2013zya,Lagouvardos:2020laf} a Lindblad equation is used, for which further approximations are employed, among these the Markov approximation and the rotating wave approximation. 
~\\
Compared to the above-mentioned open quantum mechanical models for gravitationally induced decoherence, a detailed analysis of the applicability of such approximations is much more challenging and beyond the scope of this article. An important difference to the present work is that in \cite{Anastopoulos:2013zya,Lagouvardos:2020laf} the approximations are applied on the non-renormalised one-particle master equation. Given the results of the last section, we can instead perform the Markov and rotating wave approximations for the renormalised one-particle master equation and investigate whether applying these approximations before or after renormalisation leads to differences in the final one-particle master equation, considering both the extended and non-extended one-particle projection.
~\\
In this section we consider both approximations separately, in subsection \ref{sec_Map} we discuss the Markov approximation and in subsection \ref{sec_RWA} the rotating wave approximation. In addition, for the case of an ultra-relativistic limit, we also specify some conditions when the Markov approximation can be used for the model considered here.

\subsection{Markov approximation}\label{sec_Map}
The Markov approximation consists in the assumption that the correlation functions of the environment are strongly peaked around the initial time and decay rapidly. If this is given, the integral $\int_0^{t-t_0} d\tau$ over these environmental correlation functions has the main contribution from around their peak. Thus the error obtained when shifting the initial time $t_0 \rightarrow -\infty$ and therefore the upper integration limit $t-t_0 \rightarrow \infty$ is negligibly small. As a consequence, the parameters involved in the dissipator of the final Lindblad equation will no longer depend on the temporal coordinate.
~\\
~\\
As discussed above for a field theoretical model, even in the single particle sector, to develop generic criteria for which the Markov approximation can be applied that can easily be checked for a given model, is difficult. For instance the methods developed in \cite{Nathan:2020} strongly rely on the fact that the model is formulated in a quantum mechanical context. Motivated by the physical applications in section \ref{sec:AppURel} to ultra-relativistic particles, in particular neutrinos and their oscillations, as a first step, we investigate this special case more in detail in this context and present a condition under which the Markov approximation can be applied for the model under consideration. The details are discussed in appendix \ref{sec:applic2M} by analysing the individual parts of the master equation and where we show that a suitable condition for the applicability of the Markov approximation in the ultra-relativistic case is the requirement that 
\begin{equation}\label{eq:condmarkaprurel}
    t-t_0 \gg  \hbar\beta \gg \frac{\hbar}{c u}, \frac{\hbar}{c v} \hspace{0.2in} \text{and} \hspace{0.2in} u,v\gg m\,,
\end{equation}
where $u:= |\vec u|$, $v:= |\vec v|$ denote the absolute value of the particle's momentum, $m$ its mass and $c$ the speed of light. The reason why $\beta$ and thus the temperature parameter $\Theta$, which characterises the oscillator environment, are involved here is because we use a Gibbs state to trace out the environmental and thus gravitational degrees of freedom. The specific time scales then yield that for the application in \ref{sec:AppURel} the correction terms to the Markov approximation are negligible. If another than the ultra-relativistic case is considered, the above condition could be violated, and therefore a more comprehensive analysis is needed to understand when and under what conditions the Markov approximation can be applied, which we envisage for future work.
~\\
Before applying the Markov approximation to the renormalised one-particle master equation, we briefly discuss the main steps that are involved: the first step is to  perform the $\tau$-integration using the identity
\begin{align}\label{eq:2mapdpv}
\int_0^\infty d\tau \; e^{-i\omega \tau} = \int_{-\infty}^\infty d\tau \; \Theta(\tau) e^{-i\omega \tau} = 2\pi\int_{-\infty}^\infty dx \; \Theta(x) e^{-2\pi i \omega x} = \pi \delta(\omega) - PV\left(\frac{i}{\omega}\right)\,,
\end{align}
where $PV$ denotes the Cauchy principal value, and secondly the evaluation of the $\vec{k}$-integration which simplifies due to the first step. With this, the affected terms in the master equation can then be split into two classes: one class that consists of contributions involving the delta distribution that will yield a real contribution to the master equation and hence lead to decoherence. Another class that contains terms including the principal value, which result in an imaginary contribution that affects the unitary evolution. The detailed computation for the contributions leading to decoherence  can be found in appendix \ref{app:MarkDel}. Those contributions that involve the principal value are evaluated in appendix \ref{app:MarkPV}. The final result, given in \eqref{eq:meq2Markov}, takes the form:
\begin{align}\label{eq:meqnsmap}
\frac{\partial}{\partial t} \rho(\vec{u},\vec{v},t) = &-i \rho(\vec{u},\vec{v},t) \; (\omega_u-\omega_v) \nonumber\\
&+\frac{\kappa }{16\pi\beta} \Bigg\{ -\frac{10}{3} (u^2+v^2) +2 (\omega_u^2+\omega_v^2) -\frac{2}{\omega_u u} m^4 \,\text{arctanh} \left( \frac{u}{\omega_u}\right) - \frac{2}{\omega_v v} m^4 \,\text{arctanh} \left( \frac{v}{\omega_v}\right)\nonumber\\
    &\hspace{.7in} - \frac{\omega_u}{\omega_v} \left(v^2-3\frac{(\vec u \cdot \vec v)^2}{u^2}\right) \left[ \frac{2}{3} - \frac{m^2}{u^2} + \frac{m^4}{\omega_u u^3} \text{arctanh}\left(\frac{u}{\omega_u}\right)\right] \nonumber\\
    &\hspace{.7in}-\frac{\omega_v}{\omega_u} \left(u^2-3 \frac{(\vec u \cdot \vec v)^2}{v^2}\right) \left[ \frac{2}{3} -\frac{m^2}{v^2} + \frac{m^4}{\omega_v v^3} \text{arctanh}\left(\frac{v}{\omega_v}\right)\right]\Bigg\}\rho(\vec u,\vec v,t)\nonumber\\
    &+\frac{i\kappa}{2(2\pi)^3} \frac{1}{\sqrt{\omega_u \omega_v}} \int d^3k \; P_{ijln}(\vec k) \frac{u^i u^j v^l v^n}{\sqrt{\omega_{u+k}\omega_{v+k}}} \frac{N(k)}{\Omega_k } \rho(\vec u+\vec k,\vec v +\vec k,t) \nonumber\\ &\hspace{1in}\Bigg\{ \text{PV}\left( \frac{1}{\Omega_k + \omega_u - \omega_{u+k}} \right) - \text{PV}\left( \frac{1}{\Omega_k - \omega_u + \omega_{u+k}} \right) \nonumber\\
    &\hspace{1.1in}-\text{PV}\left( \frac{1}{\Omega_k + \omega_v - \omega_{v+k}} \right) + \text{PV}\left( \frac{1}{\Omega_k - \omega_v + \omega_{v+k}} \right) \Bigg\}\nonumber\\
    &-(1-\delta_P) \frac{i \kappa }{2(2\pi)^2} \lim_{\epsilon \rightarrow 0} \bigg[\frac{u^4}{\omega_u}\int_\epsilon^\infty dk \int_0^\pi d\theta\;  \sin^5(\theta) k N(k) \frac{1 - \frac{\omega_u}{\omega_{u-k}}}{k^2 - (\omega_{u-k} - \omega_u)^2}\nonumber\\
    &\hspace{1.5in} - \frac{v^4}{\omega_v}\int_\epsilon^\infty dk \int_0^\pi d\theta\;  \sin^5(\theta) k N(k) \frac{1 - \frac{\omega_v}{\omega_{v-k}}}{k^2 - (\omega_{v-k} - \omega_v)^2}\bigg]\rho(\vec u,\vec v,t)\,.
\end{align}
The contributions in lines two to four arose from the delta distributions and are real, so they cause decoherence in the evolution of the scalar particle. The remaining terms are imaginary and therefore contribute to the unitary evolution of the density matrix. When working with the extended one-particle projection, then the expressions in the last two lines vanish. The real part in lines two to four remains unaffected by the rotating wave approximation that will be applied in the next subsection, hence it already possesses its final form. 
~\\
In the existing literature, the master equation in the one-particle picture is usually directly specified or derived for the non-relativistic (see e.g. \cite{Anastopoulos:2013zya} for scalar particles) or the ultra-relativistic case (see e.g. \cite{Lagouvardos:2020laf} for photons). In these cases, the dissipator has a simpler form and there are no arctanh-terms present as it is the case here in the general master equation, where neither the non- nor the ultra-relativistic limits have been applied yet. 

To further compare with the existing literature, in section \ref{sec:AppURel} we will consider the non- and ultra-relativistic limit of this master equation above and show that the arctanh does not appear in either limit. Thus, it indeed leads to the results obtained in the literature.

Another difference to similar work in \cite{Anastopoulos:2013zya,Lagouvardos:2020laf} is that the master equation in the present work has already been renormalised, i.e. all vacuum contributions that are independent of the temperature parameter $\Theta$ have been removed in these terms, which arise due to the gravitational influence in \eqref{eq:meqnsmap}, which in particular contains all vacuum fluctuations of the gravitational field. This can be seen by setting the temperature parameter equal to zero, as then all terms including the gravitational influence vanish. 

\subsection{Rotating wave approximation}\label{sec_RWA}
After having applied the Markov approximation, the rotating wave approximation is usually a next step in order to cast the master equation into a completely positive Lindblad form. The physical idea behind the approximation is to take into account that detectors only have a finite resolution and cannot resolve arbitrarily fast oscillations, but only measure a coarse-grained result. In the literature, there exist different ways to apply the rotating wave approximation. One possibility is, following the nomenclature in \cite{fleming2010rotating}, the pre-trace RWA, where the approximation is applied at the level of the interaction Hamiltonian by dropping counter-rotating terms. This is often employed e.g. in quantum optics and leads to the Jaynes-Cummings model, see \cite{Jaynes:1963zz,shore1993jaynes}, which is nowadays extensively studied for instance in quantum technology, see \cite{larson2021jaynes}. This pre-trace RWA, which is also applicable in closed quantum systems, has been studied from several angles yielding different results in the last years among other things on its higher order corrections and a renormalisation of the resulting series (see \cite{Wang:2023dkf}) as well as also on the bounds of its applicability (see \cite{Burgarth:2023ppw}). From the analysis in \cite{fleming2010rotating} it follows that in open quantum systems the second version of the RWA, the post-trace rotating wave approximation which is applied at the level of the master equation after tracing out the environment, yields dynamics which are expected to be closer to the true system dynamics. This analysis in \cite{fleming2010rotating} is carried out for quantum mechanical models and we expect that more work is required to extend it to the full field theoretical case. Nevertheless, we take this discussion as a motivation to apply in this work the post-trace RWA, which was also employed in similar analyses, for instance in \cite{Lagouvardos:2020laf}. This post-trace RWA is implemented by considering the master equation in the interaction picture and then dropping terms that rotate very fast compared to the other ones. 
The detailed implementation and computation can be found in appendix \ref{apRWA}, here we only state the result:
\begin{align}\label{eq:finalmeqaraaa}
\frac{\partial}{\partial t} \rho(\vec{u},\vec{v},t) = &-i \rho(\vec{u},\vec{v},t) \; (\omega_u-\omega_v) \nonumber\\
&+\frac{\kappa }{16\pi\beta} \Bigg\{ -\frac{10}{3} (u^2+v^2) +2 (\omega_u^2+\omega_v^2) -\frac{2}{\omega_u u} m^4 \,\text{arctanh} \left( \frac{u}{\omega_u}\right) - \frac{2}{\omega_v v} m^4 \,\text{arctanh} \left( \frac{v}{\omega_v}\right)\nonumber\\
    &\hspace{.7in} - \frac{\omega_u}{\omega_v} \left(v^2-3\frac{(\vec u \cdot \vec v)^2}{u^2}\right) \left[ \frac{2}{3} - \frac{m^2}{u^2} + \frac{m^4}{\omega_u u^3} \text{arctanh}\left(\frac{u}{\omega_u}\right)\right] \nonumber\\
    &\hspace{.7in}-\frac{\omega_v}{\omega_u} \left(u^2-3 \frac{(\vec u \cdot \vec v)^2}{v^2}\right) \left[ \frac{2}{3} -\frac{m^2}{v^2} + \frac{m^4}{\omega_v v^3} \text{arctanh}\left(\frac{v}{\omega_v}\right)\right]\Bigg\}\rho(\vec u,\vec v,t)\nonumber\\
    &-(1-\delta_P) \frac{i \kappa }{2(2\pi)^2} \lim_{\epsilon \rightarrow 0} \bigg[\frac{u^4}{\omega_u}\int_\epsilon^\infty dk \int_0^\pi d\theta\;  \sin^5(\theta) k N(k) \frac{1 - \frac{\omega_u}{\omega_{u-k}}}{k^2 - (\omega_{u-k} - \omega_u)^2}\nonumber\\
    &\hspace{1.5in} - \frac{v^4}{\omega_v}\int_\epsilon^\infty dk \int_0^\pi d\theta\;  \sin^5(\theta) k N(k) \frac{1 - \frac{\omega_v}{\omega_{v-k}}}{k^2 - (\omega_{v-k} - \omega_v)^2}\bigg]\rho(\vec u,\vec v,t)\,.
\end{align}
It can be seen when comparing this result to \eqref{eq:meqnsmap} that the effect of the rotating wave approximation is to remove the remaining part of the Lamb-shift in the extended projection. In the non-extended projection, there still survives one term of the Lamb-shift which corresponds to the last two lines in \eqref{eq:finalmeqaraaa}. Apart from that, the rotating wave approximation causes no further modifications on the master equation. This is due to the fact that all other terms that would be removed by the approximation were already dropped when performing the one-particle projection of the master equation. The general dissipator at the operator level can however be written in Lindblad form after the RWA, see \eqref{eq:qplbfdis} in the appendix:
\begin{equation}
\mathcal{D}_\delta[\rho_S]= \kappa\sum_{r\in\{+,-\}} \int_{\mathbb{R}^3} \frac{d^3k}{(2\pi)^2}\delta(k) \frac{N(k)}{\Omega_k} \left(L_r(\vec{k})\rho L_r(\vec{k})^\dagger - \frac{1}{2} \left\{ \rho, L_r(\vec{k})^\dagger L_r(\vec{k})\right\} \right)
\end{equation}
with Lindblad operators
\begin{equation}
    L_r(\vec k) := \int_{\mathbb{R}^3} d^3p \; \frac{1}{\sqrt{1-\frac{(k+p\cos(\theta_p))^2}{\omega_{k+p}^2}}} J_r^2(\vec k,\vec p)\,,
\end{equation}
where $\theta_p$ denotes the angle between $\vec{k}$ and $\vec{p}$ and $J_r^2(\vec{k},\vec{p}) = 2 j_r^1(\vec{k},\vec{p})$ with the latter being defined in \eqref{eq:defjundw}. With this, we have derived the final form of the renormalised one-particle master equation after Markov and rotating wave approximation. In the next section, we discuss some applications and investigate some features of the master equation at different intermediate stages before, during and after the applied approximations.

\section{Applications of the one-particle master equation}\label{sec:appl}
In this section, we discuss some applications of the one-particle master equation derived in the previous sections. We start with analysing the evolution of the populations of the one-particle density matrix with a special focus on the interplay between the renormalisation and Markov and rotating wave approximations in subsection \ref{sec_Pop} and compare the results to \cite{Oniga:2015lro} where the evolution of the populations of the non-renormalised TCL master equation is derived. Next we discuss the non-relativistic limit of the one-particle master equation in subsection \ref{ap:Nonrandurel} and compare the results to the ones in \cite{Anastopoulos:2013zya}. Furthermore, we investigate the ultra-relativistic limit in section \ref{sec_apurel}, compare it to \cite{Lagouvardos:2020laf}, and discuss the relation to the quantum mechanical model for gravitationally induced decoherence in neutrino oscillations in \cite{Domi:2024ypm} in section \ref{sec:AppURel}. This further allows to connect to phenomenological models that investigate the influence of gravity on neutrino oscillations, like for instance in \cite{Benatti:2000ph,Lisi:2000zt,Guzzo:2014jbp,BalieiroGomes:2018gtd}.

\subsection{Evolution of the populations of the one-particle master equation}\label{sec_Pop}
We start by analysing the dynamics of the populations, that is the diagonal elements, in momentum representation predicted by the master equation at different stages in the derivation of the final Lindblad equation. We have chosen this application because it is an example that allows us to discuss and compare the implications that arise depending on the stage of the calculation at which the renormalisation procedure is performed. 
~\\
To investigate the evolution of the populations, we take the different versions of the master equation and compute it for $\rho(\vec k,t) := \rho(\vec k,\vec k,t)$ before and after the renormalisation as well as after the Markov approximation. As the rotating wave approximation only affects the Lamb-shift Hamiltonian, the dynamics of the populations will not get modified after its application.

\subsubsection{Before renormalisation}\label{sec:beforeRenorm}
The dynamics of the populations in the one-particle master equation \eqref{eq:meqopf} before renormalisation and further approximations can be obtained by evaluating the master equation for $\rho(\vec k,t) := \rho(\vec k,\vec k,t)$. In this case, we have no contribution from the unitary dynamics and in the dissipator all imaginary parts will vanish\footnote{As all coefficients now enter in the form $C+C^*$.} and one obtains a dissipator that is purely real. In this subsection we adapt the notation to the one used in \cite{Oniga:2015lro} in order to better facilitate the comparison with their results:
\begin{align}\label{eq:diagterms1}
\dot{\rho}(\vec{k},t)= -\kappa \int \frac{d^3k'}{(2\pi)^3} &\frac{P_k(\vec{k}'-\vec{k})}{\Omega_{k'-k}\omega_{k'}\omega_k}\cdot\nonumber\\ \cdot \Bigg\{ &\Bigg[ [N(k'-k)+1]\:\frac{\sin({\chi} (t-t_0))}{{\chi}} +N(k'-k)\frac{\sin({\chi}' (t-t_0))}{{\chi}'}\nonumber\\ &+\delta_P [N(k'-k)+1]\:\frac{\sin({\eta} (t-t_0))}{{\eta}} +\delta_P N(k'-k)\frac{\sin({\eta}' (t-t_0))}{{\eta}'} \Bigg]\rho(\vec{k},t)\nonumber\\
&- \left[ [N(k'-k)+1]\:\frac{\sin({\chi'} (t-t_0))}{{\chi'}} +N(k'-k)\frac{\sin({\chi} (t-t_0))}{{\chi}}\right]\:\rho(\vec{k}',t)\Bigg\}
\end{align}
with $\dot{\rho}(\vec{k},t)=\partial_t\rho(\vec{k},t)$ and $\chi := \Omega_{k'-k} -\omega_k + \omega_{k'}$, $\chi' := \Omega_{k'-k} +\omega_k - \omega_{k'}$, $\eta := \Omega_{k'-k} +\omega_k + \omega_{k'}$ and $\eta' := \Omega_{k'-k} -\omega_k - \omega_{k'}$ as well as $P_k(\vec{k}'-\vec{k}) = P_{ijln}(\vec{k}'-\vec{k}) k^i k^j k^l k^n$. Using that $P_{ijln}(\vec k -\vec k') (\vec k-\vec k')^i =0$ as $P_{ijln}(\vec{k}-\vec{k}')$ projects onto the symmetric transverse traceless part and therefore removes the longitudinal part $\propto \vec{k}-\vec{k}'$, which can be seen from the definition in \eqref{eq:ttproj}, we can use
\begin{equation}
    P_{ijln}(\vec k -\vec k') k^i = P_{ijln}(\vec k -\vec k') k^{\prime i}
\end{equation}
and hence rewrite
\begin{equation}
    P_k(\vec{k}'-\vec{k}) = P_{ijln}(\vec{k}'-\vec{k}) k^i k^j k^{\prime l} k^{\prime n}\,.
\end{equation}
From equation \eqref{eq:diagterms1} one can also once more see the implication of the chosen projection, i.e. whether $\delta_P=0$ or $\delta_P=1$,  on the probability conservation, which was discussed below equation \eqref{eq:probcons}. When working with the non-extended projection $\delta_P=0$, then we have
\begin{equation}
    \int_{\mathbb{R}^3} d^3k \; \dot{\rho}(\vec{k},t) = 0
\end{equation}
due to symmetry and thus probability in the scalar particle's subsystem is conserved. If working with the extended one-particle projection $\delta_P=1$ instead, it can be seen in equation \eqref{eq:diagterms1} that the terms containing $\eta$ and $\eta'$ lack a symmetric counterpart to be cancelled and hence in that case probability conservation is not given any more, as it was also discussed below equation \eqref{eq:probcons}.
~\\
~\\
In \cite{Oniga:2015lro} the dynamics of the population for a master equation of a photon coupled to linearised gravity are discussed. We obtain an agreement with their result if we specialise to a massless scalar particle and choose as the initial time $t_0=0$. In addition we need to consider the non-extended one particle projection (i.e. $\delta_P=0$), in order to adapt to their chosen normal ordering as well as choose the temperature parameter $\Theta$ to be zero. The latter corresponds to a vacuum state of the gravitational waves environment. Inserting these assumptions in the evolution of the populations this equation becomes
\begin{align}\label{eq:diagevol4}
\dot{\rho}(\vec{k},t)=-\kappa \int \frac{d^3k'}{(2\pi)^3} \frac{P_k(\vec{k}'-\vec{k})}{\Omega_{k'-k}\omega_{k'}\omega_k}\left[  \frac{\sin({\chi} t)}{{\chi}} \rho(\vec{k},t)-\frac{\sin({\chi'} t)}{{\chi'}} \rho(\vec{k}',t)\right] \,,
\end{align}
which has a very similar form as the one in \cite{Oniga:2015lro} for a photon. The only difference arises due to the fact that for the photons in \cite{Oniga:2015lro} the polarisation vectors couple to the symmetric transverse traceless projector while here for the scalar particles, as they do not carry any polarisation, This role is taken over by the momentum, which is the only direction-dependent quantity that scalar particles possess.
~\\

\subsubsection{After renormalisation}\label{sec:afterRenorm}
As discussed in section \ref{uvren5}, the effect of the renormalisation was that the vacuum part in the one-particle master equations, these are the contributions not involving $N(\vec{k})$, vanishes. At the practical level this can be implemented by replacing everywhere $N(k'-k)+1 \rightarrow N(k'-k)$. Then the dynamics of the populations becomes
\begin{align}\label{eq:diagterms3}
\dot{\rho}(\vec{k},t)=-\kappa \int \frac{d^3k'}{(2\pi)^3} \frac{P_k(\vec{k}'-\vec{k})}{\Omega_{k'-k}\omega_{k'}\omega_k}\cdot N(k'-k) \cdot \Bigg\{ &\Bigg[ \frac{\sin({\chi} (t-t_0))}{{\chi}} + \frac{\sin({\chi}' (t-t_0))}{{\chi}'}\nonumber\\ &+\delta_P \frac{\sin({\eta} (t-t_0))}{{\eta}} +\delta_P \frac{\sin({\eta}' (t-t_0))}{{\eta}'} \Bigg]\rho(\vec{k},t)\nonumber\\
&- \left[ \frac{\sin({\chi'} (t-t_0))}{{\chi'}} +\frac{\sin({\chi} (t-t_0))}{{\chi}}\right]\:\rho(\vec{k}',t)\Bigg\}\,.
\end{align}
We realise that now all terms depend on $N(k'-k)$. As a consequence, the entire evolution of the populations is trivial, that is $\dot{\rho}(\vec{k},t)$ vanishes, if we consider the specific case of a vanishing temperature parameter $\Theta=0$ yielding directly $N(k'-k)=0$ for all $k,k'$. The comparison to the non-renormalised master equation shows that the physical properties of the two one-particle master equations are quite different as far as the dynamics of the populations is concerned. For this reason the discussions and physical implications drawn in \cite{Oniga:2015lro} based on the dynamics of the populations in the non-renormalised equation \eqref{eq:diagevol4} are problematic, as the evolution of the diagonal terms vanishes after renormalisation in the zero temperature limit.

\subsubsection{After the Markov approximation}\label{sec:afterMap}
Due to the fact that for the diagonal elements the coefficients always enter in the form $C+C^*$, only real terms in the one-particle master equation after the second Markov approximation in \eqref{eq:meqnsmap} remain:
\begin{align}
\frac{\partial}{\partial t} \rho(\vec{k},t) =&
\dot{\rho}(\vec{k},t)\nonumber \\
=&\frac{\kappa }{16\pi\beta} \Bigg\{ -\frac{20}{3} k^2 +4 \omega_k^2 -\frac{4}{\omega_k k} m^4 \,\text{arctanh} \left( \frac{k}{\omega_k}\right) \nonumber\\
    &\hspace{.7in} +4 k^2 \left[ \frac{2}{3} - \frac{m^2}{k^2} + \frac{m^4}{\omega_k k^3} \text{arctanh}\left(\frac{k}{\omega_k}\right)\right]
    \Bigg\}\rho(\vec k,t) \nonumber\\
    =& 0\,.
\end{align}
This means that the Markov approximation removes the dynamics of the populations also in the case of non-vanishing temperature and independently of the extended projection $\delta_P$. It therefore also restores probability conservation, as it removes all terms from the extended projection in the dissipator. The rotating wave approximation only affects the imaginary parts of the master equation, thus it does not change the evolution of the populations any more.\\
This result that the dynamics of the populations vanishes is also obtained in \cite{Anastopoulos:2013zya} for a non-relativistic one-particle master equation that was renormalised after the application of Markov and rotating wave approximation, and for the one of a photon after renormalisation and application of the same two approximations in \cite{Lagouvardos:2020laf}.

\subsection{Non-relativistic limit}\label{ap:Nonrandurel}
In the following, we apply the renormalised one-particle master equation after Markov and rotating wave approximation \eqref{eq:finalmeqaraaa} to non-relativistic particles in order to compare the decoherence with the one derived in \cite{Anastopoulos:2013zya}. 

In the non-relativistic limit we have  $\frac{u^2}{m^2} \ll 1$ and $\frac{v^2}{m^2} \ll 1$ and due to this the one-particle master equation simplifies. In this case we can expand the $\text{arctanh}$ as
\begin{equation}
    -2m^2\frac{m}{u}\frac{1}{\sqrt{1+\frac{u^2}{m^2}}} \text{arctanh}\left( \frac{u}{m} \frac{1}{\sqrt{1+\frac{u^2}{m^2}}}\right) = -2m^2 \frac{m}{u} \left[ \frac{u}{m} - \frac{2}{3} \frac{u^3}{m^3} + \frac{8}{15} \frac{u^5}{m^5} + O\left( \frac{u^6}{m^6}\right) \right]\, .
\end{equation}
Given this we find for the contribution from lines two to four in \eqref{eq:finalmeqaraaa} which is the part leading to decoherence  the following expression:
\begin{align}
    \frac{\kappa}{16\pi\beta} \left\{-\frac{16}{15} \frac{u^4+v^4}{m^2}-\frac{16}{15} \frac{u^2 v^2}{m^2} (1-3\cos^2(\gamma)  \right\} =-\frac{\kappa}{5\pi\beta m^2} \left[\frac{1}{3} (u^4+v^4) + u^2v^2\left( \frac{1}{3}-\cos^2(\gamma)\right) \right] \,,
\end{align}
where $\gamma$ is defined as the angle between $\vec u$ and $\vec v$, i.e. $\vec u \cdot \vec v = u v \cos(\gamma)$.
We work with the extended projection $\delta_P=1$ here, as a consequence there is no Lamb-shift contribution left and the master equation becomes
\begin{align}\label{eq:meqnonrelhere}
    \frac{\partial}{\partial t} \rho(\vec u, \vec v,t) &= -i\rho(\vec u,\vec v,t) (\omega_u-\omega_v) - \frac{\kappa}{5\pi\beta m^2} \left[\frac{1}{3} (\vec{u}^4+\vec{v}^4) + \vec{u}^2\vec{v}^2\left( \frac{1}{3}-\cos^2(\gamma)\right) \right] \rho(\vec u, \vec v,t) \nonumber\\
    &= -i\rho(\vec u,\vec v,t) \left( \frac{u^2}{2m} - \frac{v^2}{2m} \right) - \frac{\kappa}{5\pi\beta m^2} \left[\frac{1}{3} (\vec{u}^4+\vec{v}^4) + \vec{u}^2\vec{v}^2\left( \frac{1}{3}-\cos^2(\gamma)\right) \right] \rho(\vec u, \vec v,t)\,,
\end{align}
where we used in the last step $\omega_u=\sqrt{\vec{u}^2 +m^2}=m\sqrt{\frac{u^2}{m^2}+1}\approx m+\frac{u^2}{2m}$ and likewise for $\omega_v$.
The master equation (60) in \cite{Anastopoulos:2013zya}, where a Lindblad equation is used after Markov and rotating wave approximation to also describe a scalar field coupled to a linearised gravitational field, reads in momentum representation\footnote{In \cite{Anastopoulos:2013zya} a different $\kappa_{AH}$ is used that is related to the $\kappa$ used here by $\kappa_{AH} = 2 \kappa$.}
\begin{equation}\label{eq:meqnonrelAH}
    \frac{\partial}{\partial t} \rho(\vec u, \vec v,t) = -i\rho(\vec u,\vec v,t) \left(\frac{u^2}{2m_R}-\frac{v^2}{2 m_R}\right) - \frac{2\kappa }{3\beta m_R^2} \left[\frac{1}{3} (\vec{u}^4+\vec{v}^4) - \frac{1}{3} \vec{u}^2\vec{v}^2\left( 1+\cos^2(\gamma)\right) \right] \rho(\vec u, \vec v,t) \,.
\end{equation}
While in that work, they use the same underlying physical system, one of the differences is that there a gauge fixing is used while in this work the elementary physical variables were identified in \cite{Fahn:2022zql} in the relational formalism by choosing geometrical clocks with respect to which suitable Dirac observables were constructed. These observables facilitate the comparison of different gauge fixings and provide a physical meaning of the appearing physical temporal and spatial coordinates independently of a specific gauge. Additionally, the Hamiltonian used in \cite{Anastopoulos:2013zya} is not completely normal ordered, while here we worked with a completely normal ordered one. A more detailed discussion of these two points can be found in \cite{Fahn:2022zql}. Furthermore, the renormalisation is carried out in a different manner: in this work it is done before the Markov and rotating wave approximation are applied. In contrast in \cite{Anastopoulos:2013zya} the final master equation is renormalised after the application of these two approximations and after going into the non-relativistic limit. Their renormalisation procedure involves the introduction of a cutoff $\Lambda \ll m$ which is later absorbed in a redefinition of the renormalised mass $m\rightarrow m_R$, while here we found in equation \eqref{eq:renormctermsdef} that only the wave function needs to be renormalised, see section \ref{uvren4}. In \cite{Anastopoulos:2013zya} compared to our result here, there are some additional unitary terms left due to using the non-extended one-particle projection. These contributions are proportional to the UV-cutoff $\Lambda$ and to $\frac{u^4}{m_R^2}$, which is why they are dropped in \cite{Anastopoulos:2013zya} from the final master equation in the non-relativistic limit, even though in the limit $\Lambda\rightarrow \infty$ they would diverge. As our results demonstrate, using the extended projection and a renormalisation before the application of the approximations hence removes the necessity to drop diverging terms by hand. 
~\\
Additional differences between \eqref{eq:meqnonrelhere} and \eqref{eq:meqnonrelAH} are the prefactor in front of the dissipator and the structure inside the square brackets. In these two points the results derived here do not agree with the results in \cite{Anastopoulos:2013zya}. Particularly regarding the last point, our result however agrees with a similar derivation for photons in \cite{Lagouvardos:2020laf} where more intermediate steps are provided and where the final structure in the square brackets is the same as in \eqref{eq:meqnonrelhere}.

\subsection{Ultra-relativistic limit}\label{sec_apurel}
In this subsection we apply the one-particle master equation to ultra-relativistic particles. Possible applications are one-particle master equations for photons as discussed in \cite{Lagouvardos:2020laf} as well as gravitationally induced decoherence in neutrino oscillations as for instance discussed in \cite{Domi:2024ypm}, where a quantum mechanical toy model was used. 
~\\
~\\
In the ultra-relativistic limit we have $\frac{m^2}{u^2} \ll 1$ as well as $\frac{m^2}{v^2} \ll 1$. Taking this into account, we neglect all terms of order $O\left(\frac{m^2}{u^2}\right)$ and $O\left(\frac{m^2}{v^2}\right)$ respectively and higher order contributions. Note, that this also includes terms involving $\text{arctanh}$ function because 
\begin{align}
\frac{m^2}{u^2} \frac{1}{\sqrt{1+\frac{m^2}{u^2}}} \text{arctanh}\left( \frac{1}{\sqrt{1+\frac{m^2}{u^2}}}\right) = O\left(\frac{m^2}{u^2}\right)\, .\nonumber
\end{align}
This leads to the the following simplification for the decoherence term in \eqref{eq:finalmeqaraaa}:
\begin{align}
    \frac{\kappa}{16\pi\beta} \Bigg\{ &-\frac{4}{3}(u^2+v^2) -\frac{4}{3} \Big[ u v - 3 (\vec u \cdot \vec v)^2 \Big] \Bigg\} = -\frac{\kappa}{4\pi\beta} \left[ \frac{1}{3}(u^2+v^2) + u v\left( \frac{1}{3} -\cos^2(\gamma) \right) \right]\,.
\end{align}
The remaining computation of the imaginary part in the dissipator can be found in  \eqref{eq:apurelimpart} in the appendix. Combining all results, the renormalised one-particle master equation in the ultra-relativistic limit can be written in the form 
\begin{align}\label{eq:mequrelhere}
     \frac{\partial}{\partial t} \rho(\vec{u},\vec{v},t) = &-i \rho(\vec{u},\vec{v},t) \; (u-v) \nonumber\\
&- \frac{\kappa}{4\pi\beta} \left[ \frac{1}{3}(u^2+v^2) + u v\left( \frac{1}{3} -\cos^2(\gamma) \right) \right]\rho(\vec u,\vec v,t)\nonumber\\
    &-(1-\delta_P) \frac{i \kappa u^4\: \rho(\vec u,\vec v,t)}{105(2\pi)^2\omega_u} \Bigg\{4 \Bigg[ \frac{\pi^4}{15\beta^4 u^4} - \frac{7\pi^2}{6\beta^2 u^2} -6 \frac{\ln(1-e^{-\beta u})}{\beta u} + 4  \frac{\text{Li}_2(e^{-\beta u})}{\beta^2 u^2} \nonumber\\ &\hspace{1.8in}- 6 \frac{\text{Li}_3(e^{-\beta u})}{\beta^3 u^3} - 6 \frac{\text{Li}_4(e^{-\beta u})}{\beta^4 u^4} \Bigg]+ 35 - 35\frac{\ln\left(e^{\beta u}-1 \right)}{\beta u}\nonumber\\
    &\hspace{1.7in}  - 14 u \int_u^\infty dk\; \frac{N(k)}{k^2} + 3 u^3 \int_u^\infty dk\; \frac{N(k)}{k^4} \Bigg\} \nonumber\\
    &+(1-\delta_P) \frac{i \kappa v^4\: \rho(\vec u,\vec v,t)}{105(2\pi)^2\omega_v} \Bigg\{4 \Bigg[ \frac{\pi^4}{15\beta^4 v^4} - \frac{7\pi^2}{6\beta^2 v^2} -6 \frac{\ln(1-e^{-\beta v})}{\beta v} + 4  \frac{\text{Li}_2(e^{-\beta v})}{\beta^2 v^2} \nonumber\\ &\hspace{1.8in}- 6 \frac{\text{Li}_3(e^{-\beta v})}{\beta^3 v^3} - 6 \frac{\text{Li}_4(e^{-\beta v})}{\beta^4 v^4} \Bigg]+ 35 - 35\frac{\ln\left(e^{\beta v}-1 \right)}{\beta v}\nonumber\\
    &\hspace{1.7in}  - 14 v \int_v^\infty dk\; \frac{N(k)}{k^2} + 3 v^3 \int_v^\infty dk\; \frac{N(k)}{k^4} \Bigg\}\,,
\end{align}
where $\gamma$ denotes the angle between $\vec u$ and $\vec v$ and $\text{Li}_s(x)$ denotes the poly-logarithm function defined in \eqref{eq:defpolylog}. In the extended projection, i.e. for $\delta_P=1$, this becomes
\begin{align}
    \frac{\partial}{\partial t} \rho(\vec{u},\vec{v},t) = &-i \rho(\vec{u},\vec{v},t) \; (u-v) - \frac{\kappa}{4\pi\beta} \left[ \frac{1}{3}(u^2+v^2) + u v\left( \frac{1}{3} -\cos^2(\gamma) \right) \right]\rho(\vec u,\vec v,t)\,,
\end{align}
which can be rewritten in terms of an operator equation as
\begin{equation}
     \frac{\partial}{\partial t} \hat{\rho}(t) = -i [\hat{H},\hat{\rho}(t)] +\frac{\kappa}{8\pi\beta} \left( \delta^{il}\delta^{jm}-\frac{1}{3}\delta^{ij}\delta^{lm}\right) \left[ \frac{\hat{p}_i \hat{p}_j}{\hat{p}_0}, \left[\hat{\rho}(t), \frac{\hat{p}_l \hat{p}_m}{\hat{p}_0}\right]\right]\,,
\end{equation}
with $\hat{p}_0 := \sqrt{\hat{p}_n \hat{p}^n +\xi_{m^2}\mathds{1}}$. In this definition, $\xi_{m^2}$ is a small regulator that removes the eigenvalue zero from the spectrum of $\hat{p}_0$, as in that case the operator would not be invertible. For massive particles, this regulator corresponds to the mass squared $m^2$ which is still present in the ultra-relativistic limit, even though very small compared to the other summand. 
This is, up to a factor of $2$, the same result for decoherence as derived in \cite{Lagouvardos:2020laf} for gravitationally induced decoherence of photons. This difference of a factor of $2$ is already present when comparing the field-theoretical models of \cite{Lagouvardos:2020laf} and \cite{Fahn:2022zql}. Note that in \cite{Lagouvardos:2020laf} the derivation and in particular the application of the approximations is performed without a prior renormalisation of the one-particle master equation, which is done in the end to get rid of the diverging Lamb-shift term. As expected from the analysis in this work, they find a logarithmic divergence in the end. The derivation of the master equation in \cite{Lagouvardos:2020laf} is very similar to the one in \cite{Anastopoulos:2013zya}, hence we refer for a detailed comparison to the discussion in subsection \ref{ap:Nonrandurel}. The renormalisation in \cite{Lagouvardos:2020laf} is done after performing the approximations and the ultra-relativistic limit such that the detailed procedure depends on the cutoff frequency $\Lambda$ and its relation to the photon frequency $\omega_u$ (in our case the scalar particle's frequency). In the end in \cite{Lagouvardos:2020laf} the electric and magnetic fields as well as the coupling constant are renormalised.

\subsection{Application to neutrino oscillations}\label{sec:AppURel}
Finally, we want to discuss the relation of the results obtained in this work with the one presented in \cite{Domi:2024ypm}, where gravitationally induced decoherence in neutrino oscillations is investigated based on a quantum mechanical toy model \cite{Xu:2020lhc} with neutrinos as system of interest and a collection of Harmonic oscillators to model the thermal gravitational waves environment. For this purpose we consider the decoherence of neutrinos predicted by the ultra-relativistic one-particle master equation derived in this work.  
~\\
~\\
The connection to the work in \cite{Domi:2024ypm} is of interest to us in several respects: first due to the quantum mechanical nature of the the toy model used in \cite{Domi:2024ypm}, the coupling parameter encoding the strength for the coupling between system and environment cannot be determined from first principles. Instead a free parameter was introduced, that after introduction of a spectral density was denoted by\footnote{Note that this parameter $\eta$ is not related to the $\eta,\eta'$ defined below \eqref{eq:diagterms1}.} $\eta$, similar to what was done also in \cite{Xu:2020lhc}. Second, the quantum mechanical model requires the choice of a spectral density to derive the final master equation together with an appropriate cut-off function that regulates the integral over the frequency domain. In \cite{Domi:2024ypm}, four different commonly used cut-off functions were considered and it was shown that the final result of the master equations cannot distinguish between the different choices. For the spectral density, the usual linear dependence on the frequency, which is widely used in the context of quantum optics, was considered. As the one-particle master equation in this work is derived from an underlying field theory model presented in \cite{Fahn:2022zql} the situation is different here. As the matter, which is a scalar field in the present work, couples to linearised gravity, the coupling constant in the interaction Hamiltonian is naturally build into the model and given by $\kappa=8\pi G_N$, where $G_N$ is Newton's constant. Furthermore, due to the field-theoretical character of the model, it is not necessary to introduce a spectral density by hand, since the interaction Hamiltonian contains an integral over all modes from the beginning. As a third aspect we want to compare the application of the Markov approximation in the quantum mechanical toy model and in the one-particle master equation derived in this work.

~\\
 Even though the equation derived here is, taken strictly, only applicable to scalar particles, we still apply it to the case of neutrinos in order to discuss the relation with the results in \cite{Domi:2024ypm}. That this can be done in this context is due to the reason that the quantum mechanical toy model investigated in  \cite{Domi:2024ypm} treats the neutrinos as plane waves and thus does not take the full spinorial nature of neutrinos into account.
 ~\\
 We assume that the two momenta $\vec u$ and $\vec v$ are approximately parallel to each other in order to have intersection probability and to be able to measure them in a neutrino detector. With this, the one-particle master equation in the ultra-relativistic limit becomes
\begin{align}\label{eq:NApMeq1}
    \frac{\partial}{\partial t} \rho(\vec{u},\vec{v},t) = &-i \rho(\vec{u},\vec{v},t) \; (u-v) - \frac{\kappa}{12\pi\beta} (u-v)^2 \rho(\vec u,\vec v,t)\,.
\end{align}
As for ultra-relativistic particles we have $\hat{H} \hat{\rho} \approx u \hat{\rho}$, we indeed obtain the same form for the master equation as in \cite{Domi:2024ypm} in the effective mass basis\footnote{The effective mass basis is the basis in which the neutrino Hamiltonian in matter is diagonal.}:
\begin{equation}
   \frac{\partial}{\partial t} \hat{\rho}(t) = -i [\hat{H},\hat{\rho}(t)] +\frac{\kappa}{6\pi\beta} \left( \hat{H} \hat{\rho}(t) \hat{H} - \frac{1}{2}\left\{ \hat{H}^2,\hat{\rho}(t) \right\} \right)\,.
\end{equation}
The master equation above can be solved in the energy eigenbasis, where we denote the energy eigenvalues by $E_i$. With respect to this basis we denote the individual elements of $\hat{\rho}$ by $\rho_{ij}$ whose solutions read
\begin{equation}
    \rho_{ij}(t) = \rho_{ij}(0) e^{-i(E_i-E_j) t - \frac{\kappa}{12\pi\beta} (E_i-E_j)^2 t}\,.
\end{equation}
This result agrees, up to the different prefactor of $2$ in front of the decoherence term mentioned in subsection \ref{sec_apurel}, with the one obtained for the one-particle projection evaluated for motion in one dimension for a photon in \cite{Lagouvardos:2020laf}. 
~\\
~\\
We will now discuss the comparison of the three aspects mentioned above. We start with the comparison of the coupling parameter. Such a comparison can be obtained by comparing the prefactors of the decoherence terms in the model from \cite{Domi:2024ypm} and here. For the latter we have
\begin{equation}
    \frac{\kappa}{12\pi \beta} = \frac{2 G_N k_B \Theta}{3 c^4} \rightarrow \frac{2 G_N k_B \Theta}{3\hbar^2 c^5}\,,
\end{equation}
where we restored the correct units in the last step. Comparing with the decoherence rate in \cite{Domi:2024ypm}, which is $\frac{4 \eta^2 k_B \Theta}{\hbar^3}$, and introducing the Planck length $\ell_P$, we find that
\begin{equation}
    \eta^2 = \frac{\hbar G_N}{6c^5} = \frac{\ell_P^2}{6 c^2} \approx 5\cdot 10^{-88} s^2\,.
\end{equation}
An estimate for the value of $\eta^2$ inspired from field theory was already discussed in the appendix in \cite{Domi:2024ypm}. The difference in the orders of magnitude compared to the analysis in the appendix in \cite{Domi:2024ypm} arises due to the numerical prefactors that could not be determined precisely by the analogy analysis and estimate in \cite{Domi:2024ypm}. Similar results of the coupling strength can be found in \cite{Anastopoulos:2013zya,Lagouvardos:2020laf}. As the one-particle master equation considers the case of a scalar field with a thermal gravitational background, more work is needed in order to develop more sophisticated models for neutrinos or fermions in general to derive a similar master equation for a fermionic system under consideration and for more general environments. In addition the model is based on linearising gravity around a flat Minkowski background, whereas it would be interesting to also consider decoherence models for longer propagation distances and consider master equations based on a model on a cosmological background as the presence of a scale factor could modify the decoherence effect, as analysed for instance in \cite{Boriero:2017tkh,Pfenniger:2006rd,Bernardini:2012uc}. 
~\\
~\\
Compared to the quantum mechanical model \cite{Domi:2024ypm}, as mentioned above, here it was not necessary to introduce a spectral density, which is a kind of continuum limit for the frequencies of the oscillators in the environment that is typically used in similar quantum mechanical models to avoid Poincare recurrences, see for instance \cite{Breuer:2002pc}.  Furthermore, the cut-off function that needed to be used in the quantum mechanical model to regularise divergent integrals is not required for the one-particle master equation here. Instead the divergent contributions in the one-particle master equation could be linked to Feynman diagrams of a corresponding effective field theory giving a clearer physical interpretation than in the quantum mechanical toy model. With that given, the divergent contributions were treated using a standard renormalisation procedure known from quantum field theory that would be applied also in other situations in quantum field theory where such kind of diagrams play a role.
~\\
~\\
Not entirely unrelated to the latter paragraph is the discussion of the application of the Markov approximation in the model considered here and the quantum mechanical toy model in \cite{Domi:2024ypm}. In the latter, due to its simplicity compared to the one-particle master equation considered here, it was explicitly shown that the environmental correlation functions are strongly peaked around the initial time and decay rapidly after the peak. Such environmental correlation functions depend on both the chosen spectral density and the chosen cut-off function. In this work, however, none of these choices are made, but the corresponding quantities are determined and set from the beginning when formulating the model. In section \ref{sec_Map} a condition was discussed under which the Markov approximation can be applied for the ultra-relativistic limit. Considering here the application to neutrinos we can discuss whether this condition is satisfied in this application and how it relates to the application of the Markov approximation in the quantum mechanical toy model in \cite{Domi:2024ypm}. The condition applied in this work, equation \eqref{eq:condmarkaprurel}, states that the Markov approximation is justified if\footnote{Note that this is not an if and only if here.} 
\begin{equation}
    t-t_0 \gg  \frac{\hbar}{k_B \Theta} \gg \frac{\hbar}{c u}, \frac{\hbar}{c v} \hspace{0.2in} \text{and} \hspace{0.2in} u,v\gg m\,,
\end{equation}
 where $u:= |\vec u|$, $v:= |\vec v|$ denote the absolute value of the particle's momentum, $m$ its mass and $c$ the speed of light. Furthermore, $\beta = \frac{1}{k_B \Theta}$ with Boltzmann constant $k_B$ and a temperature parameter $\Theta$, which characterises the oscillator environment that is described as a Gibbs state.
At the end of Appendix \ref{sec:applic2M}, these conditions are evaluated explicitly for ultra-relativistic atmospheric neutrinos and it is shown that the neutrino energies $cu, cv$, and the temperature parameter $\Theta$ used in \cite{Domi:2024ypm} (in that work called $T$) lie inside the scope of these conditions, even with a freedom to change the parameters by several orders of magnitude to still be within the bounds of validity for the Markov approximation discussed here.
Thus, we can conclude given the proof presented in this work, the application of the Markov approximation to the physical scenario used in \cite{Domi:2024ypm} is not only justified at the level of the quantum mechanical model as shown in \cite{Domi:2024ypm} but also if one derives that model from the one-particle master equation of the QFT model.
~\\
~\\
Finally, compared to the quantum mechanical toy model in \cite{Domi:2024ypm}, the more general one-particle master equation derived here also allows to consider the generalisation to decoherence models with wave packets, whereas in \cite{Domi:2024ypm} plane waves were considered. This is due to the general form of the one particle density matrix defined in \eqref{eq:1pdm} as
\begin{equation}
    \rho_1(t) = \int_{\mathbb{R}^3} d^3u \int_{\mathbb{R}^3} d^3v \; \rho(\vec{u},\vec{v},t) \: a_u^\dagger \ket{0}\bra{0} a_v\,.
\end{equation}
When choosing suitable initial conditions for $\rho(\vec u,\vec v,0)$, one can model different descriptions for neutrinos like wave packets or plane waves, where the latter just correspond to delta distributions in this context.

\section{Conclusions}\label{sec:concl}
In this paper, we investigate the one-particle sector of the field-theoretic model in \cite{Fahn:2022zql} for gravitationally induced decoherence for a scalar field coupled to linearised gravity. One of the main foci of this work is the renormalisation of the one-particle master equation, which, in contrast to the existing literature \cite{Anastopoulos:2013zya,Lagouvardos:2020laf}, is performed before applying approximations such as the Markov or rotating wave approximation. As our results show this strategy provides a physical interpretation of the UV-divergent contributions in the one-particle master equation as being the vacuum part of the self-energy of the scalar field. To demonstrate this explicitly several steps are necessary: first the one-particle projection, where we considered two different kinds in this work, the non-extended and extended one. The latter also includes those processes in which in an intermediate steps two particles are created and annihilated afterwards or when the initial particle is left invariant and a vacuum bubble is created. Compared to the non-extended projection the last situation requires and additional renormalisation. While the non-extended projection yields a one-particle master equation with probability conservation, the extended projection does not have that property before any renormalisation, approximations or limits are taken into account. Then for both one-particle projections we identify the UV divergent contributions in the one-particle master equation. It turns out that all thermal parts of the one-particle master equation are UV finite and only the vacuum contributions in those terms which do not include other states than the one under consideration, are the divergent ones. Since in open quantum mechanical models the Lamb-shift Hamiltonians often require renormalisation and their physical interpretation is given, we wanted to address the question of the physical interpretation of the UV-divergent contributions in the one-particle master equation determined from an underlying field-theoretic model. 
~\\
~\\
For this purpose, we used the methods introduced in \cite{Burrage:2018pyg} and applied them to an open quantum model with thermal gravitational waves as environment, instead of an environment consisting of a scalar field as in \cite{Burrage:2018pyg}. These methods allow us to identify contributions in the one-particle master equation with certain Feynman diagrams of the effective field theory for the scalar field. Since the model in \cite{Fahn:2022zql} is based on a canonical quantisation of the master equation in a first step we consider the interaction part of the effective quantum field theory model in its canonical from and introduce the corresponding non-covariant Feynman rules along the lines of \cite{weinberg2005quantum,tong}, where similar methods are used in the framework of QED.  Equipped with them, we consider the (non-covariant) self-energy diagram, which can be decomposed into a thermal and a vacuum part, and show that the latter can be identified with the UV-divergent contributions in the one-particle master equation. In order to apply a standard on-shell renormalisation procedure for the self-energy of the scalar field, we relate the non-covariant Feynman rules to the covariant rules. Interestingly, a sum of two non-covariant Feynman diagrams, expressed by non-covariant propagators, can be combined into one diagram containing the corresponding covariant propagator. In the work in \cite{Burrage:2018pyg} the Feynman rules were directly available in covariant form and therefore the introduction of non-covariant Feynman rules was not necessary there. We then present the final renormalised one-particle master equation. 
~\\
As our results show, the renormalisation procedure applied to the underlying effective QFT allowed to absorb a divergence into a redefinition of the wave function in the model considered in the present work and the effect of this renormalisation is that all vacuum contributions in the Lamb-shift terms as well as in the dissipator are no longer present in the renormalised one-particle master equation.
Comparing our results to the one in \cite{Anastopoulos:2013zya}, where the renormalisation is performed after a Markov and rotating wave approximation have been applied and the non-relativistic limit has been considered, there the renormalisation procedure consists of introducing a cutoff function that is later absorbed in a redefinition of the renormalised mass. In \cite{Lagouvardos:2020laf} the effective model for the photon is also renormalised after applying the Markov and rotational wave approximation and considering the ultra-relativistic limit, and there the electric and magnetic field as well as the coupling constant are renormalised. Our results also allow a comparison to the renormalisation performed in the quantum mechanical model in \cite{Domi:2024ypm}. While there the Lamb-shift Hamiltonian, that consists in that model only of vacuum contributions, needs to be renormalised and is absent after renormalisation, in the renormalised one-particle master equation in this work, the thermal part of the Lamb-shift Hamiltonian remains. This agrees with the situation one has when QED is treated as an open quantum model \cite{Breuer:2002pc}, where only the vacuum contribution in the Lamb-shift terms is denoted as Lamb-shift and the thermal contribution as Stark shift. From the discussion of the open QED model from \cite{Breuer:2002pc}, one would expect a similar result in a quantum mechanical model where a thermal contribution in the Lamb-shift is present. We also see differences in how renormalisation affects the dissipator. While in the model presented here the renormalisation removes the vacuum contributions from the dissipator, in \cite{Domi:2024ypm} the dissipator is not changed when the renormalisation is applied and therefore the vacuum contributions remain. However, these are removed as soon as the Markov approximation is applied and therefore do not contribute to the final Lindblad equation in \cite{Domi:2024ypm}.
~\\
~\\
Given the renormalised one-particle master equation we then discuss the application of the Markov and rotation wave approximation. While a general analysis of the applicability of the Markov approximation is beyond the scope of this article, we can find a condition for the ultra-relativistic case under which the Markov approximation can be applied. For the applications to neutrino oscillations considered in \cite{Domi:2024ypm}, this condition is very mild and does not lead to severe restrictions. This fits with the fact that it could be explicitly shown for the quantum mechanical model in \cite{Domi:2024ypm} that the Markov approximation can be applied. After the Markov approximation we also applied a post-trace rotating wave approximation. Here we obtain a difference for the extended and non-extended one-particle projection. While for the extended projection, the rotating wave approximation has the effect that it removes the remaining part of the Lamb-shift term, in the case of the non-extended projection there is still one term left in the Lamb-shift contribution. The dissipator term is not affected by this approximation because those terms that would be potentially affected have been already removed by the one-particle projection.
~\\
~\\
As the first application of the renormalised one-particle master equation we discuss the evolution of the population and compare our results to the one in \cite{Oniga:2015lro}. We demonstrate that although there exists a non-trivial evolution of the populations before renormalisation, which is consistent with the results obtained in \cite{Oniga:2015lro} for the effective model of a photon if we choose the non-extended one-particle projection and specialise to the case of a vanishing temperature parameter, after renormalisation the evolution of the populations, that is the diagonal elements of the effective density matrix, becomes trivial. This shows the relevance of a renormalisation when analysing physical effects with the master equation, as the non-trivial dynamics of the populations and hence the effect discussed in \cite{Oniga:2015lro} is removed by the renormalisation applied here.
Hence, we conclude that if renormalisation is taken into account, we have no physically interesting effect in the evolution of the populations. In addition from our results for the evolution of the populations we learn that after applying the Markov approximation to the renormalised one-particle master equation, the extended and non-extended one-particle projection yield the same result.
~\\
~\\
In addition, we analyse the form of the one-particle master equation in the non- and ultra-relativistic limit. In these cases, the dissipator obtains a rather simple form which is similar to the ones in \cite{Anastopoulos:2013zya} for a non-relativistic scalar particle and in \cite{Lagouvardos:2020laf} for a photon. While the detailed structure of the dissipator in \cite{Anastopoulos:2013zya} differs from ours, \cite{Lagouvardos:2020laf} has the same one we have.
~\\
~\\
Another application we present is the comparison with the model in \cite{Domi:2024ypm}, where a quantum mechanical toy model for gravitationally induced decoherence, based on the model in \cite{Xu:2020lhc}, is investigated in the context of neutrino oscillations. Considering the renormalised one-particle master equation after applying the Markov and rotating wave approximation and in the special case of the ultra-relativistic limit, we obtain the renormalised Lindblad equation derived from the underlying quantum mechanical microscopic model in \cite{Domi:2024ypm}. This comparison allows to fix the coupling constant in the interaction Hamiltonian that was an open parameter in \cite{Domi:2024ypm}. As already discussed in \cite{Domi:2024ypm}, where a first estimate of this coupling constant was given as well as the discussions in \cite{Milburn:1991zkb,Milburn:2003zj,Diosi:2004iq,Breuer:2008rh,Lagouvardos:2020laf} the value is fairly tiny. Whether more sophisticated models for gravitationally induced decoherence will involve larger values of the coupling parameter is an interesting question to discuss in future research. In addition, in contrast to the quantum mechanical model \cite{Domi:2024ypm}, no spectral density, which can be understood as a kind of continuum limit of the frequencies of the harmonic oscillators in the environment of the model in \cite{Domi:2024ypm}, has to be introduced into the field-theoretical model in this work, but the corresponding integrals are automatically determined from the underlying action of the model. The results obtained here also allow an extension of the model in \cite{Domi:2024ypm}, where the plane wave approach for the neutrinos is used, to models including wave packets, which can be implemented by an appropriate choice of the one-particle density matrix. We plan to consider an analysis of such a model in future work.
~\\
~\\
There are several directions for further extensions and generalisations of the results obtained in this article. First, one can consider field-theoretical models where an operator ordering is chosen for which the self-interaction term of the scalar field does not vanish, as in \cite{Anastopoulos:2013zya}. In this case, we do not expect the renormalisation procedure to change, since we expect the same identification of non-covariant and covariant Feynman diagrams to hold for the self-energy, with the only difference that the second diagram in the sum of non-covariant diagrams does not vanish, as is the case in the model considered here. Instead of introducing the covariant Feynman rules in order to apply the renormalisation, one can also study if and how the methods from \cite{Adkins:1982zk} for one-loop renormalisation of QED in Coulomb gauge can be employed to perform the renormalisation directly based on the non-covariant Feynman rules. Furthermore, a more general analysis of the question under which criteria the Markov approximation can be applied would be interesting. In particular, whether it is also possible to formulate conditions that depend only on the properties of the environment, as it was done in \cite{Nathan:2020} for open quantum models with finitely many degrees of freedom. Also a study of the applicability and effect of different versions of the Markov approximation for gravitationally induced decoherence models is an interesting point, as some of them directly yield a Lindblad equation and therefore remove the necessity to apply an additional rotating wave approximation. Such modified Markov approximations is for instance the one discussed in \cite{Kaplanek:2022xrr,Colas:2022hlq}. There, in addition to the density matrix, the operators of the system in the Schrödinger picture, which appear in the double commutator in the master equation, are Taylor expanded around the final time. Another possibility is a generalisation to field theory of the approximation for quantum mechanical models in \cite{davidovic2020completely}, where one obtains a completely positive master equation by replacing an arithmetic mean of the spectral density by a geometric one.
~\\
Although there are quantum field theoretical models for gravitationally induced decoherence for scalar fields \cite{Blencowe:2012mp,Anastopoulos:2013zya,Fahn:2022zql}, photons \cite{Lagouvardos:2020laf,fgke2024photon} or generic bosons \cite{Oniga:2015lro}, a model that includes fermions is still missing in the literature. Such a model would also be interesting in connection with the application to neutrino oscillations. Another possible generalisation is to consider quantisations other than Fock quantisations for the one-particle model, see for example \cite{Giesel:2022pzh}, where a quantisation inspired by loop quantum gravity was used in the framework of polymerised quantum mechanics to formulate an open scattering model.

\SkipTocEntry\section*{Acknowledgments}
MJF and KG would like to thank Michael Kobler for the valuable discussions at an early stage of the project. MJF would like to thank Renata Ferrero and Roman Kemper for the fruitful discussions on some computational aspects of the project. MJF thanks the Heinrich-Böll foundation for financial support.

\begin{appendices}
\renewcommand{\thesection}{A.\Roman{section}}
\renewcommand{\thesubsection}{\thesection.\arabic{subsection}}
\renewcommand{\thesubsubsection}{\thesubsection.\arabic{subsubsection}}
\renewcommand{\theequation}{\thesection.\arabic{equation}}

\makeatletter
\renewcommand{\p@subsection}{}
\renewcommand{\p@subsubsection}{}
\makeatother
\section{One-particle projection of the master equation}
\label{app:OPP}
In this section, we explicitly carry out the projection of the master equation \eqref{eq:meqSPPbegin} on the one-particle sector. As discussed in the main text, we proceed by inserting the density matrix of a single particle 
\begin{equation}
\rho_1(t) = \int d^3u \int d^3v \; \rho(\vec{u},\vec{v},t)\; a_u^\dagger \ket{0}\bra{0}a_{v}\,
\end{equation}
into the master equation \eqref{eq:meqSPPbegin}, where $\vec{u}$ and $\vec{v}$ denote labels of the momentum and then by neglecting all contributions that do not preserve the one-particle subspace. The projection of the Hamiltonian $H_S$ of the free scalar field and the gravitational self-interaction of the scalar field $U$ have already been discussed in the main text. Here we focus on the remaining terms, the Lamb-shift-like Hamiltonian $H_{LS}$ and the dissipator $\mathcal{D}$. First, we give the detailed expressions for these two contributions of the master equation in subsection \ref{Apoppsu1} and then project them separately on the one-particle space in subsection \ref{Appopsu2}.

\subsection{Individual contributions to the master equation}\label{Apoppsu1}
In \cite{Fahn:2022zql} the original dissipator was decomposed into two parts by separating a purely complex contribution entering the master equation analogously to $H_S$ and $U$ in terms of a commutator with $\rho_S$. This term was then defined to be the so-called Lamb-shift Hamiltonian that contributes to the unitary evolution of the density matrix and the remaining terms were identified as a new dissipator term $\mathcal{D}$ that has a form similar to the first standard form (see e.g. \cite{Breuer:2002pc} for the different forms of dissipators). The Lamb-shift Hamiltonian is given in \cite{Fahn:2022zql} as
\begin{equation}\label{lsham}
H_{LS} := \frac{1}{2} \int \frac{d^3k\, d^3p\, d^3l}{(2\pi)^\frac{6}{2}} \sum\limits_{r;a,b} \; S_{ab}(\vec{p},\vec{l};\vec{k},t) \; j_r^a(\vec{k},\vec{p})^\dagger\, j_r^b(\vec{k},\vec{l})\,,
\end{equation}
and the dissipator as
\begin{equation}\label{Dfsf}
\mathcal{D}[\rho_S] := \frac{\kappa}{2} \int \frac{d^3k\, d^3p\, d^3l}{(2\pi)^\frac{6}{2}} \sum\limits_{r;a,b} \; {R}_{ab}(\vec{p},\vec{l};\vec{k},t) \left( j_r^b(\vec{k},\vec{l}) \rho_S(t)j^a_r(\vec{k},\vec{p})^\dagger-\frac{1}{2} \left\{ j^a_r(\vec{k},\vec{p})^\dagger j_r^b(\vec{k},\vec{l}), \rho_S(t)\right\} \right)\,,
\end{equation}
where $S_{ab}$ and $R_{ab}$ are coefficient functions and the $j_r^a$ are operators containing combinations of scalar field's creation and annihilation operators respectively. These quantities will be defined below.
From the structure of $H_{LS}$ and $\mathcal{D}$ follows that three different kinds of terms have to be evaluated for the projection:
\begin{equation}\label{eq:ap1ppIbIII}
\underbrace{j_r^a(\vec{k},\vec{p})^\dagger \, j_r^b(\vec{k},\vec{l})\, \rho_1(t)}_{(I)} \hspace{0.3in} \text{and} \hspace{0.3in} \underbrace{\rho_1(t)\,j_r^a(\vec{k},\vec{p})^\dagger \, j_r^b(\vec{k},\vec{l})}_{(II)}  \hspace{0.3in} \text{and} \hspace{0.3in}   \underbrace{ j_r^b(\vec{k},\vec{l})\, \rho_1(t) \, j_r^a(\vec{k},\vec{p})^\dagger}_{(III)}\,.
\end{equation}
The coefficient functions $S$ and $R$ are defined as
\begin{align}
S_{ab}(\vec{p},\vec{l};\vec{k},t) &:= \frac{1}{2\Omega_k} \Bigg[ [N(\Omega_k)+1] \Bigg\{ \frac{e^{-i(\Omega_k + \omega_b(\vec k,\vec l))t}-1}{\Omega_k + \omega_b(\vec k,\vec l)} +  \frac{e^{i(\Omega_k + \omega_a(\vec k,\vec p))t}-1}{\Omega_k + \omega_a(\vec k,\vec p)}\Bigg\}\nonumber\\
&\hspace{1in} - N(\Omega_k) \Bigg\{ \frac{e^{i(\Omega_k - \omega_b(\vec k,\vec l))t}-1}{\Omega_k - \omega_b(\vec k,\vec l)} +  \frac{e^{-i(\Omega_k - \omega_a(\vec k,\vec p))t}-1}{\Omega_k - \omega_a(\vec k,\vec p)}\Bigg\}\Bigg]\label{eq:SabMatrix}
\end{align}
and
\begin{align}
R_{ab}(\vec{p},\vec{l};\vec{k},t) &:=\frac{i}{\Omega_k} \Bigg[ [N(\Omega_k)+1] \Bigg\{ \frac{e^{-i(\Omega_k + \omega_b(\vec k,\vec l))t}-1}{\Omega_k + \omega_b(\vec k,\vec l)} -  \frac{e^{i(\Omega_k + \omega_a(\vec k,\vec p))t}-1}{\Omega_k + \omega_a(\vec k,\vec p)}\Bigg\}\nonumber\\
&\hspace{1in} - N(\Omega_k) \Bigg\{ \frac{e^{i(\Omega_k - \omega_b(\vec k,\vec l))t}-1}{\Omega_k - \omega_b(\vec k,\vec l)} -  \frac{e^{-i(\Omega_k - \omega_a(\vec k,\vec p))t}-1}{\Omega_k - \omega_a(\vec k,\vec p)}\Bigg\}\Bigg]\label{eq:RabMatrix}
\end{align}
with the Bose-Einstein distribution
\begin{equation}
    N(\Omega_k) =: N(k) = \frac{1}{e^{\beta k}-1}\,,
\end{equation}
where we use the graviton frequency $\Omega_k = \sqrt{\vec k^2} =: k$ and $\beta = \frac{1}{k_B \Theta}$ with the Boltzmann constant $k_B$ and a temperature parameter $\Theta$ that determines the Gibbs state of the gravitational waves environment. Furthermore we have, defining $\omega_k := \sqrt{m^2 + \vec k^2}$ with the scalar field's mass $m$:
\begin{align}\label{eq:defjundw}
    j_r^1(\vec{k},\vec{p})&:=  a_p^\dagger a_{k+p} \frac{1}{2\sqrt{\omega_p \omega_{k+p}}}  \left[p_ap^b [P^{-r}(\vec{k})]^a_b \right] & \omega_1(\vec{k},\vec{p}) &:= \omega_p - \omega_{k+p}\\
j_r^2(\vec{k},\vec{p})&:=  a_{-p-k}^\dagger a_{-p} \frac{1}{2\sqrt{\omega_p \omega_{k+p}}} \left[p_ap^b [P^{-r}(\vec{k})]^a_b \right] & \omega_2(\vec{k},\vec{p}) &:= \omega_{k+p} - \omega_{p}\\
j_r^3(\vec{k},\vec{p})&:= a_{-p} a_{k+p} \frac{1}{2\sqrt{\omega_p \omega_{k+p}}} \left[p_ap^b [P^{-r}(\vec{k})]^a_b \right] & \omega_3(\vec{k},\vec{p}) &:= -\omega_p - \omega_{k+p}\\
j_r^4(\vec{k},\vec{p})&:= a_p^\dagger a_{-k-p}^\dagger \frac{1}{2\sqrt{\omega_p \omega_{k+p}}} \left[p_ap^b [P^{-r}(\vec{k})]^a_b \right] & \omega_4(\vec{k},\vec{p}) &:= \omega_p + \omega_{k+p}\,.
\end{align}
Here, $a_k$ and $a^\dagger_k$ are annihilation and creation operator valued distributions acting on the underlying bosonic Fock space for a scalar field in the standard way. $[P^{\pm r}(\vec k)]^a_b$ are the projectors on the individual transverse modes defined via
\begin{align}
    [P^\pm(\vec{k})]_a^b &:= {m}^*_a(\pm\vec{k})\, ({m}^b)^*(\pm\vec{k})\,,
\end{align}
where $\left\{\frac{\vec k}{|\vec k|},m(\vec k),{m}^*(\vec k)\right\}$ form an orthonormal basis of $\mathbb{R}^3$, details can be found in\footnote{In that work, complex conjugation of a quantity $c$ was denoted by $\overline{c}$ instead of $c^*$ in the present work.} \cite{Fahn:2022zql}. In what follows, we only need the properties of orthonormality of the three basis elements and the fact that the symmetric transverse-traceless projector is
\begin{equation}\label{appoptvp}
    P\t{}{^a^b^c^d}(\vec{k})  = \sum\limits_{r\in\{\pm\}} [P^r(\vec{k})]^a_i [P^{-r}(\vec{k})]^c_j \delta^{ib} \delta^{cd} =  \frac{1}{2} [{P}^{ac}(\vec{k}) {P}^{bd}(\vec{k})+{P}^{ad}(\vec{k}) {P}^{bc}(\vec{k})-{P}^{ab}(\vec{k}) {P}^{cd}(\vec{k})]\,.
\end{equation}
with the transverse projector
\begin{equation}
    {P}^{ab}(\vec{k}) = \delta^{ab}-\frac{k^ak^b}{\vec{k}^2}\,.
\end{equation}
Now we can proceed to project $H_{LS}$ and $\mathcal{D}$ on the one-particle space.

\subsection{Computation of the one-particle projection}\label{Appopsu2}
We start with the evaluation of the three kinds of terms (I)-(III) and consider all possible combinations $(a,b)$ that give a one-particle state after application. To keep track of all different combination in the three cases, we do this using a table. Considering only the creation and annihilation operator valued distributions $a^{(\dagger)}_k$ in the $j$-operators, we find:
\begin{center}
\begin{tabular}{l l} \label{tab1}
 \textbf{(I)} & $j_r^a(\vec{k},\vec{p})^\dagger \, j_r^b(\vec{k},\vec{l})\, a_u^\dagger \ket{0} \bra{0} a_v$   \\
 \hline
 (1,1) & $ a^\dagger_{k+p} a_p a^\dagger_l a_{l+k} a^\dagger_u \ket{0}\bra{0} a_v = \delta^3(\vec{l}+\vec{k}-\vec{u})\, a^\dagger_{k+p} a_p a^\dagger_l \ket{0}\bra{0} a_v = \delta^3(\vec{p}+\vec{k}-\vec{u}) \, \delta^3(\vec{p}-\vec{l}) \, a^\dagger_u\ket{0}\bra{0}a_v$\\
 (1,2) & $ a^\dagger_{k+p} a_p a^\dagger_{-l-k} a_{-l} a^\dagger_u \ket{0}\bra{0} a_v = \delta^3(\vec{l}+\vec{u}) \, \delta^3(\vec{p}+\vec{l}+\vec{k}) \, a^\dagger_u\ket{0}\bra{0}a_v$\\
 (2,1) & $ a^\dagger_{-p} a_{-p-k} a^\dagger_{l} a_{k+l} a^\dagger_u \ket{0}\bra{0} a_v = \delta^3(\vec{k}+\vec{l}-\vec{u}) \, \delta^3(\vec{p}+\vec{l}+\vec{k}) \, a^\dagger_u\ket{0}\bra{0}a_v$\\
 (2,2) & $ a^\dagger_{-p} a_{-p-k} a^\dagger_{-l-k} a_{-l} a^\dagger_u \ket{0}\bra{0} a_v = \delta^3(\vec{l}+\vec{u}) \, \delta^3(\vec{p}-\vec{l}) \, a^\dagger_u\ket{0}\bra{0}a_v$\\
 (4,4) & $ \delta_P \:a_{-k-p} a_p a^\dagger_l a^\dagger_{-k-l} a^\dagger_u \ket{0}\bra{0} a_v =  \delta_P \: (a)$\hspace{0.1in} \textit{see below}\vspace{0.3in}\\

 \textbf{(II)} & $a_u^\dagger \ket{0} \bra{0} a_v\, j_r^a(\vec{k},\vec{p})^\dagger \, j_r^b(\vec{k},\vec{l})$   \\
 \hline
 (1,1) &  $a^\dagger_u \ket{0}\bra{0} a_v  a^\dagger_{k+p} a_p a^\dagger_l a_{l+k}=  \delta^3(\vec{p}+\vec{k}-\vec{v}) \, \delta^3(\vec{p}-\vec{l}) \, a^\dagger_u\ket{0}\bra{0}a_v$\\
 (1,2) & $ a^\dagger_u \ket{0}\bra{0} a_v a^\dagger_{k+p} a_p a^\dagger_{-l-k} a_{-l} = \delta^3(\vec{k}+\vec{p}-\vec{v}) \, \delta^3(\vec{p}+\vec{l}+\vec{k}) \, a^\dagger_u\ket{0}\bra{0}a_v$\\
 (2,1) & $ a^\dagger_u \ket{0}\bra{0} a_v a^\dagger_{-p} a_{-p-k} a^\dagger_{l} a_{k+l} = \delta^3(\vec{p}+\vec{v}) \, \delta^3(\vec{p}+\vec{l}+\vec{k}) \, a^\dagger_u\ket{0}\bra{0}a_v$\\
 (2,2) & $ a^\dagger_u \ket{0}\bra{0} a_v  a^\dagger_{-p} a_{-p-k} a^\dagger_{-l-k} a_{-l} = \delta^3(\vec{p}+\vec{v}) \, \delta^3(\vec{p}-\vec{l}) \, a^\dagger_u\ket{0}\bra{0}a_v$\\
 (4,4) & $\delta_P \: a^\dagger_u \ket{0}\bra{0} a_v a_{-k-p} a_p a^\dagger_l a^\dagger_{-k-l}=  \delta_P \:$ (b) \hspace{0.1in} \textit{see below}\vspace{0.3in}\\

 \textbf{(III)} & $j_r^b(\vec{k},\vec{l})\, a_u^\dagger \ket{0} \bra{0} a_v\, j_r^a(\vec{k},\vec{p})^\dagger$   \\
 \hline
 (1,1) &  $ a^\dagger_l a_{l+k} a^\dagger_u \ket{0}\bra{0} a_v  a^\dagger_{k+p} a_p=  \delta^3(\vec{l}+\vec{k}-\vec{u}) \, \delta^3(\vec{k}+\vec{p}-\vec{v}) \, a^\dagger_{u-k}\ket{0}\bra{0}a_{v-k}$\\
  (1,2) & $a^\dagger_{-l-k} a_{-l} a^\dagger_u \ket{0}\bra{0} a_v a^\dagger_{k+p} a_p  = \delta^3(\vec{k}+\vec{p}-\vec{v}) \, \delta^3(\vec{u}+\vec{l}) \, a^\dagger_{u-k}\ket{0}\bra{0}a_{v-k}$\\
 (2,1) & $ a^\dagger_{l} a_{k+l}  a^\dagger_u \ket{0}\bra{0} a_v a^\dagger_{-p} a_{-p-k} = \delta^3(\vec{p}+\vec{v}) \, \delta^3(\vec{l}+\vec{k}-\vec{u}) \, a^\dagger_{u-k}\ket{0}\bra{0}a_{v-k}$\\
 (2,2) & $ a^\dagger_{-l-k} a_{-l} a^\dagger_u \ket{0}\bra{0} a_v  a^\dagger_{-p} a_{-p-k} = \delta^3(\vec{p}+\vec{v}) \, \delta^3(\vec{u}+\vec{l}) \, a^\dagger_{u-k}\ket{0}\bra{0}a_{v-k}$\\
\end{tabular}
\end{center}
The expressions for (a) and (b) are calculated separately because they include the vacuum bubbles mentioned in the main text that require a renormalisation. By applying the commutators and commuting all annihilation operators towards the vacuum state one obtains:
\begin{align}\label{eq1ppdiva}
(a)=\, & a_{-k-p} a_p a^\dagger_l a^\dagger_{-k-l} a^\dagger_u \ket{0}\bra{0} a_v \nonumber\\ =\, & [\delta^3(\vec{p}+\vec{k}+\vec{l}) \delta^3_\xi(\vec{p}+\vec{k}+\vec{l}) + \delta^3(\vec{p}-\vec{l}) \delta^3_\xi(\vec{p}-\vec{l}) + \delta^3(\vec{p}+\vec{k}+\vec{l}) \delta^3(\vec{u}+\vec{k}+\vec{p}) \nonumber\\&+ \delta^3(\vec{p}-\vec{l}) \delta^3(\vec{u}+\vec{k}+\vec{p}) + \delta^3(\vec{p}-\vec{u}) \delta^3(\vec{p}-\vec{l}) + \delta^3(\vec{p}-\vec{u}) \delta^3(\vec{k}+\vec{p}+\vec{l}) ]\; a_u^\dagger \ket{0}\bra{0} a_v\,,\\ \label{eq1ppdivb}
(b)= \,& a^\dagger_u \ket{0}\bra{0} a_v a_{-k-p} a_p a^\dagger_l a^\dagger_{-k-l} \nonumber\\ =\,& [\delta^3(\vec{p}+\vec{k}+\vec{l}) \delta^3_\xi(\vec{p}+\vec{k}+\vec{l}) + \delta^3(\vec{p}-\vec{l}) \delta^3_\xi(\vec{p}-\vec{l}) + \delta^3(\vec{p}+\vec{k}+\vec{l}) \delta^3(\vec{v}+\vec{k}+\vec{l})\nonumber\\&+ \delta^3(\vec{p}-\vec{l}) \delta^3(\vec{v}+\vec{k}+\vec{l}) + \delta^3(\vec{v}-\vec{l}) \delta^3(\vec{p}-\vec{l}) + \delta^3(\vec{v}-\vec{l}) \delta^3(\vec{k}+\vec{p}+\vec{l}) ]\; a_u^\dagger \ket{0}\bra{0} a_v\,.
\end{align}
Note that under the map $(\vec{u},\vec{p},\vec{l}) \leftrightarrow (\vec{v},\vec{l},\vec{p})$ we can get from (a) to (b) and vice versa. An important remark here is that the first two terms in both expressions contain the square of a Dirac delta distribution. This is a problematic term since, as can be shown, the corresponding integral over this expression still diverges when for the individual delta distributions the regularised version is considered and the regulator is removed after the integration is performed. We will deal with this issue further below in this subsection where we renormalise the density matrix to handle the divergent contributions involved. For now, we replace one of the two delta distributions by a function including a regulator $\delta^3(\vec k) \rightarrow \delta^3_\xi(\vec k)$, where the regulator is sent to zero after performing the corresponding integrations. \\
~\\
In a next step, we evaluate the entire expressions appearing in the Lamb-shift Hamiltonian and the dissipator. For each term, where the delta distributions resolve two integrals, that is for all terms but the diverging ones, we choose to resolve the integrals over the range of the variables $\vec{p}$ and $\vec{l}$ respectively. Then we are left with the integration over the variables $\vec k$, $\vec u$ and $\vec v$ as well as with the sum over the polarisation labels $r$. Considering for $\mathcal{V}_{ab}$ which is either $R_{ab}$ or $S_{ab}$, that were defined above in \eqref{eq:RabMatrix} and \eqref{eq:SabMatrix}, the three expressions 
\begin{align}
    \textbf{(I):} \;  &\sum_r \int d^3p \int d^3l \;\mathcal{V}_{ab}(\vec p,\vec l;\vec k, t)\: j_r^a(\vec{k},\vec{p})^\dagger \, j_r^b(\vec{k},\vec{l})\, a_u^\dagger \ket{0} \bra{0} a_v \\
    \textbf{(II):} \; &\sum_r \int d^3p \int d^3l \;\mathcal{V}_{ab}(\vec p,\vec l;\vec k, t)\: a_u^\dagger \ket{0} \bra{0} a_v\, j_r^a(\vec{k},\vec{p})^\dagger \, j_r^b(\vec{k},\vec{l}) \\
    \textbf{(III):} \; &\sum_r \int d^3p \int d^3l \;\mathcal{V}_{ab}(\vec p,\vec l;\vec k, t)\:  j_r^a(\vec{k},\vec{p})^\dagger \, a_u^\dagger \ket{0} \bra{0} a_v\, j_r^b(\vec{k},\vec{l})\,,
\end{align}
that appear in \eqref{lsham} and \eqref{Dfsf}, we obtain:

\begin{center}
\begin{tabular}{l l} 
 \textbf{(I)} & $\sum_r \int d^3p \int d^3l \;\mathcal{V}_{ab}(\vec p,\vec l;\vec k, t)\: j_r^a(\vec{k},\vec{p})^\dagger \, j_r^b(\vec{k},\vec{l})\, a_u^\dagger \ket{0} \bra{0} a_v$   \\
 \hline
 (1,1) & $ \mathcal{V}_{11}(\vec{u}-\vec{k},\vec{u}-\vec{k};\vec{k},t)\: \rho(\vec{u},\vec{v}) \frac{1}{4\omega_u \omega_{u-k}} P_u(\vec k) \; a_u^\dagger \ket{0}\bra{0} a_v$\\
 (1,2) & $ \mathcal{V}_{12}(\vec{u}-\vec{k},-\vec{u};\vec{k},t)\: \rho(\vec{u},\vec{v}) \frac{1}{4\omega_u \omega_{u-k}} P_u(\vec k) \; a_u^\dagger \ket{0}\bra{0} a_v$\\
 (2,1) & $ \mathcal{V}_{21}(-\vec{u},\vec{u}-\vec{k};\vec{k},t)\: \rho(\vec{u},\vec{v}) \frac{1}{4\omega_u \omega_{u-k}} P_u(\vec k) \; a_u^\dagger \ket{0}\bra{0} a_v$\\
 (2,2) & $ \mathcal{V}_{22}(-\vec{u},-\vec{u};\vec{k},t)\: \rho(\vec{u},\vec{v}) \frac{1}{4\omega_u \omega_{u-k}} P_u(\vec k) \; a_u^\dagger \ket{0}\bra{0} a_v$\\
 (4,4) & $\delta_P\:$ (a) \; \textit{(see below)} \vspace{0.3in}\\
  \textbf{(II)} & $\sum_r \int d^3p \int d^3l \;\mathcal{V}_{ab}(\vec p,\vec l;\vec k, t)\: a_u^\dagger \ket{0} \bra{0} a_v\, j_r^a(\vec{k},\vec{p})^\dagger \, j_r^b(\vec{k},\vec{l})$   \\
 \hline
 (1,1) & $ \mathcal{V}_{11}(\vec{v}-\vec{k},\vec{v}-\vec{k};\vec{k},t)\: \rho(\vec{u},\vec{v}) \frac{1}{4\omega_v \omega_{v-k}} P_v(\vec k) \; a_u^\dagger \ket{0}\bra{0} a_v$\\
 (1,2) & $ \mathcal{V}_{12}(\vec{v}-\vec{k},-\vec{v};\vec{k},t)\: \rho(\vec{u},\vec{v}) \frac{1}{4\omega_v \omega_{v-k}} P_v(\vec k) \; a_u^\dagger \ket{0}\bra{0} a_v$\\
 (2,1) & $ \mathcal{V}_{21}(-\vec{v},\vec{v}-\vec{k};\vec{k},t)\: \rho(\vec{u},\vec{v}) \frac{1}{4\omega_v \omega_{v-k}} P_v(\vec k) \; a_u^\dagger \ket{0}\bra{0} a_v$\\
 (2,2) & $ \mathcal{V}_{22}(-\vec{v},-\vec{v};\vec{k},t)\: \rho(\vec{u},\vec{v}) \frac{1}{4\omega_v \omega_{v-k}} P_v(\vec k) \; a_u^\dagger \ket{0}\bra{0} a_v$\\
 (4,4) &$ \delta_P\:$ (b) \; \textit{(see below)} \vspace{0.3in}\\

 \textbf{(III)} & $\sum_r \int d^3p \int d^3l \;\mathcal{V}_{ab}(\vec p,\vec l;\vec k, t)\:  j_r^a(\vec{k},\vec{p})^\dagger \, a_u^\dagger \ket{0} \bra{0} a_v\,j_r^b(\vec{k},\vec{l})$   \\
 \hline
 (1,1) & $ \mathcal{V}_{11}(\vec{v}-\vec{k},\vec{u}-\vec{k};\vec{k},t)\: \rho(\vec{u},\vec{v}) \frac{1}{4\sqrt{\omega_v \omega_{v-k} \omega_u \omega_{u-k}}} P_{u,v}(\vec k) \; a_{u-k}^\dagger \ket{0}\bra{0} a_{v-k}$\\
 (1,2) & $  \mathcal{V}_{12}(\vec{v}-\vec{k},-\vec{u};\vec{k},t)\: \rho(\vec{u},\vec{v}) \frac{1}{4\sqrt{\omega_v \omega_{v-k} \omega_u \omega_{u-k}}}P_{u,v}(\vec k) \; a_{u-k}^\dagger \ket{0}\bra{0} a_{v-k}$\\
 (2,1) & $ \mathcal{V}_{11}(-\vec{v},\vec{u}-\vec{k};\vec{k},t)\: \rho(\vec{u},\vec{v}) \frac{1}{4\sqrt{\omega_v \omega_{v-k} \omega_u \omega_{u-k}}} P_{u,v}(\vec k)\; a_{u-k}^\dagger \ket{0}\bra{0} a_{v-k}$\\
 (2,2) & $ \mathcal{V}_{22}(-\vec{v},-\vec{u};\vec{k},t)\: \rho(\vec{u},\vec{v}) \frac{1}{4\sqrt{\omega_v \omega_{v-k} \omega_u \omega_{u-k}}} P_{u,v}(\vec k) \; a_{u-k}^\dagger \ket{0}\bra{0} a_{v-k}$\,,\\
\end{tabular}
\end{center}
where we defined\footnote{The equivalence of this definition and the one in \eqref{appoptvp} can  be seen by expanding $\vec u$ in the basis $\left\{\frac{\vec k}{|\vec k|},\vec m(\vec k),\vec m(-\vec k)\right\}$. As the contribution parallel to $\vec k$ always vanishes the important part is $\vec u = \mu\vec m(\vec k) +\nu \vec m(-\vec k)$. Then one obtains: $P^{ab}(\vec k) u_a u_b = 2 \mu\nu\vec m(\vec k) \vec m(-\vec k)=2\mu\nu$, and thus $P^{abcd}(\vec k) u_au_bu_cu_d = \frac{1}{2} (4+4-4)\mu^2\nu^2=2\mu^2\nu^2$ and $[\vec u \cdot \vec m(\vec k)]^2[\vec u \cdot \vec m(-\vec k)]^2=\mu^2\nu^2$. }
\begin{align}
P_u(\vec k) &:= \sum_r [\vec{u}\cdot \vec{m}(-r\vec{k})]^2 [\vec{u}\cdot \vec{m}(r\vec{k})]^2= 2[\vec{u}\cdot \vec{m}(-\vec{k})]^2 [\vec{u}\cdot \vec{m}(\vec{k})]^2= P^{abcd}(\vec k) u_au_bu_cu_d  \\ 
P_{u,v}(\vec k) &:=\sum_r [\vec{u}\cdot \vec{m}(-r\vec{k})]^2 [\vec{v}\cdot \vec{m}(r\vec{k})]^2= P^{abcd}(\vec k) u_au_bv_cv_d\,.    
\end{align}
Next, we evaluate in more detail the coefficient functions $S$ and $R$ which were defined in \eqref{eq:SabMatrix} and \eqref{eq:RabMatrix}. Firstly excluding all terms from the extended projection, i.e. the ones arising from the combination $(4,4)$, it turns out that the coefficients appearing in $(I)-(III)$ are equal to each other in each group:  
\begin{align}
&S_{11}(\vec{u}-\vec{k},\vec{u}-\vec{k};\vec{k},t) = S_{12}(\vec{u}-\vec{k},-\vec{u};\vec{k},t) = S_{21}(-\vec{u},\vec{u}-\vec{k};\vec{k},t) = S_{22}(-\vec{u},-\vec{u};\vec{k},t) =:S_{1P}^{(-)}(\vec{u},\vec{k},t) \nonumber\\
&= \frac{1}{\Omega_k}\Bigg\{ \frac{N(k)+1}{\Omega_k + \omega_{u-k}-\omega_u} \left( \cos\left[ (\Omega_k + \omega_{u-k}-\omega_u)t\right] -1 \right) -\frac{N(k)}{\Omega_k - \omega_{u-k}+\omega_u} \left( \cos\left[ (\Omega_k - \omega_{u-k}+\omega_u)t\right] -1 \right)\Bigg\}\,,
\end{align}
\begin{align}
&R_{11}(\vec{u}-\vec{k},\vec{u}-\vec{k};\vec{k},t) = R_{12}(\vec{u}-\vec{k},-\vec{u};\vec{k},t) = R_{21}(-\vec{u},\vec{u}-\vec{k};\vec{k},t) = R_{22}(-\vec{u},-\vec{u};\vec{k},t)=: 2R_{1P}^{(-)}(\vec{u},\vec{k},t)\nonumber\\
&= \frac{2}{\Omega_k}\Bigg\{ \frac{N(k)+1}{\Omega_k + \omega_{u-k}-\omega_u} \sin\left[ (\Omega_k + \omega_{u-k}-\omega_u)t\right]+\frac{N(k)}{\Omega_k - \omega_{u-k}+\omega_u}  \sin\left[ (\Omega_k - \omega_{u-k}+\omega_u)t\right]\Bigg\}
\end{align}
and
\begin{align}
&R_{11}(\vec{v}-\vec{k},\vec{u}-\vec{k};\vec{k},t) = R_{12}(\vec{v}-\vec{k},-\vec{u};\vec{k},t) = R_{21}(-\vec{v},\vec{u}-\vec{k};\vec{k},t) = R_{22}(-\vec{v},-\vec{u};\vec{k},t) =:2R_{1P}^{(2)}(\vec{u},\vec{v},\vec{k},t)\nonumber\\
&= \frac{i}{\Omega_k} \Bigg\{ [N(k)+1] \left( \frac{e^{-i(\Omega_k+\omega_{u-k}-\omega_u)t}-1}{\Omega_k+\omega_{u-k}-\omega_u} - \frac{e^{i(\Omega_k+\omega_{v-k}-\omega_v)t}-1}{\Omega_k+\omega_{v-k}-\omega_v} \right)\nonumber\\ &\hspace{2in}-N(k) \left( \frac{e^{i(\Omega_k-\omega_{u-k}+\omega_u)t}-1}{\Omega_k-\omega_{u-k}+\omega_u} - \frac{e^{-i(\Omega_k-\omega_{v-k}+\omega_v)t}-1}{\Omega_k-\omega_{v-k}+\omega_v} \right)\Bigg\}\,.
\end{align}
With these expressions, we can rewrite the equations for the Lamb-shift Hamiltonian and the dissipator that contain the contributions of $(a,b)\in\{1,2\}$:
\begin{align}
-i[H_{LS}^{\{1,2\}},\rho_1(t)]=&-\frac{i\kappa}{2} \int \frac{d^3k \, d^3u\, d^3v}{(2\pi)^3} \rho(\vec{u},\vec{v},t) \; a_u^\dagger \ket{0}\bra{0}a_v \; \frac{P_u(\vec k)}{ \omega_{u-k}\omega_u}\; S_{1P}^{(-)}(\vec{u},\vec{k},t)\nonumber\\
&+\frac{i\kappa}{2} \int \frac{d^3k \, d^3u\, d^3v}{(2\pi)^3} \rho(\vec{u},\vec{v},t) \; a_u^\dagger \ket{0}\bra{0}a_v \; \frac{P_v(\vec k)}{\omega_{v-k}\omega_v}\; S_{1P}^{(-)}(\vec{v},\vec{k},t)
\end{align}
as well as
\begin{align}
\mathcal{D}^{\{1,2\}}[\rho_1(t)]=&-\frac{\kappa}{2} \int \frac{d^3k \, d^3u\, d^3v}{(2\pi)^3} \rho(\vec{u},\vec{v},t) \; a_u^\dagger \ket{0}\bra{0}a_v \; \frac{P_u(\vec k)}{\omega_{u-k}\omega_u}\; R_{1P}^{(-)}(\vec{u},\vec{k},t)\nonumber\\
&-\frac{\kappa}{2} \int \frac{d^3k \, d^3u\, d^3v}{(2\pi)^3} \rho(\vec{u},\vec{v},t) \; a_u^\dagger \ket{0}\bra{0}a_v \; \frac{P_v(\vec k)}{\omega_{v-k}\omega_v}\; R_{1P}^{(-)}(\vec{v},\vec{k},t)\nonumber\\
&+\kappa\int \frac{d^3k \, d^3u\, d^3v}{(2\pi)^3} \rho(\vec{u},\vec{v},t) \; a_{u-k}^\dagger \ket{0}\bra{0} a_{v-k} \; \frac{P_{u,v}(\vec k)}{\sqrt{\omega_{u-k}\omega_u\omega_{v-k}\omega_v}}\; R_{1P}^{(2)}(\vec{u},\vec{v},\vec{k},t)\,.
\end{align}
It remains to deal with the $(4,4)$-terms that arise when using the extended projection. The four summands from (a) and (b) without the terms containing the $\xi$-regulator yield
\begin{align}
-i[H_{LS}^{\{4,4\}},\rho_1(t)]=&-\frac{i\kappa}{2} \int \frac{d^3k \, d^3u\, d^3v}{(2\pi)^3} \rho(\vec{u},\vec{v},t) \; a_u^\dagger \ket{0}\bra{0}a_v \; \frac{P_u(\vec k)}{\omega_{u+k}\omega_u}\; \widetilde{S}_{1P}^{(+)}(\vec{u},\vec{k},t)\: \delta_P\nonumber\\
&+\frac{i\kappa}{2} \int \frac{d^3k \, d^3u\, d^3v}{(2\pi)^3} \rho(\vec{u},\vec{v},t) \; a_u^\dagger \ket{0}\bra{0}a_v \; \frac{P_v(\vec k)}{\omega_{v+k}\omega_v}\;\widetilde{S}_{1P}^{(+)}(\vec{v},\vec{k},t)\: \delta_P
\end{align}
with
\begin{align}
&\widetilde{S}_{1P}^{(+)}(\vec{u},\vec{k},t) := S_{44}(-\vec{u}-\vec{k},\vec{u};\vec{k},t) = S_{44}(-\vec{u}-\vec{k},-\vec{u}-\vec{k};\vec{k},t) = S_{44}(\vec{u},\vec{u};\vec{k},t) = S_{44}(\vec{u},-\vec{u}-\vec{k};\vec{k},t)\nonumber\\
&= \frac{1}{\Omega_k}\Bigg\{ \frac{N(k)+1}{\Omega_k + \omega_{u+k}+\omega_u} \left( \cos\left[ (\Omega_k + \omega_{u+k}+\omega_u)t\right] -1 \right) -\frac{N(k)}{\Omega_k - \omega_{u+k}-\omega_u} \left( \cos\left[ (\Omega_k - \omega_{u+k}-\omega_u)t\right] -1 \right)\Bigg\}\,.
\end{align}
After a substitution $\vec{k}\rightarrow -\vec{k}$ in the integration we find:
\begin{align}
-i[H_{LS}^{\{4,4\}},\rho_1(t)]=&-\frac{i\kappa}{2} \int \frac{d^3k \, d^3u\, d^3v}{(2\pi)^3} \rho(\vec{u},\vec{v},t) \; a_u^\dagger \ket{0}\bra{0}a_v \; \frac{P_u(\vec k)}{\omega_{u-k}\omega_u}\; S_{1P}^{(+)}(\vec{u},\vec{k},t) \:\delta_P\nonumber\\
&+\frac{i\kappa}{2} \int \frac{d^3k \, d^3u\, d^3v}{(2\pi)^3} \rho(\vec{u},\vec{v},t) \; a_u^\dagger \ket{0}\bra{0}a_v \; \frac{P_v(\vec k)}{\omega_{v-k}\omega_v}\; S_{1P}^{(+)}(\vec{v},\vec{k},t)\:\delta_P
\end{align}
where 
\begin{equation}
S_{1P}^{(+)}(\vec{u},\vec{k},t) := \widetilde{S}_{1P}^{(+)}(\vec{u},-\vec{k},t)\,.
\end{equation}
The same substitution and the definition
\begin{align}
R_{1P}^{(+)}(\vec{u},\vec{k},t) := \frac{1}{\Omega_k}\Bigg\{ \frac{N(k)+1}{\Omega_k + \omega_{u-k}+\omega_u} \sin\left[ (\Omega_k + \omega_{u-k}+\omega_u)t\right] +\frac{N(k)}{\Omega_k - \omega_{u-k}-\omega_u}  \sin\left[ (\Omega_k - \omega_{u-k}-\omega_u)t\right]\Bigg\}
\end{align}
lead to:
\begin{align}
\mathcal{D}^{\{4,4\}}[\rho_1]=&-\frac{\kappa}{2} \int \frac{d^3k \, d^3u\, d^3v}{(2\pi)^3} \rho(\vec{u},\vec{v},t) \; a_u^\dagger \ket{0}\bra{0}a_v \; \frac{P_u(\vec k)}{\omega_{u-k}\omega_u}\; R_{1P}^{(+)}(\vec{u},\vec{k},t) \:\delta_P\nonumber\\
&-\frac{\kappa}{2} \int \frac{d^3k \, d^3u\, d^3v}{(2\pi)^3} \rho(\vec{u},\vec{v},t) \; a_u^\dagger \ket{0}\bra{0}a_v \; \frac{P_v(\vec k)}{\omega_{v-k}\omega_v}\; R_{1P}^{(+)}(\vec{v},\vec{k},t)\:\delta_P\,.
\end{align}
Defining additionally
\begin{align}
S_{1P}^{(1)}(\vec{u},\vec{k},t)&:= S_{1P}^{(-)}(\vec{u},\vec{k},t)+S_{1P}^{(+)}(\vec{u},\vec{k},t)\:\delta_P\\
R_{1P}^{(1)}(\vec{u},\vec{k},t)&:= R_{1P}^{(-)}(\vec{u},\vec{k},t)+ R_{1P}^{(+)}(\vec{u},\vec{k},t)\:\delta_P
\end{align}
gives us the opportunity to rewrite
\begin{align}
-i[H_{LS},\rho_1(t)]=&-\frac{i\kappa}{2} \int \frac{d^3k \, d^3u\, d^3v}{(2\pi)^3} \rho(\vec{u},\vec{v},t) \; a_u^\dagger \ket{0}\bra{0}a_v \; \frac{P_u(\vec k)}{\omega_{u-k}\omega_u}\; S_{1P}^{(1)}(\vec{u},\vec{k},t)\nonumber\\
&+\frac{i\kappa}{2} \int \frac{d^3k \, d^3u\, d^3v}{(2\pi)^3} \rho(\vec{u},\vec{v},t) \; a_u^\dagger \ket{0}\bra{0}a_v \; \frac{P_v(\vec k)}{\omega_{v-k}\omega_v}\; S_{1P}^{(1)}(\vec{v},\vec{k},t)
\end{align}
as well as
\begin{align}
\mathcal{D}[\rho_1]=&-\frac{\kappa}{2} \int \frac{d^3k \, d^3u\, d^3v}{(2\pi)^3} \rho(\vec{u},\vec{v},t) \; a_u^\dagger \ket{0}\bra{0}a_v \; \frac{P_u(\vec k)}{\omega_{u-k}\omega_u}\; R_{1P}^{(1)}(\vec{u},\vec{k},t)\nonumber\\
&-\frac{\kappa}{2} \int \frac{d^3k \, d^3u\, d^3v}{(2\pi)^3} \rho(\vec{u},\vec{v},t) \; a_u^\dagger \ket{0}\bra{0}a_v \; \frac{P_v(\vec k)}{\omega_{v-k}\omega_v}\; R_{1P}^{(1)}(\vec{v},\vec{k},t)\nonumber\\
&+\kappa \int \frac{d^3k \, d^3u\, d^3v}{(2\pi)^3} \rho(\vec{u},\vec{v},t) \; a_{u-k}^\dagger \ket{0}\bra{0} a_{v-k} \; \frac{P_{u,v}(\vec k)}{\sqrt{\omega_{u-k}\omega_u\omega_{v-k}\omega_v}}\; R_{1P}^{(2)}(\vec{u},\vec{v},\vec{k},t)\,.
\end{align}
This can be summarised as
\begin{align}
-i[H_{LS},&\rho_1(t)]+\mathcal{D}[\rho_1]=\nonumber\\=&-\frac{\kappa}{2} \int \frac{d^3k \, d^3u\, d^3v}{(2\pi)^3} \rho(\vec{u},\vec{v},t) \; a_u^\dagger \ket{0}\bra{0}a_v \; \frac{P_u(\vec k)}{ \omega_{u-k}\omega_u}\; [R_{1P}^{(1)}(\vec{u},\vec{k},t)+iS_{1P}^{(1)}(\vec{u},\vec{k},t)]\nonumber\\
&-\frac{\kappa}{2} \int \frac{d^3k \, d^3u\, d^3v}{(2\pi)^3} \rho(\vec{u},\vec{v},t) \; a_u^\dagger \ket{0}\bra{0}a_v \; \frac{P_v(\vec k)}{\omega_{v-k}\omega_v}\; [R_{1P}^{(1)}(\vec{v},\vec{k},t)-iS_{1P}^{(1)}(\vec{u},\vec{k},t)]\nonumber\\
&+\kappa \int \frac{d^3k \, d^3u\, d^3v}{(2\pi)^3} \rho(\vec{u},\vec{v},t) \; a_{u-k}^\dagger \ket{0}\bra{0} a_{v-k} \; \frac{P_{u,v}(\vec k)}{\sqrt{\omega_{u-k}\omega_u\omega_{v-k}\omega_v}}\; R_{1P}^{(2)}(\vec{u},\vec{v},\vec{k},t)\,.
\end{align}
Finally, the terms in the extended projection that contain the $\xi$-regulator, that is the expressions arising from the first two terms in \eqref{eq1ppdiva} and \eqref{eq1ppdivb}, are analysed. These terms are equal for (a) and (b) and read:
\begin{align}
&\delta_P \lim_{\xi\rightarrow 0} \int \frac{d^3p\, d^3l\, d^3k\, d^3u\, d^3v}{(2\pi)^\frac{3}{2}} \; a_u^\dagger\ket{0}\bra{0} a_v\; \mathcal{V}_{44}(\vec{p},\vec{l};\vec{k},t) \frac{P_{p,l}(\vec k)}{4\sqrt{\omega_p \omega_{k+p} \omega_l \omega_{k+l}}} \rho(\vec u,\vec v,t) \nonumber\\ &\hspace{2.4in}\cdot\{\delta^3(\vec{p}+\vec{k}+\vec{l})\delta^3_\xi(\vec{p}+\vec{k}+\vec{l})+ \delta^3(\vec{p}-\vec{l})\delta^3_\xi(\vec{p}-\vec{l})\}\nonumber\\
&= \delta_P \lim_{\xi\rightarrow 0}\int\frac{d^3p \, d^3k}{(2\pi)^\frac{3}{2}} \; \mathcal{V}_{44}(\vec{p},\vec{p};\vec{k},t) \frac{P_p(\vec k)}{2\omega_p \omega_{k+p}} \; \delta^3_\xi(\vec{0})\; \rho_1(t) \,,
\end{align}
where we used that $\mathcal{V}_{44}(\vec{p},\vec{p};\vec{k},t) = \mathcal{V}_{44}(\vec{p},\vec{-k-p};\vec{k},t)$. Due to the equality of these terms for (a) and (b), they drop out of the Lamb-shift Hamiltonian and are only left in the dissipator term:
\begin{equation}
\mathcal{D}^{div}[\rho_1] = -\delta_P\frac{\kappa}{4} \lim_{\xi\rightarrow 0}\int \frac{d^3k\, d^3p}{(2\pi)^\frac{3}{2}} R_{44}(\vec{p},\vec{p};\vec{k},t)\frac{P_p(\vec k)}{2\omega_p \omega_{k+p}} \delta^3_\xi(\vec{0})\; \rho_1(t) =: Z(t) \rho_1(t)\,.
\end{equation}
Written in this form, it becomes evident that they do act as a multiplicative constant and do not modify the state $\rho_1$. Therefore they are nothing but vacuum bubbles expressed in QFT language. With the definition of $Z(t)$ above the entire master equation for the single particle is given by
\begin{equation}
\frac{\partial}{\partial t}\rho_1(t) = -i [H_{S}  + H_{LS},\rho_1(t)] + \mathcal{D}[\rho_1] + Z(t) \rho_1(t)\,,
\end{equation}
where from now on the terms absorbed in $Z(t)$ are dropped from the definition of $\mathcal{D}[\rho_1]$. We can see that the diverging term $Z(t)$ can be absorbed by a renormalisation of the density matrix, likewise to a renormalisation of the wave function known from QFT:
\begin{equation}
\rho_1(t) \rightarrow \rho^{(ren)}_{1}(t) := \exp\left( \int_0^t dt' \; Z(t') \right) \rho_1(t)\,.
\end{equation}
In terms of the renormalised density matrix, the one-particle master equation then reads:
\begin{align}
\frac{\partial}{\partial t} \rho^{(ren)}_{1}(t) = &-i \int d^3u\, d^3v\; \rho^{(ren)}(\vec{u},\vec{v},t) \ket{\vec{u}}\bra{\vec{v}} \; (\omega_u-\omega_v)\nonumber\\
&-\frac{\kappa}{2} \int \frac{d^3k \, d^3u\, d^3v}{(2\pi)^3} \rho^{(ren)}(\vec{u},\vec{v},t) \; \ket{\vec{u}}\bra{\vec{v}} \; \frac{P_u(\vec k)}{\omega_{u-k}\omega_u}\; [R_{1P}^{(1)}(\vec{u},\vec{k},t)+iS_{1P}^{(1)}(\vec{u},\vec{k},t)]\nonumber\\
&-\frac{\kappa}{2} \int \frac{d^3k \, d^3u\, d^3v}{(2\pi)^3} \rho^{(ren)}(\vec{u},\vec{v},t) \; \ket{\vec{u}}\bra{\vec{v}} \; \frac{P_v(\vec k)}{\omega_{v-k}\omega_v}\; [R_{1P}^{(1)}(\vec{v},\vec{k},t)-iS_{1P}^{(1)}(\vec{v},\vec{k},t)]\nonumber\\
&+\kappa \int \frac{d^3k \, d^3u\, d^3v}{(2\pi)^3} \rho^{(ren)}(\vec{u},\vec{v},t) \; \ket{\vec{u}-\vec{k}}\bra{\vec{v}-\vec{k}} \; \frac{P_{u,v}(\vec k)}{ \sqrt{\omega_{u-k}\omega_u\omega_{v-k}\omega_v}}\; R_{1P}^{(2)}(\vec{u},\vec{v},\vec{k},t)\,.
\end{align}
To simplify the notation, we drop the label $(ren)$ from the density matrix from this point on. The complex combinations $R\pm iS$ can be evaluated further and one finds
\begin{align}
R_{1P}^{(1)}&(\vec{u},\vec{k},t)\pm iS_{1P}^{(1)}(\vec{u},\vec{k},t)=\nonumber\\ &= \pm\frac{i}{\Omega_k} \Bigg\{ \frac{N(k)+1}{\Omega_k+\omega_{u-k}-\omega_u} \left( e^{\mp i(\Omega_k+\omega_{u-k}-\omega_u)t} -1 \right) - \frac{N(k)}{\Omega_k-\omega_{u-k}+\omega_u} \left( e^{\pm i(\Omega_k-\omega_{u-k}+\omega_u)t} -1 \right)\nonumber\\
 &\hspace{0.5in}+\delta_P\frac{N(k)+1}{\Omega_k+\omega_{u-k}+\omega_u} \left( e^{\mp i(\Omega_k+\omega_{u-k}+\omega_u)t} -1 \right) -\delta_P \frac{N(k)}{\Omega_k-\omega_{u-k}-\omega_u} \left( e^{\pm i(\Omega_k-\omega_{u-k}-\omega_u)t} -1 \right)\Bigg\}\nonumber\\
 &= \int_0^t \frac{d\tau}{\Omega_k} \Big\{ [N(k)+1]\: e^{\mp i(\Omega_k+\omega_{u-k}-\omega_u)\tau} +N(k)\;e^{\pm i(\Omega_k-\omega_{u-k}+\omega_u)\tau} \nonumber\\ 
 &\hspace{0.55in}+\delta_P [N(k)+1]\: e^{\mp i(\Omega_k+\omega_{u-k}+\omega_u)\tau} +\delta_P \: N(k)\;e^{\pm i(\Omega_k-\omega_{u-k}-\omega_u)\tau}\Big\}\,.
\end{align}
For $R^{(2)}$ we have:
\begin{align}
R_{1P}^{(2)}(\vec{u},\vec{v},\vec{k},t)=& \frac{i}{2\Omega_k} \Bigg\{ [N(k)+1] \left( \frac{e^{-i(\Omega_k+\omega_{u-k}-\omega_u)t}-1}{\Omega_k+\omega_{u-k}-\omega_u} - \frac{e^{i(\Omega_k+\omega_{v-k}-\omega_v)t}-1}{\Omega_k+\omega_{v-k}-\omega_v} \right)\nonumber\\ 
&\hspace{0.6in}-N(k) \left( \frac{e^{i(\Omega_k-\omega_{u-k}+\omega_u)t}-1}{\Omega_k-\omega_{u-k}+\omega_u} - \frac{e^{-i(\Omega_k-\omega_{v-k}+\omega_v)t}-1}{\Omega_k-\omega_{v-k}+\omega_v} \right)\Bigg\}\nonumber\\
=&\int_0^t \frac{d\tau}{2\Omega_k} \Big\{ [N(k)+1]\:  e^{-i(\Omega_k+\omega_{u-k}-\omega_u)\tau} +N(k)\; e^{i(\Omega_k-\omega_{u-k}+\omega_u)\tau}\nonumber\\ 
&\hspace{0.4in}+ [N(k)+1]\: e^{i(\Omega_k+\omega_{v-k}-\omega_v)\tau} + N(k)\; e^{-i(\Omega_k-\omega_{v-k}+\omega_v)\tau}\Big\}\,.
\end{align}
Defining then
\begin{align}
C(\vec{u},\vec{k},t) &= \int_0^{t-t_0} \frac{d\tau}{\Omega_k} \Big\{ [N(k)+1]\: e^{- i(\Omega_k+\omega_{u-k}-\omega_u)\tau} +N(k)\;e^{ i(\Omega_k-\omega_{u-k}+\omega_u)\tau}\Big\}\\
C_P(\vec{u},\vec{k},t) &= \int_{0}^{t-t_0} \frac{d\tau}{\Omega_k} \Big\{ [N(k)+1]\: e^{- i(\Omega_k+\omega_{u-k}+\omega_u)\tau} +N(k)\;e^{ i(\Omega_k-\omega_{u-k}-\omega_u)\tau}\Big\}\,,
\end{align}
where we have restored the initial time\footnote{Originally, the integration is of the form $\int_{t_0}^t d\tau e^{i \Omega (t-\tau)}$. The version given here can be achieved by substituting $t-\tau \rightarrow \tau$.} $t_0$ that was set to $0$ in \cite{Fahn:2022zql}, the one-particle master equation in momentum representation has the form
\begin{align}\label{eq:meqopff}
\frac{\partial}{\partial t} \rho(\vec{u},\vec{v},t) = &-i \rho(\vec{u},\vec{v},t) \; (\omega_u-\omega_v) \nonumber\\
&-\frac{\kappa}{2} \int \frac{d^3k}{(2\pi)^3} \bigg\{ \frac{ P_u(\vec k)}{\omega_{u-k}\omega_u}\; \left[C(\vec{u},\vec{k},t)+\delta_P C_P(\vec{u},\vec{k},t)\right] \nonumber\\ &\hspace{1.1in} +\frac{P_v(\vec k)}{\omega_{v-k}\omega_v}\; \left[C^*(\vec{v},\vec{k},t)+\delta_P C_P^*(\vec{v},\vec{k},t)\right]\bigg\}\;\rho(\vec{u},\vec{v},t)\nonumber\\
&+\frac{\kappa}{2} \int \frac{d^3k}{(2\pi)^3}\frac{P_{u,v}(\vec k)}{\sqrt{\omega_{u+k}\omega_u\omega_{v+k}\omega_v}}\; \left\{C(\vec{u}+\vec{k},\vec{k},t)+ C^*(\vec{v}+\vec{k},\vec{k},t)\right\}\: \rho(\vec{u}+\vec{k},\vec{v}+\vec{k},t) \,.
\end{align}

\section{Equivalence of the non-covariant and covariant Feynman rules for the Coulomb-like scattering diagram}\label{ap:equivFR}
Here we show that the non-covariant and the covariant Feynman rules introduced in sections \ref{uvren2} and \ref{uvren3} respectively yield the same result for the Coulomb-like scattering diagram
\\
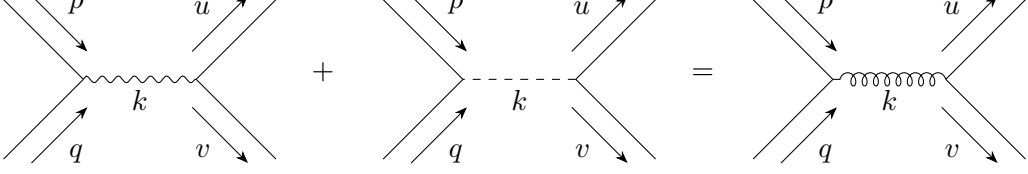
\begin{tikzpicture}
  \begin{feynman}
    \vertex (a);
    \vertex [left=of a] (b);
    \vertex [above left=of b] (g);
    \vertex [below left=of b] (c); 
    \vertex [above right=of a] (h);
    \vertex [below right=of a] (f);  
    \diagram* {
      (a) -- [boson, edge label=\(k\)] (b) -- [plain, reversed momentum'=\(p\)] (g),
      (b) -- [plain, reversed momentum=\(q\)] (c),
      (a) -- [plain, momentum=\(u\)] (h), (a) -- [plain, momentum'=\(v\)] (f)
    };
  \end{feynman} \hspace{4em} $ + $\end{tikzpicture} \hspace{3em}
  \begin{tikzpicture}
  \begin{feynman}
    \vertex (a);
    \vertex [left=of a] (b);
    \vertex [above left=of b] (g);
    \vertex [below left=of b] (c); 
    \vertex [above right=of a] (h);
    \vertex [below right=of a] (f);  
    \diagram* {
      (a) -- [scalar, edge label=\(k\)] (b) -- [plain, reversed momentum'=\(p\)] (g),
      (b) -- [plain, reversed momentum=\(q\)] (c),
      (a) -- [plain, momentum=\(u\)] (h), (a) -- [plain, momentum'=\(v\)] (f)
    };
  \end{feynman} \hspace{4em} $ = $
\end{tikzpicture}\hspace{3em}
\begin{tikzpicture}
  \begin{feynman}
    \vertex (a);
    \vertex [left=of a] (b);
    \vertex [above left=of b] (g);
    \vertex [below left=of b] (c); 
    \vertex [above right=of a] (h);
    \vertex [below right=of a] (f);  
    \diagram* {
      (a) -- [gluon, edge label=\(k\)] (b) -- [plain, reversed momentum'=\(p\)] (g),
      (b) -- [plain, reversed momentum=\(q\)] (c),
      (a) -- [plain, momentum=\(u\)] (h), (a) -- [plain, momentum'=\(v\)] (f)
    };
  \end{feynman}
\end{tikzpicture}\\
The left hand side of the equation is expressed in terms of the non-covariant Feynman rules and the first diagram, which we label \textbf{A}, reads
\begin{align}
    \mathbf{A}=&\frac{1}{2}(-i\kappa)^2 \widetilde{T}_{ab}(p,q) \widetilde{T}_{cd}(-u,-v) \frac{-i}{\kappa} \frac{1}{k^2} P^{abcd}(\vec k) \nonumber\\
    &= \frac{i\kappa}{2k^2}\widetilde{T}_{ab}(p,q) \widetilde{T}_{cd}(-u,-v) \Bigg[ \left(\delta^{ac} - \frac{k^a k^c}{\vec{k}^2} \right) \left(\delta^{bd} - \frac{k^b k^d}{\vec{k}^2} \right) -\frac{1}{2} \left(\delta^{ab} - \frac{k^a k^b}{\vec{k}^2} \right) \left(\delta^{cd} - \frac{k^c k^d}{\vec{k}^2} \right)\Bigg]\nonumber\\
    &= \frac{i\kappa}{k^2} \Bigg[\frac{1}{2}\widetilde{T}_{ab}(p,q) \widetilde{T}^{ab}(-u,-v) - \frac{1}{4} \widetilde{T}_a^a(p,q)\widetilde{T}_b^b(-u,-v) + \frac{1}{4\vec{k}^4} k^a k^b k^c k^d \widetilde{T}_{ab}(p,q) \widetilde{T}_{cd}(-u,-v) \nonumber\\ &\hspace{0.42in}- \frac{1}{\vec{k}^2} \Big\{ \widetilde{T}_{ab}(p,q) \widetilde{T}^a_c(-u,-v) k^b k^c -\frac{1}{4} \widetilde{T}_{ab}(p,q) \widetilde{T}_c^c(-u,-v) k^a k^b -\frac{1}{4} \widetilde{T}_a^a(p,q) \widetilde{T}_{cd}(-u,-v) k^c k^d\Big\} \Bigg]\,,
\end{align}
where the overall factor of $\frac{1}{2}$ arises due to the fact that it is a diagram of second order in the expansion of the Dyson series. Next, we can make use of energy-momentum conservation, as we assumed the scalar particles to be on-shell, which reads, as introduced in the main text in \eqref{eq:momcons1} and \eqref{eq:momcons2} for $k = p + q = u + v$:
\begin{align}
k^\mu \widetilde{T}_{\mu 0}(p,q) &= k^0 \widetilde{T}_{00}(p,q) + k^a \widetilde{T}_{a0}(p,q) = \frac{1}{2} [q_0 (p_0^2 -\vec{p}^2 -m^2) + p_0 (q_0^2 -\vec{q}^2 -m^2)]=0\\
k^\mu \widetilde{T}_{\mu a}(p,q) &= k^0 \widetilde{T}_{0a}(p,q) + k^b \widetilde{T}_{ba}(p,q) = \frac{1}{2} [q_a (p_0^2 -\vec{p}^2 -m^2) + p_a (q_0^2 -\vec{q}^2 -m^2)]=0\,.
\end{align}
From these we find
\begin{align}\label{eq:momconsfeyndg1}
    k^ak^b\widetilde{T}_{ab}(p,q) &= - k^0 k^b \widetilde{T}_{0b}(p,q)= (k^0)^2 \widetilde{T}_{00}(p,q)\\
    k^a \widetilde{T}_{ab}(p,q) &= -k^0 \widetilde{T}_{0b}(p,q)\\
    k^a \widetilde{T}_{0a}(p,q) &= - k^0 \widetilde{T}_{00}(p,q)\label{eq:momconsfeyndg3}
\end{align}
which leads to
\begin{align}
\mathbf{A} &= \frac{i\kappa}{k^2} \Bigg[\frac{1}{2}\widetilde{T}_{ab}(p,q) \widetilde{T}^{ab}(-u,-v) - \frac{1}{4} \widetilde{T}_a^a(p,q)\widetilde{T}_b^b(-u,-v) + \frac{(k^0)^4}{4\vec{k}^4}  \widetilde{T}_{00}(p,q) \widetilde{T}_{00}(-u,-v) \nonumber\\ &\hspace{0.42in}- \frac{(k^0)^2}{\vec{k}^2} \Big\{ \widetilde{T}_{a0}(p,q) \widetilde{T}^a_0(-u,-v) -\frac{1}{4} \widetilde{T}_{00}(p,q) \widetilde{T}_c^c(-u,-v) -\frac{1}{4} \widetilde{T}_a^a(p,q) \widetilde{T}_{00}(-u,-v) \Big\} \Bigg]\,.
\end{align}
Combining this with the expression for the second term in the above Feynman diagram for Coulomb-like scattering, which we call \textbf{B} and which reads
\begin{align}
    \mathbf{B} =&- \frac{i\kappa}{\vec{k}^2} \Bigg[ -\frac{1}{4} \tilde{T}_{00}(p,q) \tilde{T}_{00}(-u,-v) + \tilde{T}_{0a}(p,q) \tilde{T}_{0b}(-u,-v) \left( \delta^{ab} -\frac{k^a k^b}{4\vec{k}^2} \right)  \nonumber\\
  &\hspace{2in}-\frac{1}{4} \widetilde{T}_{00}(p,q) \widetilde{T}_a^a(-u, -v) -\frac{1}{4} \widetilde{T}_a^a(p,q) \widetilde{T}_{00}(-u,-v)\Bigg]\nonumber\\
  =& - \frac{i\kappa}{k^2} \Bigg[ -\frac{k^2}{4\vec{k}^2} \tilde{T}_{00}(p,q) \tilde{T}_{00}(-u,-v) + \frac{k^2}{\vec{k}^2} \tilde{T}_{0a}(p,q) \tilde{T}_{0}^a(-u,-v) - \frac{k^2(k^0)^2}{4\vec{k}^4} \tilde{T}_{00}(p,q) \tilde{T}_{00}(-u,-v) \nonumber\\
  &\hspace{2in}-\frac{k^2}{4\vec{k}^2} \widetilde{T}_{00}(p,q) \widetilde{T}_a^a(-u, -v) -\frac{k^2}{4\vec{k}^2} \widetilde{T}_a^a(p,q) \widetilde{T}_{00}(-u,-v)\Bigg]\,,
\end{align}
one can obtain
\begin{align}
    \mathbf{A + B} &= \frac{i\kappa}{k^2} \Bigg[\left( \frac{(k^0)^4 + k^2 \vec{k}^2 + k^2 (k^0)^2}{4\vec{k}^4} \right) \tilde{T}_{00}(p,q) \tilde{T}_{00}(-u,-v) - \left( \frac{(k^0)^2+k^2}{\vec{k}^2} \right) \tilde{T}_{0a}(p,q) \tilde{T}_{0}^a(-u,-v)\nonumber\\
    &\hspace{0.7in} + \frac{1}{2}\widetilde{T}_{ab}(p,q) \widetilde{T}^{ab}(-u,-v) - \frac{1}{4} \widetilde{T}_a^a(p,q)\widetilde{T}_b^b(-u,-v) \nonumber\\
    &\hspace{0.7in} +\left( \frac{(k^0)^2+k^2}{4\vec{k}^2} \right) \widetilde{T}_{00}(p,q) \widetilde{T}_c^c(-u,-v) + \left( \frac{(k^0)^2+k^2}{4\vec{k}^2} \right)\widetilde{T}_a^a(p,q) \widetilde{T}_{00}(-u,-v) \Bigg]\nonumber\\
    &= \frac{i\kappa}{k^2}\Bigg[ \frac{1}{4}\tilde{T}_{00}(p,q) \tilde{T}_{00}(-u,-v) - \tilde{T}_{0a}(p,q) \tilde{T}_{0}^a(-u,-v) + \frac{1}{2}\widetilde{T}_{ab}(p,q) \widetilde{T}^{ab}(-u,-v)\nonumber\\
    &\hspace{0.7in} - \frac{1}{4} \widetilde{T}_a^a(p,q)\widetilde{T}_b^b(-u,-v)+\frac{1}{4} \widetilde{T}_{00}(p,q) \widetilde{T}_c^c(-u,-v) + \frac{1}{4}\widetilde{T}_a^a(p,q) \widetilde{T}_{00}(-u,-v)  \Bigg]\,.\label{eq:feyndiagAB}
\end{align}
On the other hand, we obtain with the covariant Feynman rules for the right side of the Coulomb-like scattering diagram above\footnote{Note that this diagram is again of second order in the expansion of the Dyson series, hence we obtain a factor of $\frac{1}{2}$.}, which we name \textbf{C}:
\begin{align}
    \mathbf{C} &=\frac{1}{2} \frac{i\kappa}{k^2} \widetilde{T}_{\mu\nu}(p,q) \widetilde{T}_{\rho\sigma}(-u,-v) \left(\eta^{\mu\rho} \eta^{\nu\sigma} -\frac{1}{2} \eta^{\mu\nu} \eta^{\rho\sigma}\right)\nonumber\\\
    &= \frac{i\kappa}{2k^2} \Bigg[\widetilde{T}_{\mu\nu}(p,q) \widetilde{T}^{\mu\nu}(-u,-v) -\frac{1}{2} \widetilde{T}_{\mu}^\mu(p,q) \widetilde{T}_{\rho}^\rho(-u,-v)  \Bigg]\nonumber\\
    &= \frac{i\kappa}{4k^2} \Bigg[2\widetilde{T}_{00}(p,q) \widetilde{T}^{00}(-u,-v) + 2\widetilde{T}_{ab}(p,q) \widetilde{T}^{ab}(-u,-v) + 4 \widetilde{T}_{0a}(p,q) \widetilde{T}^{0a}(-u,-v)\nonumber\\
    &\hspace{0.5in} - \widetilde{T}_{0}^0(p,q) \widetilde{T}_{0}^0(-u,-v) - \widetilde{T}_{0}^0(p,q) \widetilde{T}_{c}^c(-u,-v)- \widetilde{T}_{a}^a(p,q) \widetilde{T}_{0}^0(-u,-v)- \widetilde{T}_{a}^a(p,q) \widetilde{T}_{b}^b(-u,-v) \Bigg]\nonumber\\
    &= \frac{i\kappa}{k^2} \Bigg[\frac{1}{4}\widetilde{T}_{00}(p,q) \widetilde{T}_{00}(-u,-v) \widetilde{T}_{0a}(p,q) \widetilde{T}^{0a}(-u,-v)  + \frac{1}{2}\widetilde{T}_{ab}(p,q) \widetilde{T}^{ab}(-u,-v) -\nonumber\\
    &\hspace{0.5in}- \frac{1}{4}\widetilde{T}_{a}^a(p,q) \widetilde{T}_{b}^b(-u,-v) +\frac{1}{4} \widetilde{T}_{00}(p,q) \widetilde{T}_{c}^c(-u,-v)+\frac{1}{4} \widetilde{T}_{a}^a(p,q) \widetilde{T}_{00}(-u,-v) \Bigg]\,.\label{eq:feyndiagC}
\end{align}
Note that we use the mostly plus signature of the metric, hence pulling a temporal index results in a sign change. By comparing \eqref{eq:feyndiagAB} and \eqref{eq:feyndiagC} we can see that they are identical, therefore we indeed have that the non-covariant and the covariant Feynman rules produce the same result for Coulomb-like scattering. \\
When considering a loop diagram as discussed at the end of section \ref{uvren3}, then the momentum inside the loop is not on-shell. Due to this, in a similar calculation as the one shown in this appendix, there will remain correction terms to the relations in \eqref{eq:momconsfeyndg1} - \eqref{eq:momconsfeyndg3} that will prevent one from directly seeing the equivalence. However, as we discuss in the main text, as both sets can be derived from the same underlying Lagrangian, we expect that they yield the same physics.

\section{Detailed computation of the UV-renormalisation}
\label{sec:apLoopInt}
In this appendix we present the evaluation of the vacuum self-energy diagram from equation \eqref{eq:selfendiag},
\begin{align}\label{eq:apL1}
\Pi_{vac}(u^2) = \frac{\kappa}{2} \mu^\epsilon \int d^dk \frac{u^2 k^2 +2m^2 uk -2m^4 \left( 1+\frac{\epsilon}{4}\right)}{[(k+u)^2+\lambda^2-i\epsilon] [k^2+m^2-i\epsilon]} \,,
\end{align}
which encodes the self-energy in vacuum of the scalar particle of mass $m$ caused by an internal triad particle of small auxiliary mass $\lambda$. For the evaluation, we follow the strategy in \cite{Hatfield:2019sox} for the regularisation of the QED self-energy at one-loop level. 

\subsection{Computation of the Loop integral}\label{secapren1}
In a first step we use the identity
\begin{equation}
    \frac{1}{x-i\epsilon} = i \int_0^\infty dz \; e^{-i z(x-i\epsilon)}
\end{equation}
in order to rewrite equation \eqref{eq:apL1} as
\begin{align}
    \Pi_{vac}(u^2) &= -\frac{\kappa}{2} \mu^\epsilon \int d^d k \left[ u^2k^2+2m^2uk-2m^4\left( 1+\frac{\epsilon}{4}\right)\right]\nonumber\\
    & \hspace{2.15in} \cdot\int_0^\infty dz_1 \int_0^\infty dz_2\; e^{-i z_1 (k^2+m^2-i\epsilon)} e^{-i z_2 ((k+u)^2+\lambda^2-i\epsilon)}\nonumber\\
    &= -\frac{\kappa}{2} \mu^\epsilon\int_0^\infty dz_1 \int_0^\infty dz_2 \int d^d k \left[ u^2k^2+2m^2uk-2m^4\left( 1+\frac{\epsilon}{4}\right)\right] \nonumber\\ &\hspace{2.15in} \cdot e^{-i(z_1+z_2) \left( k+\frac{z_2 u}{z_1+z_2} \right)^2 -\left[ \epsilon (z_1+z_2) + i\lambda^2 z_2 +i m^2 z_1 +i u^2 z_2 -i \frac{z_2^2 u^2}{z_1+z_2} \right]}\,.
\end{align}
Substituting $k \rightarrow k - \frac{z_2 u}{z_1+z_2}$ then yields
\begin{align}
     \Pi_{vac}(u^2) &= -\frac{\kappa}{2} \mu^\epsilon \int_0^\infty dz_1 \int_0^\infty dz_2 \int d^d k \Big[ u^2k^2 -2u^2 \frac{z_2 u k}{z_1+z_2} + \frac{z_2^2 u^4}{(z_1+z_2)^2}+ 2m^2 u k - 2m^2 \frac{z_2 u^2}{z_1+z_2} \nonumber\\ &\hspace{4.7in}-2m^4 \left( 1+\frac{\epsilon}{4}\right) \Big]\nonumber\\
     &\hspace{2.1in} \cdot e^{-i(z_1+z_2) k^2 - \left[ \epsilon(z_1+z_2) +i(\lambda^2+u^2)z_2 + i m^2 z_1 - i\frac{z_2^2 u^2}{z_1+z_2} \right]}\,.
\end{align}
Due to symmetry, all terms linear in $k$ vanish and only terms of two different kinds remain for which the $k$-integrations can be performed directly:
\begin{align}
    \int d^d k\; e^{-i (z_1+z_2) k^2} &=  \sqrt{\frac{-\pi i}{z_1+z_2}}^d \\
    \int d^d k\; k^2\, e^{-i (z_1+z_2) k^2} &= i \frac{\partial}{\partial (z_1+z_2)}  \int d^d k\; e^{-i (z_1+z_2) k^2} = \frac{d}{2\pi} \sqrt{\frac{-\pi i}{z_1+z_2}}^{d+2}\,.
\end{align}
Employing these, one can rewrite the self-energy as:
\begin{align}
    \Pi_{vac}(u^2) &= -\frac{\kappa}{2} \mu^\epsilon \int_0^\infty dz_1 \int_0^\infty dz_2 \; e^{-\left[ \epsilon(z_1+z_2) + i(\lambda^2+u^2)z_2 + im^2 z_1 -i\frac{z_2^2 u^2}{z_1+z_2} \right]} \nonumber\\ &\hspace{1.6in}\cdot \Bigg\{ \sqrt{\frac{-\pi i}{z_1+z_2}}^d \left[ \frac{z_2^2 u^4}{(z_1+z_2)^2} -2m^2 \frac{z_2 u^2}{z_1+z_2} -2m^4 \left( 1+\frac{\epsilon}{4} \right) \right] \nonumber\\ &\hspace{4.4in}+ \frac{u^2 d}{2\pi} \sqrt{\frac{-\pi i}{z_1+z_2}}^{d+2} \Bigg\}\,.
\end{align}
To continue, we use\footnote{This equality can be shown by using that $\delta(f(x))=\sum_i \frac{\delta(x-x_i)}{|f'(x_i)|}$, where $x_i$ are the points for which $f(x_i)=0$, $f'(x_i)\neq 0$ holds. That carries here over to $ \int_\xi^\infty d\beta \; \frac{z_1+z_2}{\beta} \delta(\beta-(z_1+z_2))$, where we used that $z_1+z_2 \geq 0$ which allowed us to drop the absolute value.}
\begin{equation}
   1 =\lim_{\xi \rightarrow 0} \int_\xi^\infty \frac{d\beta}{\beta} \; \delta \left( 1-\frac{z_1+z_2}{\beta}\right)\,.
\end{equation}
By substituting $z_1 \rightarrow z_1 \beta$ and $z_2 \rightarrow z_2 \beta$ we then obtain
\begin{align}
    \Pi_{vac}(u^2) &= -\frac{\kappa}{2} \mu^\epsilon \int_0^\infty dz_1 \int_0^\infty dz_2\; \lim_{\xi\rightarrow 0} \int_\xi^\infty d\beta \; \delta(1-(z_1+z_2)) \beta \, e^{-\beta\left[ \epsilon(z_1+z_2) + i(\lambda^2+u^2)z_2 + im^2 z_1 -i\frac{z_2^2 u^2}{z_1+z_2} \right]}
    \nonumber\\ &\hspace{1.4in}\cdot \Bigg\{\sqrt{\frac{-\pi i}{\beta(z_1+z_2)}}^d \left[ \frac{z_2^2 u^4}{(z_1+z_2)^2} -2m^2 \frac{z_2 u^2}{z_1+z_2} -2m^4 \left( 1+\frac{\epsilon}{4} \right) \right] \nonumber\\ &\hspace{4.4in}+ \frac{u^2 d}{2\pi} \sqrt{\frac{-\pi i}{\beta(z_1+z_2)}}^{d+2} \Bigg\}\nonumber\\
    &=-\frac{\kappa}{2}  \mu^\epsilon \int_0^1 dz \lim_{\xi\rightarrow 0}\int_\xi^\infty d\beta \; \beta \, e^{-\beta\left[ \epsilon+ i(\lambda^2+u^2)z + im^2 (1-z) -i z^2 u^2 \right]}
    \nonumber\\ &\hspace{1.4in}\cdot \left\{\sqrt{\frac{-\pi i}{\beta}}^d \left[z^2 u^4 -2m^2 z u^2-2m^4 \left( 1+\frac{\epsilon}{4} \right) \right] + \frac{u^2 d}{2\pi} \sqrt{\frac{-\pi i}{\beta}}^{d+2} \right\}\nonumber\\
    &=-\frac{\kappa}{2} \mu^\epsilon \int_0^1 dz \lim_{\xi\rightarrow 0} \int_\xi^\infty d\beta \;  e^{-i\beta\left[ -i\epsilon+ (\lambda^2+u^2)z + m^2 (1-z) - z^2 u^2 \right]}
    \nonumber\\ &\hspace{1.4in}\cdot \Bigg\{ -\pi^2 \frac{\beta^\frac{\epsilon}{2}}{\beta} \sqrt{-i\pi}^{-\epsilon} \left[z^2 u^4 -2m^2 z u^2-2m^4 \left( 1+\frac{\epsilon}{4} \right) \right] \nonumber\\ &\hspace{4in}+ i \pi^2 \frac{4-\epsilon}{2} u^2 \sqrt{-i\pi}^{-\epsilon} \frac{\beta^\frac{\epsilon}{2}}{\beta^2} \Bigg\}\,,
\end{align}
where in the last step we used $d=4-\epsilon$. Next, we perform the $\beta$-integration, that is 
\begin{equation}
    \lim_{\xi\rightarrow 0}\int_\xi^\infty d\beta \; e^{-i\beta a} \;\frac{\beta^{\frac{\epsilon}{2}}}{\beta^n}\,,
\end{equation}
where we already set $\epsilon=0$ in the exponential, with $a=(\lambda^2+u^2)z + m^2 (1-z) - z^2 u^2 $ and $n\in \{1,2\}$. To obtain the result, we apply the residue theorem. For $a >0$ the contour can be closed by a quarter circle from $\infty$ to $-i \infty+\xi$ and a line from $-i\infty+\xi$ to $\xi$. With this, the pole at $\beta=0$ can be avoided. The integral then becomes
\begin{equation}
     \lim_{\xi\rightarrow 0}\int_\xi^\infty d\beta \; e^{-i\beta a} \;\frac{\beta^{\frac{\epsilon}{2}}}{\beta^n} = -  \lim_{\xi\rightarrow 0}\int_0^\infty dt \; e^{-t} (t+i\xi a)^{\frac{\epsilon}{2}-n} (-i)^{\frac{\epsilon}{2}-n+1} a^{-\frac{\epsilon}{2}+n-1}\,,
\end{equation}
where we substituted $t = \beta a$. Expanding the term $(t+i\xi a)^{\frac{\epsilon}{2}-n}$ for small $\xi$ yields 
\begin{equation}
    (t+i\xi a)^{\frac{\epsilon}{2}-n} = t^{\frac{\epsilon}{2}-n} + i\xi a \left(\frac{\epsilon}{2}-n\right) t^{\frac{\epsilon}{2}-n-1} + O(\xi^2)\,.
\end{equation}
With the definition of the Gamma function
\begin{equation}
    \Gamma(z) = \int_0^\infty dt \; t^{z-1} e^{-t}
\end{equation}
it then follows that
\begin{equation}
    \lim_{\xi\rightarrow 0}\int_\xi^\infty d\beta \; e^{-i\beta a} \;\frac{\beta^{\frac{\epsilon}{2}}}{\beta^n} =  (i a)^{n-1-\frac{\epsilon}{2}} \left[ \Gamma\left(\frac{\epsilon}{2}-n+1\right) + \lim_{\xi\rightarrow 0} O(\xi) \right] =  (i a)^{n-1-\frac{\epsilon}{2}} \Gamma\left(\frac{\epsilon}{2}-n+1\right)\,.
\end{equation}
For $a<0$, the contour is closed by a quarter circle from $\infty$ to $i\infty+\xi$ and a line from $i\infty+\xi$ to $\xi$. The result turns out to be the same as for the case $a>0$. This thus yields
\begin{align}
    \Pi_{vac}(u^2) &= -\frac{\kappa}{2} \mu^\epsilon \int_0^1 dz \; \Bigg\{ -\pi^2 \sqrt{-i\pi}^{-\epsilon} \left[ z^2u^4-2m^2zu^2-2m^4\left(1+\frac{\epsilon}{4}\right) \right] i^{-\frac{\epsilon}{2}}\nonumber\\ &\hspace{2in} \cdot [(\lambda^2+u^2)z + m^2 (1-z) - z^2 u^2]^{-\frac{\epsilon}{2}} \Gamma\left( \frac{\epsilon}{2} \right)\nonumber\\
    &\hspace{1in} + i\pi^2 \left( 2-\frac{\epsilon}{2}\right) u^2 \sqrt{-i\pi}^{-\epsilon} i^{1-\frac{\epsilon}{2}} [(\lambda^2+u^2)z + m^2 (1-z) - z^2 u^2]^{1-\frac{\epsilon}{2}} \Gamma\left( \frac{\epsilon}{2}-1 \right) \Bigg\}\,.
\end{align}
Next, we expand all terms in $\epsilon$ and then perform the $z$-integration which results in
\begin{align}\label{apLoop:Pi1}
    \Pi_{vac}(u^2) = &- \frac{\pi^2 \kappa}{\epsilon} \left[ 2m^2(m^2+u^2) + \lambda^2 u^2 \right] \nonumber\\
    &+ \frac{\pi^2 \kappa}{2 u^2} \Bigg\{ u^2[2 m^2(m^2+u^2)(\gamma_E-2) + \lambda^2 (u^2 (\gamma_E-1)-m^2)]\nonumber\\
    &\hspace{0.5in} +m^2 (m^2+2u^2+\lambda^2) \sqrt{(m^2+u^2)^2 - 2\lambda^2(m^2-u^2)+\lambda^4} \nonumber\\ &\hspace{0.8in} \cdot\Bigg[ \arctanh \left( \frac{m^2+u^2-\lambda^2}{\sqrt{(m^2+u^2)^2 - 2\lambda^2 (m^2-u^2) +\lambda^4}} \right) \nonumber\\ &\hspace{1in} - \arctanh \left( \frac{m^2-u^2-\lambda^2}{\sqrt{(m^2+u^2)^2 - 2\lambda^2 (m^2-u^2) +\lambda^4}} \right)\Bigg]\nonumber\\
    &\hspace{0.5in} + \ln(\pi) [u^4 \lambda^2 + 2m^2u^2(m^2+u^2)] + \ln\left( \frac{\lambda}{m}\right) [m^6-3m^2u^2 \lambda^2 - m^2 \lambda^4]\nonumber\\
    &\hspace{0.5in} + \ln(\lambda) [3u^2 m^4 +2u^4 (m^2+\lambda^2)] + \ln(m) [u^2m^2 (2u^2+m^2)]\nonumber\\ &\hspace{3in} - \ln(\mu^2) [2u^2m^2(m^2+u^2)+\lambda^2 u^4]\Bigg\}\nonumber\\
    &+ O(\epsilon)\,,
\end{align}
where $\gamma_E$ denotes the Euler-Mascheroni constant.
Given this, it becomes evident that the pole in $\epsilon$ arises from the term $- \frac{\pi^2 \kappa}{\epsilon} \left[ 2m^2(m^2+u^2) + \lambda^2 u^2 \right]$, which yields in the limit of vanishing graviton mass $- \frac{\pi^2 \kappa}{\epsilon} \left[ 2m^2(m^2+u^2) \right]$. A suitable counter term should remove this divergence, which is discussed in the main text, and we are left with the regularised version
\begin{align}\label{apLoop:Pireg}
    \Pi_{vac}^{\text{reg}}(u^2) = & \frac{\pi^2 \kappa}{2 u^2} \Bigg\{ u^2[2 m^2(m^2+u^2)(\gamma_E-2) + \lambda^2 (u^2 (\gamma_E-1)-m^2)]\nonumber\\
    &\hspace{0.5in} +m^2 (m^2+2u^2+\lambda^2) \sqrt{(m^2+u^2)^2 - 2\lambda^2(m^2-u^2)+\lambda^4} \nonumber\\ &\hspace{0.8in} \cdot\Bigg[ \arctanh \left( \frac{m^2+u^2-\lambda^2}{\sqrt{(m^2+u^2)^2 - 2\lambda^2 (m^2-u^2) +\lambda^4}} \right) \nonumber\\ &\hspace{1in} - \arctanh \left( \frac{m^2-u^2-\lambda^2}{\sqrt{(m^2+u^2)^2 - 2\lambda^2 (m^2-u^2) +\lambda^4}} \right)\Bigg]\nonumber\\
    &\hspace{0.5in} + \ln(\pi) [u^4 \lambda^2 + 2m^2u^2(m^2+u^2)] + \ln\left( \frac{\lambda}{m}\right) [m^6-3m^2u^2 \lambda^2 - m^2 \lambda^4]\nonumber\\
    &\hspace{0.5in} + \ln(\lambda) [3u^2 m^4 +2u^4 (m^2+\lambda^2)] + \ln(m) [u^2m^2 (2u^2+m^2)]\nonumber\\ &\hspace{3in} - \ln(\mu^2) [2u^2m^2(m^2+u^2)+\lambda^2 u^4]\Bigg\}\,.
\end{align}
In order to determine the (arbitrary) finite part of the counter term, we pick the on-shell renormalisation condition. This requires the pole in the propagator to be at $u^2=-m^2$, that is it imposes the condition $\Pi(u^2=-m^2) \mbeq 0$, and the fixing of its residue such that $\frac{\partial}{\partial u^2}\Pi(u^2=-m^2) \mbeq 0$. Together, these two conditions imply the following formula for the renormalised $\Pi^R_{vac}$:
\begin{equation}
    \Pi^R_{vac}(u^2) := \Pi_{vac}^{\text{reg}}(u^2) - \Pi_{vac}^{\text{reg}}(-m^2) - (u^2+m^2) \frac{\partial}{\partial u^2} \Pi_{vac}^{\text{reg}}(-m^2)\,.
\end{equation}

Applied to \eqref{apLoop:Pi1}, one obtains for the additional terms:
\begin{align}\label{apploop:piregm2}
    \Pi_{vac}^{\text{reg}}(-m^2) = -\frac{\pi^2 \kappa}{2} \lambda \Bigg\{ &(m^2-\lambda^2) \sqrt{\lambda^2-4m^2} \left[ \arctanh\left( \frac{2m^2-\lambda^2}{\lambda\sqrt{\lambda^2-4m^2}} \right) + \arctanh\left( \frac{\lambda}{\sqrt{\lambda^2-4m^2}}\right) \right] \nonumber\\ 
    &+\lambda^3 \ln\left( \frac{m}{\lambda}\right) +m^2\lambda \left[ \gamma_E + \ln\left(\frac{\pi \lambda^5}{\mu^2 m^3}\right) \right] \Bigg\}\,.
\end{align}
as well as
\begin{align}
    \frac{\partial}{\partial u^2} \Pi_{vac}^{\text{reg}}(-m^2) =\frac{\pi^2\kappa}{2m^2 } \Bigg\{ &\frac{5m^4\lambda +2m^2\lambda^3 -\lambda^5}{\sqrt{\lambda^2-4m^2}} \left[ \arctanh\left( \frac{\lambda^2-2m^2}{\lambda\sqrt{\lambda^2-4m^2}}\right) -\arctanh\left( \frac{\lambda}{\sqrt{\lambda^2-4m^2}}\right) \right]\nonumber\\
    &+ (2\gamma_E-3) m^4 +m^2\lambda^2 (\gamma_E-2 +\ln(\pi) ) +(m^4-\lambda^4) \ln\left(\frac{m}{\lambda}\right) \nonumber\\
    &+2m^2\lambda^2 \ln(\lambda) +2m^4 \ln(m\pi \lambda) - m^2(2m^2+\lambda^2) \ln\left(\mu^2\right) \Bigg\}\,.
\end{align}
If we had not introduced the small triad mass $\lambda$, then this last expression would be divergent, see for instance \cite{Arteaga:2003we,Hatfield:2019sox}. Hence we continue to work now with $\Pi^R_{vac}(u^2)$. This concludes the discussion of the renormalisation of the self-energy diagram. 

\subsection{Contribution to the master equation}\label{secapren2}
In order to see the effect of the renormalisation on the master equation, one has to evaluate the following expression, as discussed at the end of section \ref{uvren2}:
\begin{equation}
    \Xi_R(\omega_u,\vec u,t_0,t) = \int_{t_0}^t d\tau \int_\mathbb{R} du^0 \; \Pi^R_{vac}(u^2) \: \cos[(u^0-\omega_u)(t-\tau)]\,.
\end{equation}
Substituting $t-\tau \rightarrow \tau$ yields
\begin{equation}
    \Xi_R(\omega_u,\vec u,t_0,t) = \int_{0}^{t-t_0} d\tau \int_\mathbb{R} du^0 \; \Pi^R_{vac}(u^2) \: \cos[(u^0-\omega_u)\tau]
\end{equation}
and due to symmetry it holds that
\begin{align}
    \int_\mathbb{R} du^0 \; \Pi^R_{vac}(u^2) \: \cos[(u^0-\omega_u)\tau] &=\int_\mathbb{R} du^0 \; \Pi^R_{vac}(\vec{u}^2 - u_0^2) \: [\cos(u^0 \tau) \cos(\omega_u \tau) + \sin(u^0 \tau) \sin(\omega_u \tau)] \nonumber\\
    &=\cos(\omega_u \tau) \int_\mathbb{R} du^0 \; \Pi^R_{vac}(\vec{u}^2 - u_0^2) \: \cos(u^0 \tau) \,.
\end{align}
To solve the integrations, we first consider all terms that depend on $u^0$ in the form $(u^2+m^2)= (\omega_u^2-u_0^2)$ with $\omega_u=\sqrt{\vec{u}^2+m^2}$. \\
To evaluate this, we would like to use the distributional integration $\int_\mathbb{R} du^0 \; \cos(u^0 \tau) = \pi \delta(\tau)$. However, in order for this to be true, we would need to have a Schwartz function paired with the distribution under the $\tau$ integration and its integration domain should be $\mathbb{R}$. To have this, we modify the cosine slightly and we will see when evaluating the integration that this modification does not affect the final value. We introduce the Schwartz function $S_c(\omega_u \tau)$ which coincides on the interval $[\epsilon, t-t_0-\epsilon]$ with $\cos(\omega_u\tau)$, where $\epsilon \ll 1$. On the interval $[-\epsilon,\epsilon]$ it is a smooth function constructed in such a way that $S_c(0) = \frac{1}{2}\cos(0)=\frac{1}{2}$, $\frac{d^2}{d\tau^2} S_c(\omega_u \tau) |_{\tau=0} =\frac{1}{2} \frac{d^2}{d\tau^2} \cos(\omega_u \tau)|_{\tau=0} =- \frac{\omega_u^2}{2}$ and $S_c(\omega_u \tau)=0$ for $\tau < \epsilon$. For $[t-t_0-\epsilon,t-t_0+\epsilon]$ it is also constructed as a smooth function from $S_c(\omega_u (t-t_0-\epsilon)) = \cos(t-t_0-\epsilon)$ to $0$ for $S_c(\omega_u \tau)$ for $\tau > t-t_0+\epsilon$. The construction is in such a way, that $S_c(\omega_u \tau)$ is a Schwartz function on $\mathbb{R}$. Then we have: 
\begin{align}
    &\int_{0}^{t-t_0} d\tau \cos(\omega_u \tau) \int_\mathbb{R} du^0 \; (\omega_u^2-u_0^2) \: \cos(u^0\tau)\nonumber\\
    &= \int_{\mathbb{R}} d\tau\; S_c(\omega_u \tau) \left[ \pi \delta(\tau)\, \omega_u^2 + \frac{d^2}{d \tau^2} \int_\mathbb{R} du^0 \; \cos(u^0\tau) \right] \nonumber\\
    &= \int_{\mathbb{R}} d\tau\; S_c(\omega_u \tau) \left[ \pi \delta(\tau)\, \omega_u^2 + \pi \frac{d^2}{d \tau^2} \delta(\tau) \right]\nonumber\\
    &= \int_{\mathbb{R}} d\tau\;  \left[ \pi \delta(\tau)\, \omega_u^2 \, S_c(\omega_u \tau) + \pi \delta(\tau) \, \frac{d^2}{d \tau^2} S_c(\omega_u \tau)  \right]\nonumber\\
    & = \pi \left[ \frac{\omega_u^2}{2}- \frac{\omega_u^2}{2} \right] = 0\,.
\end{align}
Using this\footnote{And once also using that hence $\ln(m) u^2 m^2 \rightarrow -\ln(m) m^4$}, what remains is
\begin{align}
   \frac{\pi^2\kappa}{2} \int_0^{t-t_0} d\tau \; \cos(\omega_u\tau) \int_\mathbb{R} du^0\; \cos(u^0 \tau) \, \Bigg\{& (m^4-\lambda^4) \frac{m^2+u^2}{u^2} \ln\left(\frac{\lambda}{m}\right) \nonumber\\
   &+ m^2 \frac{m^2+2u^2+\lambda^2}{u^2} \sqrt{(m^2+u^2)^2-2\lambda^2(m^2-u^2)+\lambda^4} 
    \cdot\nonumber\\
    &\hspace{0.2in}\cdot \Bigg[ \arctanh\left( \frac{m^2+u^2-\lambda^2}{\sqrt{(m^2+u^2)^2-2 \lambda^2(m^2-u^2)+\lambda^4}} \right) \nonumber\\
    &\hspace{.4in} - \arctanh\left( \frac{m^2-u^2-\lambda^2}{\sqrt{(m^2+u^2)^2-2 \lambda^2(m^2-u^2)+\lambda^4}} \right)\Bigg] \nonumber\\
    &+\lambda (m^2-\lambda^2) \sqrt{\lambda^2 -4m^2} \Bigg[\arctanh\left( \frac{2m^2-\lambda^2}{\lambda\sqrt{\lambda^2-4m^2}} \right) \nonumber\\
    &\hspace{1.7in} +\arctanh\left( \frac{\lambda}{\sqrt{\lambda^2-4m^2}} \right)  \Bigg] \Bigg\}\,,
\end{align}
which is independent of $\mu$. Before explicitly evaluating the integrations, we simplify the integrands by taking the limit $\lambda\rightarrow 0$ where possible, as $\lambda$ was only introduced as artificial small graviton mass to be able to fix the residuum of the pole in the propagator. As $\arctanh(i 0)$ and $\arctanh(i \infty)$ are finite, the last two lines vanish. Also, $\lim_{\lambda\rightarrow 0}\lambda^4 \ln(\lambda)=0$ and, when expanding the square root in the second line, we find that $\lim_{\lambda\rightarrow 0} \lambda \arctanh(1+\lambda)=0$ as $\arctanh(x) = \frac{1}{2} \ln(1+x) -\frac{1}{2} \ln(1-x)$.

This leaves us with
\begin{align}
   \frac{\pi^2\kappa}{2} \int_0^{t-t_0} d\tau \; \cos(\omega_u\tau) \int_\mathbb{R} du^0\; \cos(u^0 \tau) \, \Bigg\{& m^4\frac{m^2+u^2}{u^2} \ln\left(\frac{\lambda}{m}\right) \nonumber\\
   &+ m^2 \frac{m^2+2u^2}{u^2} (m^2+u^2) 
    \cdot\nonumber\\
    &\hspace{0.2in}\cdot \Bigg[ \arctanh\left( \frac{m^2+u^2-\lambda^2}{\sqrt{(m^2+u^2)^2-2 \lambda^2(m^2-u^2)+\lambda^4}} \right) \nonumber\\
    &\hspace{.4in} - \arctanh\left( \frac{m^2-u^2}{m^2+u^2} \right)\Bigg]\Bigg\}\,.
\end{align}
From the above named relation $\arctanh(x) = \frac{1}{2} \ln(1+x) -\frac{1}{2} \ln(1-x)$ follows that
\begin{equation}
    \arctanh \left(\frac{x}{y}\right) = \frac{1}{2}\ln\left(\frac{x+y}{y-x}\right)\,.
\end{equation}
This yields for the last line:
\begin{equation}
    \arctanh\left( \frac{m^2-u^2}{m^2+u^2} \right) = \frac{1}{2} \ln\left(\frac{m^2}{u^2}\right) \,.
\end{equation}
For the line before, we expand the argument in second order for small $\lambda$ and obtain
\begin{align}
    \arctanh\left( \frac{m^2+u^2-\lambda^2}{\sqrt{(m^2+u^2)^2-2 \lambda^2(m^2-u^2)+\lambda^4}} \right) \approx \arctanh\left( 1 - \frac{2\lambda^2 u^2}{(m^2+u^2)^2}\right)\,.
\end{align}
Higher orders will not contribute in the final limit $\lambda \rightarrow 0$. Expressing the $\arctanh$ again in terms of logarithms, we obtain
\begin{align}
     \arctanh\left( 1 - \frac{2\lambda^2 u^2}{(m^2+u^2)^2}\right) &= \frac{1}{2} \ln\left( 2 - \frac{2\lambda^2 u^2}{(m^2+u^2)^2}\right) -\frac{1}{2} \ln\left( \frac{2\lambda^2 u^2}{(m^2+u^2)^2}\right) \nonumber\\ &\approx  \frac{1}{2} \ln\left( 2\right) -\frac{1}{2} \ln\left( \frac{2\lambda^2 u^2}{(m^2+u^2)^2}\right)\,, 
\end{align}
where we neglected terms of order $\lambda^2$ and higher. With these simplifications one then finds that
\begin{align}
    \Xi_R(\omega_u,\vec u,t_0,t) = \frac{\pi^2\kappa}{2} \int_0^{t-t_0} d\tau \; \cos(\omega_u\tau) \int_\mathbb{R} du^0\; \cos(u^0 \tau) \, \Bigg\{& m^4\frac{m^2+u^2}{u^2} \ln\left(\frac{\lambda}{m}\right) \nonumber\\
   &- \frac{m^2}{2} \frac{m^2+2u^2}{u^2} (m^2+u^2) 
    \cdot\nonumber\\
    &\hspace{0.2in}\cdot \Bigg[ \ln\left( \frac{2\lambda^2 u^2}{(m^2+u^2)^2}\right) +\ln\left(\frac{m^2}{2 u^2}\right)\Bigg]\Bigg\}\nonumber\\
    = \frac{\pi^2\kappa}{2} \int_0^{t-t_0} d\tau \; \cos(\omega_u\tau) \int_\mathbb{R} du^0\; \cos(u^0 \tau) \, \Bigg\{& m^4\frac{m^2+u^2}{u^2} \ln\left(\frac{\lambda}{m}\right) \nonumber\\
   &- \frac{m^2}{2} \frac{m^2+2u^2}{u^2} (m^2+u^2) 
    \cdot\nonumber\\
    &\hspace{0.2in}\cdot \ln\left( \frac{\lambda^2m^2}{(m^2+u^2)^2}\right) \Bigg\}\,.
\end{align}
The next step is to solve the following two integrations:
\begin{align}
    \textbf{(A) } \hspace{0.2in}& m^4 \ln\left(\frac{\lambda}{m}\right) \int_\mathbb{R} du^0\; \cos(u^0 \tau) \, \frac{\omega_u^2-u_0^2}{\vec{u}^2-u_0^2}\\
    \textbf{(B) } \hspace{0.2in}& -\frac{m^2}{2} \int_\mathbb{R} du^0\; \cos(u^0 \tau) \, \frac{m^2+2 \vec{u}^2-2 u_0^2}{\vec{u}^2-u_0^2} (\omega_u^2-u_0^2) \ln\left( \frac{\lambda^2 m^2}{(\omega_u^2-u_0^2)^2} \right)\,.
\end{align}
We start with \textbf{(A)}:
\begin{align}
    \int_\mathbb{R} du^0\; \cos(u^0 \tau) \, \frac{\omega_u^2-u_0^2}{\vec{u}^2-u_0^2} = -\int_\mathbb{R} du^0\; e^{i u^0 \tau} \, \frac{\omega_u^2-u_0^2}{(u^0-|\vec{u}|)(u^0+|\vec{u}|)}\,.
\end{align}
Applying the residue theorem by closing the contour with a semi-circle in the upper half plane leaves us with the contributions of the poles that lie on the contour, hence contribute $+\frac{1}{2}$ their residua, yielding
\begin{align}
    -\int_\mathbb{R} du^0\; e^{i u^0 \tau} \, \frac{\omega_u^2-u_0^2}{(u^0-|\vec{u}|)(u^0+|\vec{u}|)} = - \pi i \left[ e^{i|\vec u| \tau} \frac{m^2}{2|\vec u|} -  e^{-i|\vec u| \tau} \frac{m^2}{2|\vec u|}\right] = \frac{\pi m^2}{|\vec u|} \sin(|\vec u| \tau)\,.
\end{align}
Therefor we have
\begin{equation}
    \textbf{(A) } = \frac{\pi m^6}{|\vec u|} \sin(|\vec u| \tau)\ln\left(\frac{\lambda}{m}\right)\,.
\end{equation}
We proceed with \textbf{(B)}:
\begin{align}
    &\int_\mathbb{R} du^0\; \cos(u^0 \tau) \, \frac{m^2+2 \vec{u}^2-2 u_0^2}{\vec{u}^2-u_0^2} (\omega_u^2-u_0^2) \ln\left( \frac{\lambda^2 m^2}{(\omega_u^2-u_0^2)^2} \right)\nonumber\\
    &= -\int_\mathbb{R} du^0\; e^{i u^0 \tau} \, \frac{m^2+2 \vec{u}^2-2 u_0^2}{(u^0-|\vec{u}|)(u^0+|\vec u|)} (\omega_u^2-u_0^2) \ln\left( \frac{\lambda^2 m^2}{(\omega_u^2-u_0^2)^2} \right)\,.
\end{align}
We apply the residue theorem once again. The integrand has singularities at $u^0=\pm |\vec u|$. Working with the principal value logarithm, i.e. the complex logarithm with branch cut at the negative real axis, the logarithm here does not have any branch cuts given that $\omega_u>0$, hence the equation $\frac{\lambda^2 m^2}{(\omega_u^2-u_0^2)^2} \mbeq - \alpha$ with $\alpha\in \mathbb{R}, \alpha > 0$ does not have any solution for $u^0 \in \mathbb{C}$. Due to the prefactor $(\omega_u^2-u_0^2)$, there is no singularity in $u^0 = \pm \omega_u$. We pick a closed integration contour from $-\infty$ to $-|\vec u| - \epsilon$, then go in a semi-circle clockwise around the pole to $-|\vec u| + \epsilon$, continue to $|\vec u| - \epsilon$, again go around the singularity in a semi-circle clockwise to $|\vec u| + \epsilon$, continue to $+\infty$ and close it with a semicircle in the upper half-plane. The closed contour does not contain any singularities, hence its contribution vanishes. Due to the exponential $e^{i u^0 \tau}$, also the semi-circle at infinite radius in the upper half-plane vanishes. Hence only the contributions of the singularities at $u^0 = \pm |\vec u|$ remain. As we went for the closed contour around them clockwise, the singularity contributions have to be evaluated counter-clockwise and added to the closed contour. We start by investigating the one at $u^0 = +|\vec u|$ and replace $u^0 = |\vec u| + \epsilon e^{i\phi}$:
\begin{align}
     &-\lim_{\epsilon\rightarrow 0} \int_0^\pi d\phi\; i \epsilon e^{i\phi} e^{i\tau \epsilon e^{i\phi} +i\tau |\vec u|} \frac{m^2 - 4 |\vec u| \epsilon e^{i\phi} - 2\epsilon^2 e^{2i\phi}}{\epsilon e^{i\phi} (2|\vec u| +\epsilon e^{i\phi})} (m^2 - 2 |\vec u| \epsilon e^{i\phi} - \epsilon^2 e^{2i\phi}) \nonumber\\ &\hspace{4in} \cdot \ln\left( \frac{\lambda^2 m^2}{(m^2 - 2 |\vec u| \epsilon e^{i\phi} - \epsilon^2 e^{2i\phi})^2} \right)\nonumber\\
     &= -i \lim_{\epsilon\rightarrow 0} \int_0^\pi d\phi\; e^{i\tau |\vec u|} \frac{m^2 - 4 |\vec u| \epsilon e^{i\phi} - 2\epsilon^2 e^{2i\phi}}{2|\vec u| +\epsilon e^{i\phi}} (m^2 - 2 |\vec u| \epsilon e^{i\phi} - \epsilon^2 e^{2i\phi}) \ln\left( \frac{\lambda^2 m^2}{(m^2 - 2 |\vec u| \epsilon e^{i\phi} - \epsilon^2 e^{2i\phi})^2} \right)\nonumber\\
     &= -i \lim_{\epsilon\rightarrow 0} \int_0^\pi d\phi\; e^{i\tau |\vec u|} m^2  \frac{1}{2|\vec u|} m^2\ln\left( \frac{\lambda^2 m^2}{(m^2 - 2 |\vec u| \epsilon e^{i\phi} - \epsilon^2 e^{2i\phi})^2} \right)\nonumber\\
     &= -i \frac{m^4}{2 |\vec u|} e^{i\tau |\vec u|} \lim_{\epsilon\rightarrow 0} \int_0^\pi d\phi\;\ln\left(\frac{\lambda^2}{m^2}\right)\nonumber\\
     &=-i \frac{\pi m^4}{|\vec u|} e^{i\tau |\vec u|} \ln\left(\frac{\lambda}{m}\right)\,,
\end{align}
where in the first step we expanded $e^{i\tau \epsilon e^{i\phi} +i\tau |\vec u|}  = e^{i\tau |\vec u|} (1 + O(\epsilon)) $ and neglected all but the zeroth order due to the limit, in the second step we expanded $\frac{1}{2|\vec u| +\epsilon e^{i\phi}} =  \frac{1}{2|\vec u|} \left(1-\frac{\epsilon}{2|\vec u|} e^{i\phi} + O(\epsilon) \right)$ and applied the limit to the terms depending on $\epsilon$ where possible. In the third step, we expanded
\begin{equation}
    \ln\left( \frac{\lambda^2 m^2}{(m^2 - 2 |\vec u| \epsilon e^{i\phi} - \epsilon^2 e^{2i\phi})^2} \right) = \ln\left(\frac{\lambda^2}{m^2}\right) + \epsilon \frac{4 |\vec u|}{m^2} e^{i\phi} + O(\epsilon^2)\,.
\end{equation}
For the other singularity at $u^0=-|\vec u|$ we find analogously for $u^0 = -|\vec u|+\epsilon e^{i\phi}$:
\begin{align}
     &-\lim_{\epsilon\rightarrow 0} \int_0^\pi d\phi\; i \epsilon e^{i\phi} e^{i\tau \epsilon e^{i\phi} -i\tau |\vec u|} \frac{m^2 + 4 |\vec u| \epsilon e^{i\phi} - 2\epsilon^2 e^{2i\phi}}{\epsilon e^{i\phi} (-2|\vec u| +\epsilon e^{i\phi})} (m^2 + 2 |\vec u| \epsilon e^{i\phi} - \epsilon^2 e^{2i\phi})\nonumber\\ &\hspace{4in}\cdot \ln\left( \frac{\lambda^2 m^2}{(m^2 + 2 |\vec u| \epsilon e^{i\phi} - \epsilon^2 e^{2i\phi})^2} \right)\nonumber\\
     &= +i \lim_{\epsilon\rightarrow 0} \int_0^\pi d\phi\; e^{-i\tau |\vec u|} m^2  \frac{1}{2|\vec u|} m^2 \ln\left( \frac{\lambda^2 m^2}{(m^2 + 2 |\vec u| \epsilon e^{i\phi} - \epsilon^2 e^{2i\phi})^2} \right)\nonumber\\
     &= i \frac{m^4}{2 |\vec u|} e^{-i\tau |\vec u|} \lim_{\epsilon\rightarrow 0} \int_0^\pi d\phi\;  \ln\left(\frac{\lambda^2}{m^2}\right)\nonumber\\
     &=i \frac{\pi m^4}{|\vec u|} e^{-i\tau |\vec u|}  \ln\left(\frac{\lambda}{m}\right)\,.
\end{align}
Combining these two yields
\begin{align}
    \textbf{(B)} =- \frac{\pi m^6}{|\vec u|} \sin(|\vec u| \tau) \ln\left(\frac{\lambda}{m}\right) = -\textbf{(A)}\,.
\end{align}
From this follows that we have
\begin{equation}
    \Xi_R(\omega_u,\vec u,t_0,t) = 0\,.
\end{equation}
Due to the way the renormalised quantities entered in the master equation, the result is now independent of the scale $\mu$ as well as of the artificial graviton mass $\lambda$, whose limit to zero can therefore be taken without problems.

\section{Application of the Markov approximation}
\label{app2Mark}
In this appendix, we apply the Markov approximation to the renormalised one-particle master equation \eqref{eq:meqren}. Applying the formula given in \eqref{eq:2mapdpv}, we obtain two classes of terms, one class that contains the $\delta$-distributions and the other one containing the Cauchy principal value. Before we evaluate them in subsections \ref{app:MarkDel} and \ref{app:MarkPV}, we first discuss the applicability of the Markov approximation for ultra-relativistic particles with focus on neutrinos in subsection \ref{sec:applic2M}. 

\subsection{Applicability of the Markov approximation for ultra-relativistic particles}\label{sec:applic2M}
In general, the Markov approximation can be applied if the timescales $\tau_B$ on which the correlation functions decay are much smaller than the timescales $\tau_R$ on which the state of the system varies (see \cite{Breuer:2002pc}). The identification of these timescales is however hard without solving the one-particle master equation before the application of the approximation. As the Markov approximation corresponds to sending $t_0 \rightarrow -\infty$ and hence $\int_0^{t-t_0} d\tau \longrightarrow \int_0^\infty d\tau$, we will analyse the error one makes when extending the integration domain from $t-t_0$ to $\infty$. If the integrand is strongly peaked around $\tau=0$, which is usually assumed when deriving Markovian master equations, then the error of the additional contribution should be negligible. For this, we analyse the different parts of the renormalised one-particle master equation in \eqref{eq:meqren}:
\begin{align}\label{eq:meqrenWH}
\frac{\partial}{\partial t} \rho(\vec{u},\vec{v},t) = &-i \rho(\vec{u},\vec{v},t) \; (\omega_u-\omega_v) \nonumber\\
&-\frac{\kappa}{2} \int \frac{d^3k}{(2\pi)^3} \bigg\{ \frac{ P_u(\vec k)}{\omega_{u-k}\omega_u}\; \left[C^R(\vec{u},\vec{k},t)+\delta_P C_P^R(\vec{u},\vec{k},t)\right] \nonumber\\ &\hspace{1.1in} +\frac{P_v(\vec k)}{\omega_{v-k}\omega_v}\; \left[\left(C^R(\vec{v},\vec{k},t)\right)^*+\delta_P \left(C_P^R(\vec{v},\vec{k},t)\right)^*\right]\bigg\}\rho(\vec{u},\vec{v},t)\nonumber\\
&+\frac{\kappa}{2} \int \frac{d^3k}{(2\pi)^3} \frac{P_{ijln}(\vec{k})\:u^i u^j  v^l v^n}{\sqrt{\omega_{u+k}\omega_u\omega_{v+k}\omega_v}}\; \left\{C^R(\vec{u}+\vec{k},\vec{k},t)+ \left(C^R(\vec{v}+\vec{k},\vec{k},t)\right)^*\right\} \rho(\vec{u}+\vec{k},\vec{v}+\vec{k},t) 
\end{align}
with
\begin{align}
C^R(\vec{u},\vec{k},t) &= 2 \int_0^{t-t_0} \frac{d\tau}{\Omega_k} N(k)\:\cos[\Omega_k \tau] e^{ -i(\omega_{u-k}-\omega_u)\tau}\\
C_P^R(\vec{u},\vec{k},t) &= 2\int_0^{t-t_0} \frac{d\tau}{\Omega_k} N(k)\:\cos[\Omega_k \tau] e^{ -i(\omega_{u-k}+\omega_u)\tau}\,.
\end{align}
We start with $C^R$ in the second line of equation \eqref{eq:meqrenWH}, the other terms will be discussed below equation \eqref{eq:rev1Aplmarkli}. The term we have to take into account is
\begin{align}\label{eq:appAplmeq1}
    &-\kappa \int\frac{d^3k}{(2\pi)^3} \frac{P_u(\vec k)}{\omega_{u-k}\omega_u} \int_0^{t-t_0} \frac{d\tau}{\Omega_k} N(k)\: \cos[\Omega_k \tau] e^{-i (\omega_{u-k} - \omega_u)\tau}  \,.
\end{align}

To draw general statements on the effect of sending $t-t_0\to\infty$ is rather complicated, so we will restrict our discussion of the applicability of the Markov approximation to ultra-relativistic scalar particles $u\gg m$ that fulfil additionally
\begin{equation}\label{eq:rev1condubeta}
    u \gg \frac{1}{c\beta}\,,
\end{equation}
where we stated this condition in SI units. This can equally be interpreted as an upper bound on the temperature, which might seem a bit unusual given the Markov limit being often referred to an high-temperature limit, see e.g. \cite{Breuer:2002pc}. Our strategy here is however to show that under these circumstances the result of the $\tau$-integration will, after performing the $\vec{k}$-integration, only be negligibly affected by sending $t-t_0\rightarrow\infty$ once the particle has propagated a suitably ling time interval $t-t_0$, which we will specify below. We refer to the term Markov approximation here in this sense. \\
An inspection of the numerator in \eqref{eq:appAplmeq1}, excluding $N(k)$, reveals that for large values of $|\vec{k}|$ it can always be bounded from above, given that $P_u(\vec{k})$ only depends on the angle between $\vec{u}$ and $\vec{k}$ and is thus independent of $|\vec{k}|$. The denominator causes the entire integrand to vanish for large $|\vec{k}|$ due to the exponential in $N(k)$, which goes as $\sim \frac{1}{e^{\beta |\vec{k}|}}$ for large values of $\beta |\vec{k}|$. To simplify the entire integrand, we approximate
\begin{equation}
    \omega_{u-k} = \sqrt{m^2 + (\vec{u}-\vec{k})^2} \approx \sqrt{m^2 + |\vec{u}|^2} = \omega_u\,.
\end{equation}
This approximation is justified in the regime we are considering, as once $|\vec{k}|$ has the same order of magnitude as $|\vec{u}|$, then condition \eqref{eq:rev1condubeta} gives $k \beta \gg 1$, which means that the contribution to the integral of these terms is exponentially suppressed. \\
This simplifies \eqref{eq:appAplmeq1} and allows an evaluation in spherical coordinates:
\begin{align}
    &-\frac{\kappa}{\omega_u^2} \int_0^{t-t_0} d\tau \int\frac{d^3k}{(2\pi)^3} P_u(\vec k) N(k) \frac{\cos(\Omega_k\tau)}{\Omega_k}\nonumber\\
    &= -\frac{u^4 \kappa}{8\pi^2\omega_u^2} \int_0^{t-t_0} d\tau \int_0^\pi d\theta\; \sin^5\theta \int_0^\infty dk\; k \frac{\cos(k \tau)}{e^{\beta k}-1} \nonumber\\
    &=-\frac{2 u^4  \kappa}{15 \pi^2 \omega_u^2} \int_0^{t-t_0} d\tau \int_0^\infty dk\; k \frac{\cos(k \tau)}{e^{\beta k}-1}\,.
\end{align}
The $k$-integration can now be solved using the Hurwitz Zeta function defined for $\text{Re}(n)>1$ and $\text{Re}(a)>0$ as
\begin{equation}
    \zeta(n,a) = \frac{1}{\Gamma(n)} \int_0^\infty dk\; k^{n-1} \frac{e^{-a k}}{1-e^{-k}} 
\end{equation}
with $\Gamma(n) = (n-1)!$ for $n\in\mathbb{N}$, by rewriting 
\begin{align}\label{eq:rev1Aplmarkli}
    &-\frac{ u^4  \kappa}{15 \pi^2\omega_u^2} \int_0^{t-t_0} d\tau \int_0^\infty dk\; k \frac{e^{ik \tau} + e^{-ik\tau} }{e^{\beta k}-1}\nonumber\\
    &=-\frac{u^4  \kappa}{15\pi^2 \beta^2 \omega_u^2} \int_0^{t-t_0} d\tau \int_0^\infty d\tilde{k}\; \tilde{k} \frac{e^{i\tilde{k}\frac{\tau}{\beta}} + e^{-i\tilde{k}\frac{\tau}{\beta}} }{e^{\tilde{k}}-1}\nonumber\\ 
    &=-\frac{u^4  \kappa}{15\pi^2 \beta^2 \omega_u^2} \int_0^{t-t_0} d\tau \int_0^\infty d\tilde{k}\;  \left[ \tilde{k} \frac{e^{-\left(1-i \frac{\tau}{\beta} \right)\tilde k} }{1-e^{-\tilde k}} +\tilde{k} \frac{e^{-\left(1+i\frac{\tau}{\beta} \right)\tilde k} }{1-e^{- \tilde k}}\right]\nonumber\\
    &= -\frac{u^4  \kappa}{15\pi^2 \beta^2 \omega_u^2} \int_0^{t-t_0} d\tau  \left[ \zeta\left(2, 1-i\frac{\tau}{\beta}\right) + \zeta\left(2,1+i\frac{\tau}{\beta} \right) \right]\nonumber\\
    &= -\frac{u^4  \kappa}{15\pi^2 \beta^2 \omega_u^2} \int_0^{t-t_0} d\tau  \left[ \frac{\beta^2}{\tau^2} - \frac{\pi^2}{\sinh^2\left( \frac{\pi\tau}{\beta}\right)} \right]\nonumber\\
    &= -\frac{u^4  \kappa}{15\pi^2 \beta \omega_u^2} \left[ \pi \coth\left( \frac{\pi (t-t_0)}{\beta} \right) - \frac{\beta}{t-t_0} \right]\,.
\end{align}
where we substituted $\tilde{k} := \beta k$. \\
For the second term in the second line of equation \eqref{eq:meqrenWH}, one has in the $\tau$-integration an additional factor of $e^{-2i\omega_u \tau}$ contributing. Given that the remaining integrand is non-negative on the integration domain, the absolute value of the integrand for the $\tau$-integral does no change compared to \eqref{eq:rev1Aplmarkli} and hence the contribution from the second term in line two of equation \eqref{eq:meqrenWH} is less or equal to equation \eqref{eq:rev1Aplmarkli}. The argumentation used so far provides the same results to the remaining terms in equation \eqref{eq:meqrenWH} when assuming that also $v\gg \frac{1}{c\beta}$. This allows then to replace $\vec{u} +\vec{k}\approx \vec{u}$ and $\vec{v} +\vec{k}\approx \vec{v}$. Therefore, it suffices to discuss only the contribution in equation \eqref{eq:rev1Aplmarkli}, as it generalises to all the other contributions in equation \eqref{eq:meqrenWH}.\\

Application of the Markov approximation now means replacing $t-t_0 \rightarrow\infty$. From equation \eqref{eq:rev1Aplmarkli} it is visible that the approximation does change the result negligibly if the propagation time of the scalar particle under consideration is much larger than $\beta$ of the environment, i.e. in SI units if 
\begin{equation}
    t-t_0 \gg \hbar \beta\,.
\end{equation}
Thus the Markov approximation is applicable on the one-particle master equation \eqref{eq:meqrenWH} if the following conditions hold for the parameters of the total system under consideration:
\begin{equation}
    t-t_0 \gg  \hbar\beta \gg \frac{\hbar}{c u} \hspace{0.2in} \text{and} \hspace{0.2in} u\gg m\,.
\end{equation}
Note that this is not a necessary, but only a sufficient condition for the applicability of the Markov approximation.\\
Considering typical parameters used in decoherence experiments with ultra-relativistic atmospheric neutrinos (see e.g. \cite{Lisi:2000zt,MORGAN2006311,coloma2018decoherence,Domi:2024ypm,icecube2024search,aiello2025search}), we see that in this case the conditions are fulfilled: There, typical baselines are starting at the order of $10^7 \rm m$, which correspond to propagation times (assuming as speed the speed of light) of $t-t_0 \gtrsim 10^{-2}s$. On the other hand, the neutrino energies of interest start usually from the MeV scale and go into the TeV scale. To use conservative bounds, we assume that $c u \gtrsim 1 \rm MeV$, which yields $\frac{\hbar}{c u} \lesssim 10^{-21}s$. Thus for this class of particles, the Markov approximation is a good approximation for environmental temperatures $\Theta$ in the range
\begin{equation}
    10^{10}\rm K \gg \Theta \gg 10^{-9}\rm K\,.
\end{equation}
This includes in particular the case of $\Theta=0.9\rm K$ used in \cite{Domi:2024ypm} for cosmological gravitational waves.

\subsection{Evaluation of the delta terms}\label{app:MarkDel}

The main step in the application of the Markov approximation consists in replacing 
\begin{equation}\label{eq:appMapIdpv}
    \int_0^{t-t_0} d\tau \; e^{-i\omega \tau} \longrightarrow \int_0^{\infty} d\tau \; e^{-i\omega \tau} = \pi \delta(\omega) - PV\left( \frac{i}{\omega}\right)\,,
\end{equation}
as argued in detail in the main text. In this subsection, we investigate the terms containing the delta distribution and in the next subsection we evaluate the principal value terms. The delta distribution parts of the terms in lines two and three of \eqref{eq:meqrenWH} become
\begin{align}
    -\frac{\pi \kappa}{2} \int \frac{d^3k}{(2\pi)^3} \bigg\{& \frac{ P_u(\vec k)}{\omega_{u-k}\omega_u}\; \frac{N(k)}{\Omega_k} \Bigg[ \delta(\Omega_k -\omega_{u-k} +\omega_u) + \delta (\Omega_k +\omega_{u-k} -\omega_u) \nonumber\\ &\hspace{1.1in}+\delta_P \delta(\Omega_k -\omega_{u-k} -\omega_u) +  \delta_P \delta(\Omega_k +\omega_{u-k} +\omega_u)\Bigg]\nonumber\\
    & +\frac{ P_v(\vec k)}{\omega_{v-k}\omega_v}\; \frac{N(k)}{\Omega_k} \Bigg[ \delta(\Omega_k -\omega_{v-k} +\omega_v) + \delta (\Omega_k +\omega_{v-k} -\omega_v) \nonumber\\ &\hspace{1.1in}+\delta_P \delta(\Omega_k -\omega_{v-k} -\omega_v) +  \delta_P \delta(\Omega_k +\omega_{v-k} +\omega_v)\Bigg]\Bigg\}\,.
\end{align}
To evaluate the delta distributions, we first have to determine the zeroes of the arguments. As the part for $v$ is exactly the same as the one for $u$, we focus on the latter one. The four equations to solve then read, expressing the $\vec k$-integration in spherical coordinates $(k,\theta,\phi)$ and picking them such that $\vec u || \vec k_z$:
\begin{align}
   \pm k &= \sqrt{\omega_u^2 +k^2 - 2 u k \cos(\theta)} - \omega_u \\
   \pm k &= \sqrt{\omega_u^2 +k^2 - 2 u k \cos(\theta)} + \omega_u\,,\label{eq:deltaEval2}
\end{align}
where $u:= |\vec u|$. With the limits set by the spherical coordinates, i.e. $k \in \{0,\infty\}$ and $\cos(\theta) \in \{-1,1\}$ it is evident that the right hand side of the equation in the second line is always positive, hence $\delta(\Omega_k +\omega_{u-k} +\omega_u)=0$. For the positive sign in the second line we find
\begin{align}
    (k-\omega_u) &= \sqrt{\omega_u^2 +k^2 - 2 u k \cos(\theta)} \nonumber\\
    \iff k \omega_u &= uk\cos(\theta)\,,
\end{align}
which is solved for $k=0$. There is no other solution, as the remaining equation $\omega_u = \sqrt{m^2+u^2} = u \cos(\theta)$ is never fulfilled for $m>0$. Note however that $k=0$ is here only a solution of the squared equation and not of the original one\footnote{If $x=x_s$ is a solution to the equation $f(x) = g(x)$, then it is also to the squared version $f(x)^2=g(x)^2$, but not necessarily vice versa.} in \eqref{eq:deltaEval2}, hence we also have $\delta(-\Omega_k +\omega_{u-k} +\omega_u)=0$. This means that here all additional terms arising from the extended projection vanish. For the two equations in the first line we have
\begin{align}
    (\pm k +\omega_u) &=  \sqrt{\omega_u^2 +k^2 - 2 u k \cos(\theta)} \nonumber\\
    \iff \mp k\omega_u &= uk\cos(\theta)\,.
\end{align}
This is solved by $k=0$, which is also a solution to both non-squared equations. Apart from that, there is no other solution as again $|\mp \omega_u| = |\mp \sqrt{m^2+u^2} | > u$ while $|u \cos(\theta)| \leq u$. From this we obtain
\begin{equation}\label{eq:evalDelta2Mka}
     \delta(\pm\Omega_k -\omega_{u-k} +\omega_u) = \frac{\delta(k)}{\left|\pm 1 + \frac{u\cos(\theta)}{\omega_u}\right|} = \frac{\delta(k)}{1 \pm \frac{u\cos(\theta)}{\omega_u}} \,.
\end{equation}
Applying this to the original expression in spherical coordinates, we find
\begin{align}
    -\frac{\pi \kappa}{2}  \frac{1}{(2\pi)^2} \int_0^\infty dk \int_0^\pi d\theta \; k \sin(\theta) \bigg\{& \frac{ P_u(\vec k)}{\omega_{u-k}\omega_u}\; N(k) \Bigg[ \frac{\delta(k)}{1 + \frac{u\cos(\theta)}{\omega_u}}+\frac{\delta(k)}{1 - \frac{u\cos(\theta)}{\omega_u}} \Bigg]\nonumber\\
    & +\frac{ P_v(\vec k)}{\omega_{v-k}\omega_v}\; N(k) \Bigg[ \frac{\delta(k)}{1 + \frac{v\cos(\theta)}{\omega_v}} + \frac{\delta(k)}{1 - \frac{v\cos(\theta)}{\omega_v}}\Bigg]\Bigg\}\,,
\end{align}
where in the second line we picked different spherical coordinates with $\vec v || \vec k_z$ and defined $v:= |\vec v|$. Evaluation of the delta yields, using 
\begin{equation}\label{eq:apPusin}
    P_u(k,\theta,\phi) = \frac{1}{2} \left[  u^2 - \frac{(u k \cos(\theta))^2}{k^2} \right]^2 = \frac{u^4}{2} \left[  1 -\cos^2(\theta) \right]^2 = \frac{u^4}{2} \sin^4(\theta) = P_u(\theta,\phi)
\end{equation}
and with l'Hospital's limit
\begin{equation}\label{eq:apLHosp}
    \lim_{k\rightarrow 0} k \, N(k) = \lim_{k\rightarrow 0} \frac{k}{e^{\beta k}-1} = \lim_{k\rightarrow 0} \frac{1}{\beta e^{\beta k}} = \frac{1}{\beta}\,,
\end{equation}
the following result\footnote{Note that here the terms that were renormalised would have dropped out as they have $1$ instead of $N(k)$ and hence one would have gotten $\lim_{k\rightarrow 0} k \cdot 1 = 0$. This is consistent with the literature, e.g. with \cite{Anastopoulos:2013zya,Lagouvardos:2020laf}, where the renormalisation after application of the Markov approximation in the one-particle master equation does not affect the decoherence part which comes from the delta terms.}, where we get an additional factor of $\frac{1}{2}$ as the point $k=0$, where the delta distribution does not vanish, lies at the edge of the integration area:
\begin{align}
    &-\frac{\pi \kappa}{8\beta}  \frac{1}{(2\pi)^2} \int_0^\pi d\theta \;\sin^5(\theta) \bigg\{ \frac{ u^4}{\omega_u^2} \Bigg[ \frac{1}{1 + \frac{u\cos(\theta)}{\omega_u}}+\frac{1}{1 - \frac{u\cos(\theta)}{\omega_u}} \Bigg] +\frac{ v^4}{\omega_v^2}\Bigg[ \frac{1}{1 + \frac{v\cos(\theta)}{\omega_v}} + \frac{1}{1 - \frac{v\cos(\theta)}{\omega_v}}\Bigg]\Bigg\}\nonumber\\
    &=\frac{\pi \kappa }{4\beta}  \frac{1}{(2\pi)^2} \int_0^\pi d\theta \;\sin^5(\theta) \bigg\{\frac{u^2}{\cos^2(\theta)-\frac{\omega_u^2}{u^2}} +\frac{v^2}{\cos^2(\theta)-\frac{\omega_v^2}{v^2}} \Bigg\}\nonumber\\
    &=\frac{\pi \kappa }{4\beta}  \frac{1}{(2\pi)^2} \Bigg[ -\frac{10}{3} (u^2+v^2) + 2(\omega_u^2+\omega_v^2) -\frac{2}{\omega_u u} m^4 \,\text{arccoth} \left( \frac{\omega_u}{u}\right) - \frac{2}{\omega_v v} m^4 \,\text{arccoth} \left( \frac{\omega_v}{v}\right)\Bigg] \,.
\end{align}
Next, we compute the last line of equation \eqref{eq:meqrenWH}. The terms containing the delta distributions read
\begin{align}
    \frac{\pi \kappa}{2} \int \frac{d^3k}{(2\pi)^3} P_{ijln}(\vec k)& \frac{u^iu^jv^lv^n}{\sqrt{\omega_{u+k} \omega_u \omega_{v+k} \omega_v}} \frac{N(k)}{\Omega_k} \nonumber\\
    &\cdot \Bigg\{\delta(\Omega_k +\omega_{u} - \omega_{u+k}) + \delta(\Omega_k - \omega_{u} + \omega_{u+k})\nonumber\\
    &\hspace{0.2in}+\delta(\Omega_k +\omega_{v} - \omega_{v+k}) + \delta(\Omega_k - \omega_{v} + \omega_{v+k}) \Bigg\}\: \rho(\vec u+\vec k,\vec v + \vec k,t)\,.
\end{align}
Proceeding the same way as above it follows that
\begin{align}
    \delta(\pm\Omega_k +\omega_{u} - \omega_{u+k}) &= \frac{\delta(k)}{\left| 1 \mp \frac{u \cos(\alpha_{uk})}{\omega_u} \right|} = \frac{\omega_u \delta(k)}{\omega_u \mp u \cos(\alpha_{uk})} \,,
\end{align}
where $\alpha_{uk}$ is the angle between $\vec u$ and $\vec k$. Due to the presence of the two directions $\vec u$ and $\vec v$ in the prefactor, it is not possible to pick $\vec k_z$ parallel to both $\vec u$ and $\vec v$, as in general $\vec u || \vec v$ does not hold. Given that $P_{ijln}(\vec k) = P_{ijln}(\theta,\phi)$, the delta distribution can be applied and one obtains\footnote{Note that a factor $\frac{1}{2}$ arises due to the fact that the point $k=0$ where the delta distribution is not equal to zero is at the edge of the integration interval.}
\begin{align}
    \frac{\pi \kappa}{2(2\pi)^3 \beta} \int_0^\pi d\theta \int_0^{2\pi} d\phi \sin(\theta) P_{ijln}(\theta,\phi)& \frac{u^iu^jv^lv^n}{\omega_u \omega_v}  \nonumber\\
    &\cdot \Bigg\{ \frac{\omega_u^2}{\omega_u^2 - u^2 \cos^2(\alpha_{uk})}+\frac{\omega_v^2}{\omega_v^2 - v^2 \cos^2(\alpha_{vk})} \Bigg\}\: \rho(\vec u,\vec v,t)\,.
\end{align}
Now we choose in the first term the spherical coordinates such that $\vec u || \vec k_z$ and in the second one $\vec v || \vec k_z$ and obtain
\begin{align}
    \frac{\pi \kappa}{2(2\pi)^3 \beta} \int_0^\pi d\theta \int_0^{2\pi} d\phi \sin(\theta) P_{ijln}(\theta,\phi)& \frac{u^iu^jv^lv^n}{\omega_u \omega_v}  \Bigg\{ \frac{\omega_u^2}{\omega_u^2 - u^2 \cos^2(\theta)}+\frac{\omega_v^2}{\omega_v^2 - v^2 \cos^2(\theta)} \Bigg\}\: \rho(\vec u,\vec v,t)\,.
\end{align}
The the only quantity depending on $\phi$ is $P_{ijln}(\theta,\phi)$. The appearing contraction is, assuming $\vec u ||\vec k_z$:
\begin{align}
    P_{ijln}(\theta,\phi)u^iu^jv^lv^n &= \left[ \vec{u} \cdot \vec v - \frac{(\vec u \cdot \vec k)(\vec v \cdot \vec k)}{\vec{k}^2} \right]^2 -\frac{1}{2} \left[  \vec{u}^2 - \frac{(\vec u \cdot \vec k)^2}{\vec{k}^2} \right] \left[  \vec{v}^2 - \frac{(\vec v \cdot \vec k)^2}{\vec{k}^2} \right]\nonumber\\
    &= \left[ \vec{u} \cdot \vec v - u \cos(\theta)(\vec v \cdot \hat{\vec{k}}) \right]^2 - \frac{1}{2}\left[ u^2- u^2 \cos^2(\theta)\right] \left[ v^2 - (\vec v \cdot \hat{\vec k})^2\right]\nonumber\\
    &= (\vec u \cdot \vec v)^2 + u^2\cos^2(\theta) (\vec v \cdot \hat{\vec k})^2 - 2u \cos(\theta)(\vec u \cdot \vec v)(\vec v \cdot \hat{\vec k}) - \frac{1}{2}u^2\sin^2(\theta) \left[v^2 - (\vec v \cdot \hat{\vec k})^2\right] \,,
\end{align}
where the unit vector $\hat{\vec k}$ is defined as
\begin{equation}
    \hat{\vec k} = \begin{pmatrix}
        \sin(\theta) \cos(\phi) \\ \sin(\theta)\sin(\phi) \\ \cos(\theta)
    \end{pmatrix}\,.
\end{equation}
From this follows that
\begin{align}
    \int_0^{2\pi} d\phi \; P_{ijln}(\theta,&\phi)u^iu^jv^lv^n \nonumber\\=& 2\pi (\vec u \cdot \vec v)^2 + u^2\cos^2(\theta) [2\pi \cos^2(\theta) v_z^2 + \pi \sin^2(\theta) (v_x^2+v_y^2)] - 4\pi u v_z \cos^2(\theta) (\vec u \cdot \vec v)\nonumber\\ &- \frac{\pi u^2}{2}\sin^2(\theta) [2 v^2 -2 v_3^2\cos^2(\theta) - \sin^2(\theta) (v_1^2+v_2^2)] \nonumber\\
    =& \pi\Big[ 2(\vec u\cdot \vec v)^2 - \frac{u^2}{2}(v^2+v_z^2) \Big] +\pi \cos^2(\theta) \Big[  u^2(v^2+v_z^2) - 4 u v_z (\vec u \cdot \vec v)\Big] \nonumber\\ &+ \pi \cos^4(\theta) \Big[ \frac{u^2}{2} (3v_z^2-v^2) \Big]\nonumber\\
    =&  -\pi \frac{u^2}{2}(v^2-3v_z^2) +\pi \cos^2(\theta) u^2(v^2-3v_z^2)- \pi \cos^4(\theta) \frac{u^2}{2} (v^2-3v_z^2) \nonumber\\
    =& -\frac{\pi}{2} u^2 (v^2-3v_z^2) \Big[ 1 - \cos^2(\theta)\Big]^2 \nonumber\\
    =& -\frac{\pi}{2} u^2 (v^2-3v_z^2) \sin^4(\theta) \,,\label{eqPuvsin}
\end{align}
where we used in the second step that we chose the coordinate system such that $\vec u = u \vec{e}_z$, hence $\vec u \cdot \vec v = u v_z$. 
The $\theta$-integration then has the form
\begin{align}
    &-\frac{\pi}{2} u^2  (v^2-3v_z^2)\int_0^\pi d\theta\; \sin^5(\theta) \frac{\omega_u^2}{\omega_u^2-u^2\cos^2(\theta)}\nonumber\\
    &= -\pi \omega_u (v^2-3v_z^2) \Big[ \frac{5}{3}\omega_u- \frac{\omega_u^3}{u^2} + \frac{m^4}{u^3} \text{arctanh}\left(\frac{u}{\omega_u}\right)  \Big]\,.
\end{align}
Due to symmetry, we get the same result for the other term just with the replacement $u \leftrightarrow v$, as for the corresponding terms we can analogously choose $\vec v = v \vec{e}_z$. The contribution of the last line of equation \eqref{eq:meqrenWH} is therefore\footnote{Here it is important to use the coordinate independent expressions $u v_z = (\vec u \cdot \vec v)$ and $v u_z = (\vec u \cdot \vec v)$, as we picked different coordinate systems when evaluating the two sets of terms.}:
\begin{align}
    \frac{\kappa}{16\pi \beta} \Bigg\{ & \frac{\omega_u}{\omega_v} \left(v^2-3 \frac{(\vec u \cdot \vec v)^2}{u^2}\right) \left[ \frac{5}{3} - \left(1+\frac{m^2}{u^2} \right) + \frac{m^4}{\omega_u u^3} \text{arctanh}\left(\frac{u}{\omega_u}\right)\right] \nonumber\\
    &+\frac{\omega_v}{\omega_u} \left(u^2-3 \frac{(\vec u \cdot \vec v)^2}{v^2}\right) \left[ \frac{5}{3} - \left(1+\frac{m^2}{v^2} \right) + \frac{m^4}{\omega_v v^3} \text{arctanh}\left(\frac{v}{\omega_v}\right)\right]\Bigg\}\,.
\end{align}

Collecting all contributions from the delta terms yields
\begin{align}\label{eq:delPart2Mark}
    \frac{\kappa }{16\pi\beta} \Bigg\{ &-\frac{10}{3} (u^2+v^2) +2 (\omega_u^2+\omega_v^2) -\frac{2}{\omega_u u} m^4 \,\text{arctanh} \left( \frac{u}{\omega_u}\right) - \frac{2}{\omega_v v} m^4 \,\text{arctanh} \left( \frac{v}{\omega_v}\right)\nonumber\\
    & - \frac{\omega_u}{\omega_v} \left(v^2-3\frac{(\vec u \cdot \vec v)^2}{u^2}\right) \left[ \frac{2}{3} - \frac{m^2}{u^2} + \frac{m^4}{\omega_u u^3} \text{arctanh}\left(\frac{u}{\omega_u}\right)\right] \nonumber\\
    &-\frac{\omega_v}{\omega_u} \left(u^2-3 \frac{(\vec u \cdot \vec v)^2}{v^2}\right) \left[ \frac{2}{3} -\frac{m^2}{v^2} + \frac{m^4}{\omega_v v^3} \text{arctanh}\left(\frac{v}{\omega_v}\right)\right]\Bigg\}\,.
\end{align}
This is the final form of the real part of the dissipator that causes decoherence. The rotating wave approximation which is carried out as a next step leaves this part of the master equation invariant.

\subsection{Evaluation of the Cauchy principal value contributions}\label{app:MarkPV}
For the Markov approximation, it remains to compute the terms that contain the Cauchy principal value in \eqref{eq:meqrenWH} after the approximation \eqref{eq:appMapIdpv}. The terms in line two of \eqref{eq:meqrenWH} read
\begin{align}
    \frac{i \kappa}{2} \int \frac{d^3k}{(2\pi)^3} \frac{u^4}{2} \sin^4(\alpha_{uk}) \frac{1}{\omega_{u-k}\omega_u} \frac{N(k)}{\Omega_k} \Bigg[ & PV\left(\frac{1}{\Omega_k+\omega_{u-k}-\omega_u} \right) - PV\left(\frac{1}{\Omega_k-\omega_{u-k}+\omega_u} \right) \nonumber\\ &+ \delta_P PV\left(\frac{1}{\Omega_k+\omega_{u-k}+\omega_u} \right) - \delta_P PV\left(\frac{1}{\Omega_k-\omega_{u-k}-\omega_u} \right) \Bigg]\,,
\end{align}
where $\alpha_{uk}$ is the angle between $\vec u$ and $\vec k$. It can be seen that the term inside the principal value causes problems for $\vec k \rightarrow \vec 0$. Thus we exclude a small region of radius $\epsilon$ around $\vec k = \vec 0$, perform the integration and take the limit $\epsilon \rightarrow 0$ in the end. We then obtain in spherical coordinates\footnote{Note that the change of coordinates could have been done before applying the second Markov approximation, hence also before introducing the principal value.} with $\vec{k}_z || \vec u$:
\begin{align}
      \lim_{\epsilon \rightarrow 0}\frac{i \kappa u^4}{4(2\pi)^2\omega_u} \int_\epsilon^\infty dk \int_0^\pi d\theta\;  \sin^5(\theta) \frac{1}{\omega_{u-k}} k N(k) \Bigg[ & \frac{1}{\Omega_k+\omega_{u-k}-\omega_u} - \frac{1}{\Omega_k-\omega_{u-k}+\omega_u} \nonumber\\ &+ \delta_P \frac{1}{\Omega_k+\omega_{u-k}+\omega_u}  - \delta_P\frac{1}{\Omega_k-\omega_{u-k}-\omega_u}  \Bigg]\,.
\end{align}
For $\delta_P=1$ this can be simplified to
\begin{align}
    &\lim_{\epsilon \rightarrow 0}\frac{i \kappa u^4}{4(2\pi)^2\omega_u} \int_\epsilon^\infty dk \int_0^\pi d\theta\;  \sin^5(\theta) \frac{1}{\omega_{u-k}} k N(k) \Bigg[ \frac{2u \cos(\theta) \omega_{u-k}}{k \omega_u^2 - k u^2 \cos^2(\theta)} \Bigg] \nonumber\\
    &=\lim_{\epsilon \rightarrow 0}\frac{i \kappa u^5}{2(2\pi)^2\omega_u} \Bigg[ \int_\epsilon^\infty dk\; N(k) \Bigg]  \underbrace{\Bigg[\int_0^\pi d\theta \frac{\cos(\theta) \sin^5(\theta)}{\omega_u^2 - u^2 \cos^2(\theta)} \Bigg]}_{=0} \nonumber\\
    &= \lim_{\epsilon \rightarrow 0} 0 = 0\,.
   \end{align}
Without the principal value the $k$-integration would diverge:
\begin{equation}
    \int_0^\infty dk\; N(k) = \int_0^\infty dk\; \frac{1}{e^{\beta k}-1} = \frac{i\pi}{\beta} - \lim_{\epsilon \rightarrow 0} \frac{2}{\beta} \text{arctanh}\left( 1- 2 e^{\epsilon\beta} \right)\,.
\end{equation}
Without prior renormalisation, some terms arising due to the additional term present in the non-renormalised coefficients $C(\vec u, \vec k, t)$ in \eqref{eq:Cunrenor} compared to the renormalised one would remain here and lead to logarithmic divergences, as expected from the discussion in section \ref{uvren1}.\\
For the non-extended projection, i.e. for $\delta_P=0$, the situation is more complicated. In that case we find, again using spherical coordinates and implementing the principal value by excluding a sphere of radius $\epsilon$ around the critical point $\vec k=\vec 0$:
\begin{align}
     & \lim_{\epsilon \rightarrow 0}\frac{i \kappa u^4}{4(2\pi)^2\omega_u} \int_\epsilon^\infty dk \int_0^\pi d\theta\;  \sin^5(\theta) \frac{1}{\omega_{u-k}} k N(k) \Bigg[  \frac{1}{\Omega_k+\omega_{u-k}-\omega_u} - \frac{1}{\Omega_k-\omega_{u-k}+\omega_u}\Bigg]\nonumber\\
     &=-\lim_{\epsilon \rightarrow 0}\frac{i \kappa u^4}{2(2\pi)^2\omega_u} \int_\epsilon^\infty dk \int_0^\pi d\theta\;  \sin^5(\theta) k N(k) \frac{1 - \frac{\omega_u}{\omega_{u-k}}}{k^2 - (\omega_{u-k} - \omega_u)^2}\,.
\end{align}
The $\theta$-integration leads to a complicated result that can be simplified when considering e.g. the ultra-relativistic limit. Then it yields, where the limit $\epsilon\rightarrow 0$ can be taken also before the integration:
\begin{align}\label{eq:apurelimpart}
    &-\frac{i \kappa u^4}{105(2\pi)^2\omega_u} \Bigg\{- 4 \int_0^u dk \; N(k) \left( \frac{k^3}{u^4} - 7\frac{k}{u^2} \right) + \int_u^\infty dk\; N(k) \left( 35\frac{1}{u} - 14\frac{u}{k^2} + 3\frac{u^3}{k^4} \right) \Bigg\}\nonumber\\
    &= -\frac{i \kappa u^4}{105(2\pi)^2\omega_u} \Bigg\{4 \left[ \frac{\pi^4}{15\beta^4 u^4} - \frac{7\pi^2}{6\beta^2 u^2} -6 \frac{\ln(1-e^{-\beta u})}{\beta u} + 4  \frac{\text{Li}_2(e^{-\beta u})}{\beta^2 u^2} - 6 \frac{\text{Li}_3(e^{-\beta u})}{\beta^3 u^3} - 6 \frac{\text{Li}_4(e^{-\beta u})}{\beta^4 u^4} \right]\nonumber\\
    &\hspace{1.1in} + 35 - 35\frac{\ln\left(e^{\beta u}-1 \right)}{\beta u} - 14 u \int_u^\infty dk\; \frac{N(k)}{k^2} + 3 u^3 \int_u^\infty dk\; \frac{N(k)}{k^4} \Bigg\}\,.
\end{align}
Here, $\text{Li}_s(x)$ denotes the poly-logarithm function defined by
\begin{equation}\label{eq:defpolylog}
    \text{Li}_s(x) := \sum_{n=1}^\infty \frac{x^n}{n^s}\,.
\end{equation}
The remaining two integrations cannot be performed analytically, but they can be solved numerically given a specific temperature $\Theta$ and a value for $u$, and are finite as long as $u>0$.\\
For the terms in line three of \eqref{eq:meqrenWH} we get the same results when replacing $\vec u \rightarrow \vec v$ and applying complex conjugation. For the terms in line four we obtain:
\begin{align}
    \frac{i\kappa}{2(2\pi)^3} \frac{1}{\sqrt{\omega_u \omega_v}} \int d^3k &\; P_{ijln}(\vec k) \frac{u^i u^j v^l v^n}{\sqrt{\omega_{u+k}\omega_{v+k}}}  \frac{N(k)}{\Omega_k }\rho(\vec u+\vec k,\vec v +\vec k,t) \nonumber\\ \Bigg\{ &\text{PV}\left( \frac{1}{\Omega_k + \omega_u - \omega_{u+k}} \right) - \text{PV}\left( \frac{1}{\Omega_k - \omega_u + \omega_{u+k}} \right) \nonumber\\
    &-\text{PV}\left( \frac{1}{\Omega_k + \omega_v - \omega_{v+k}} \right) +\text{PV}\left( \frac{1}{\Omega_k - \omega_v + \omega_{v+k}} \right) \Bigg\}\,. 
\end{align}
Without further specification of $\rho$, this cannot be simplified further at this point. \\In summary, after the second Markov approximation we hence have
\begin{align}\label{eq:meq2Markov}
\frac{\partial}{\partial t} \rho(\vec{u},\vec{v},t) = &-i \rho(\vec{u},\vec{v},t) \; (\omega_u-\omega_v)\nonumber\\
&+\frac{\kappa }{16\pi\beta} \Bigg\{ -\frac{10}{3} (u^2+v^2) +2 (\omega_u^2+\omega_v^2) -\frac{2}{\omega_u u} m^4 \,\text{arctanh} \left( \frac{u}{\omega_u}\right) - \frac{2}{\omega_v v} m^4 \,\text{arctanh} \left( \frac{v}{\omega_v}\right)\nonumber\\
    &\hspace{.7in} - \frac{\omega_u}{\omega_v} \left(v^2-3\frac{(\vec u \cdot \vec v)^2}{u^2}\right) \left[ \frac{2}{3} - \frac{m^2}{u^2} + \frac{m^4}{\omega_u u^3} \text{arctanh}\left(\frac{u}{\omega_u}\right)\right] \nonumber\\
    &\hspace{.7in}-\frac{\omega_v}{\omega_u} \left(u^2-3 \frac{(\vec u \cdot \vec v)^2}{v^2}\right) \left[ \frac{2}{3} -\frac{m^2}{v^2} + \frac{m^4}{\omega_v v^3} \text{arctanh}\left(\frac{v}{\omega_v}\right)\right]\Bigg\}\rho(\vec u,\vec v,t)\nonumber\\
    &+\frac{i\kappa}{2(2\pi)^3} \frac{1}{\sqrt{\omega_u \omega_v}} \int d^3k \; P_{ijln}(\vec k) \frac{u^i u^j v^l v^n}{\sqrt{\omega_{u+k}\omega_{v+k}}} \frac{N(k)}{\Omega_k} \rho(\vec u+\vec k,\vec v +\vec k,t) \nonumber\\ &\hspace{1in}\Bigg\{ \text{PV}\left( \frac{1}{\Omega_k + \omega_u - \omega_{u+k}} \right) - \text{PV}\left( \frac{1}{\Omega_k - \omega_u + \omega_{u+k}} \right) \nonumber\\
    &\hspace{1.1in}-\text{PV}\left( \frac{1}{\Omega_k + \omega_v - \omega_{v+k}} \right) + \text{PV}\left( \frac{1}{\Omega_k - \omega_v + \omega_{v+k}} \right) \Bigg\}\nonumber\\
    &-(1-\delta_P) \frac{i \kappa }{2(2\pi)^2} \lim_{\epsilon \rightarrow 0} \bigg[\frac{u^4}{\omega_u}\int_\epsilon^\infty dk \int_0^\pi d\theta\;  \sin^5(\theta) k N(k) \frac{1 - \frac{\omega_u}{\omega_{u-k}}}{k^2 - (\omega_{u-k} - \omega_u)^2}\nonumber\\
    &\hspace{1.5in} - \frac{v^4}{\omega_v}\int_\epsilon^\infty dk \int_0^\pi d\theta\;  \sin^5(\theta) k N(k) \frac{1 - \frac{\omega_v}{\omega_{v-k}}}{k^2 - (\omega_{v-k} - \omega_v)^2}\bigg]\rho(\vec u,\vec v,t)\,,
\end{align}
where the last two lines vanish when working with the extended projection.

\section{Application of the Rotating Wave Approximation (RWA)}\label{apRWA}
In this appendix the detailed implementation of the rotating wave approximation for the renormalised Markovian master equation under consideration in this work is discussed. We proceed in the standard way by considering the master equation in interaction picture and removing all terms that oscillate fast (see e.g. \cite{Breuer:2002pc,Lagouvardos:2020laf}. For this we start in the field theory and consider the full dissipator from (4.60) in \cite{Fahn:2022zql}:
\begin{align}
\mathcal{D}[\rho_S]= - \frac{\kappa}{2}\sum_{r\in\{+,-\}} \sum_{a,b=1}^4 \int_{\mathbb{R}^3} \frac{d^3k\: d^3p\: d^3l}{(2\pi)^3}\frac{1}{\Omega_k} \Bigg\{ \left[ j^a_r(\vec{k},\vec{p})^\dagger,j_r^b(\vec{k},\vec{l})\rho_S(t)\right] \; f(\Omega_k+\omega_b(\vec{k},\vec{l}))+h.c.\nonumber \\ + N(k) \; \left[j_r^a(\vec{k},\vec{p})^\dagger,\left[ j^b_r(\vec{k},\vec{l}),\rho_S(t)\right]\right]\; f(\Omega_k+\omega_b(\vec{k},\vec{l})) +h.c.\Bigg\}\,,
\end{align}
where $f(\omega;t) := \int_0^{t-t_0} d\tau\; e^{-i\omega \tau}$ and we assume that the Markov approximation has already been applied, thus the former functions $f(\omega;t)$ are now distributions $f(\omega)= \pi \delta(\omega) - iPV\left( \frac{1}{\omega}\right)$ independent of time. The different $\omega_a$ and $j^r_a$ were defined starting in \eqref{eq:defjundw}. Taking into account that the renormalisation removed the terms that are independent of $N(k)$, the dissipator becomes (see \eqref{eq:renodisoplvl}):
\begin{align}
\mathcal{D}[\rho_S]= - \frac{\kappa}{2}\sum_{r\in\{+,-\}} \sum_{a,b=1}^4 \int_{\mathbb{R}^3} \frac{d^3k\: d^3p\: d^3l}{(2\pi)^3}\frac{N(k)}{\Omega_k} \Bigg\{ \left[j_r^a(\vec{k},\vec{p})^\dagger,\left[ j^b_r(\vec{k},\vec{l}),\rho_S(t)\right]\right]\; f(\Omega_k+\omega_b(\vec{k},\vec{l})) +h.c.\Bigg\}\,.
\end{align}
The basis for the RWA is the dissipator in interaction picture, which is obtained by substituting
\begin{equation}
j_r^a(\vec{k},\vec{p}) \longrightarrow j_r^a(\vec{k},\vec{p}) e^{i\omega_a(\vec{k},\vec{p})t}\,.
\end{equation}
Thus we get as time-dependent frequencies that cause oscillations terms of the form
\begin{equation}
e^{\pm i[\omega_a(\vec{k},\vec{p})-\omega_b(\vec{k},\vec{l})]t}\,.
\end{equation}
Next we apply the RWA, which means that we discard all the rapidly oscillating terms and only keep those where $\omega_a(\vec{k},\vec{p})=\omega_b(\vec{k},\vec{l})$ holds. For $a=3$ and $a=4$ this means that only $b=3$ and $b=4$ survive. However, from the definition of the $\omega_a(\vec{k},\vec{p})$ follows that $\omega_1(\vec{k},\vec{p})=\omega_2(\vec{k},-\vec{k}-\vec{p})$, so also terms of the form $a=1$, $b=2$ and vice versa will remain. To simplify this, we introduce $J_r^A(\vec{k},\vec{p})$ and $\omega^A(\vec{k},\vec{p})$ with $A\in\{2,3,4\}$, similar as in \cite{Lagouvardos:2020laf}, such that
\begin{align}
J_r^2(\vec{k},\vec{p}) &:=2j_r^1(\vec{k},\vec{p}) & \omega^2(\vec{k},\vec{p}) &:=  \omega_1(\vec{k},\vec{p}) = \omega_2(\vec{k},-\vec{k}-\vec{p}) = \omega_p - \omega_{k+p}\\
J_r^3(\vec{k},\vec{p}) &:= j_r^3(\vec{k},\vec{p}) & \omega^3(\vec{k},\vec{p}) &:= \omega_3(\vec{k},\vec{p}) = -\omega_p - \omega_{k+p} \\
J_r^4(\vec{k},\vec{p}) &:= j_r^4(\vec{k},\vec{p}) & \omega^4(\vec{k},\vec{p}) &:= \omega_4(\vec{k},\vec{p}) = \omega_p + \omega_{k+p}\,.
\end{align}
Making use of the fact that $j_r^1(\vec{k},\vec{p})= j_r^2(\vec{k},-\vec{k}-\vec{p})$, we can rewrite the dissipator in terms of a sum over capital letters $A,B$:
\begin{align}
\mathcal{D}[\rho_S]= - \frac{\kappa}{2}\sum_{r\in\{+,-\}} \sum_{A,B=2}^4 \int_{\mathbb{R}^3} \frac{d^3k\: d^3p\: d^3l}{(2\pi)^3}\frac{N(k)}{\Omega_k} \Bigg\{ \left[J_r^A(\vec{k},\vec{p})^\dagger,\left[ J^B_r(\vec{k},\vec{l}),\rho_S(t)\right]\right]\; f(\Omega_k+\omega^B(\vec{k},\vec{l})) +h.c.\Bigg\}\,.
\end{align}
The RWA-requirement to keep only the terms where $\omega^A(\vec{k},\vec{p}) = \omega^B(\vec{k},\vec{l})$ then keeps the following summands
\begin{align}
A=B=2 && A=B=3 && A=B=4
\end{align}
 completely, which were also exactly the same summands that survived the one-particle projection, while for the remaining six summands it yields the following conditions:
\begin{align}
&A=2,\; B=3: \hspace{0.2in} \omega_p-\omega_{k+p}+\omega_l+\omega_{k+l}=0\label{inequ1} \\
&A=2,\; B=4: \hspace{0.2in} \omega_p-\omega_{k+p}-\omega_l-\omega_{k+l}=0\label{inequ2} \\
&A=3,\; B=4: \hspace{0.2in} -\omega_p-\omega_{k+p}-\omega_l-\omega_{k+l}=0\,,
\end{align}
and for the other three summands the same conditions with the role of $\vec{p}$ and $\vec{l}$ swapped. For mass $m>0$, which implies $\omega_k>0$, the last condition is never fulfilled. For $m=0$ there is a solution, namely $\vec{k}=\vec{p}=\vec{l}=0$. The first condition reads
\begin{equation}
\omega_{k+p} = \omega_p+\omega_{l}+\omega_{k+l}\,.
\end{equation}
This is never fulfilled, as we show in the following (where we define $p:= ||\vec{p}||, l:=||\vec{l}||, k:=||\vec{k}||$ and assume\footnote{Afterwards we comment on the case $m=0$.} $m>0$):
\begin{align}
\omega_{k+p} \leq \sqrt{(k+p)^2+m^2} < \sqrt{p^2+m^2}+\sqrt{l^2+m^2}+\sqrt{(k-l)^2+m^2} \leq \omega_p+\omega_{l}+\omega_{k+l}\,.
\end{align}
The inequality in the middle can be proven by considering the square of both sides (as each summand individually is positive, the direction of the inequality remains unaffected), which yields:
\begin{align}
kp+kl < m^2+l^2+\sqrt{p^2+m^2}\sqrt{(k-l)^2+m^2}+\sqrt{p^2+m^2}\sqrt{l^2+m^2}+\sqrt{l^2+m^2}\sqrt{(k-l)^2+m^2}\,.
\end{align}
We can now estimate the right hand side downwards as
\begin{multline}
m^2+l^2+p|k-l|+pl+l|k-l| \\< m^2+l^2+\sqrt{p^2+m^2}\sqrt{(k-l)^2+m^2}+\sqrt{p^2+m^2}\sqrt{l^2+m^2}+\sqrt{l^2+m^2}\sqrt{(k-l)^2+m^2}\,,
\end{multline}
which yields 
\begin{equation}
kp+kl <m^2+l^2+p|k-l|+pl+l|k-l|
\end{equation}
to be proven. Due to the absolute value, we consider two different cases: Let's first assume that $k>l$. Then the inequality reads
\begin{equation}
kp+kl <m^2+ kp+kl\,,
\end{equation}
which is true as long as $m>0$. In the second case, i.e. for $l\geq k$, we are left with
\begin{equation}
kp+kl < m^2+ 2l^2+ 2pl -pk -kl\,,
\end{equation}
or equivalently
\begin{equation}
0 < m^2 + 2[l^2-kl] + 2[pl-pk]\,.
\end{equation}
However, as we are considering the case $l \geq k$, both brackets yield non-negative results and thus the inequality is fulfilled for $m>0$. Hence \eqref{inequ1} does also not have any solutions as long as $m>0$. For $m=0$ we have to solve the following equality:
\begin{equation}\label{eq_rwacond1}
    |\vec k + \vec p| - p \mbeq |\vec k + \vec l| -l\,.
\end{equation}
However, we also have
\begin{equation}
    |\vec k + \vec p| - p \leq k \leq |k-l| + l \leq |\vec k + \vec l| - \vec l\,,
\end{equation}
hence in order for equation \eqref{eq_rwacond1} to hold, all $\leq$ signs must become equalities. For the first one this is the case if $\vec{k} \parallel \vec{p}$ and for the last one if $\vec{k} \parallel -\vec{l}$. For the one in the middle, we have to consider two cases on how to resolve $|k-l|$. If $k\geq l$, then we can directly drop the absolute value and the middle $\leq$ becomes an equality. In case $k < l$ we find $|k-l|+l = 2l-k >2k-k = k$, so there is no equality. Hence for $m=0$ the following solutions exist:
\begin{equation}
\vec{k} \parallel \vec{p} \hspace{0.2in} \land \hspace{0.2in} \vec{k}\parallel -\vec{l} \hspace{0.2in} \land \hspace{0.2in} k \geq l\,.
\end{equation}
It remains to investigate \eqref{inequ2}. Isolating $\omega_p$ on one side and following the same argumentation as above (squaring the inequality and estimating downwards the rights hand side) we end up with
\begin{equation}
kp+kl < k^2+m^2+l^2 +l |k-p| + |k-p| |k-l| + l|k-l|\,.
\end{equation}
Here we have to consider four different cases:
\begin{itemize}
\item $k\geq p$ and $ k\geq l$: We obtain $2kp < 2k^2+m^2$, which has no solution for $m>0$ and for $m=0$ and $k=p$ we get solutions, thus $k\geq l, k=p, \vec{k} \parallel -\vec{l}, \vec{k}\parallel -\vec{p}$.
\item $k\geq p$ and $k < l$: We obtain $0<m^2 +2l^2-2pl$ which has no solution for $m\geq 0$.
\item $k<p$ and $k\geq l$: We obtain $0<m^2$, which has only for $m=0$ a solution, thus there $k\geq l, k<p, \vec{k} \parallel -\vec{l}, \vec{k}\parallel -\vec{p}$.
\item $k<p$ and $k<l$: We obtain $0<m^2 + 2[(p-k)(l-k)+l(l-k)]$ and thus no solution as every bracket is positive.
\end{itemize}
Summarising, for equality \eqref{inequ2} we get again no solution if $m>0$ and for $m=0$ we have 
\begin{equation}
\vec{k} \parallel -\vec{p} \hspace{0.2in} \land \hspace{0.2in} \vec{k}\parallel -\vec{l} \hspace{0.2in} \land \hspace{0.2in} p\geq k \geq l\,.
\end{equation}
So in total, for a positive mass non of the non-diagonal terms survives the rotating wave approximation. If the mass is zero, the following non-diagonal terms survive:
\begin{align}
&A=2, B=3: && \vec{k} \parallel \vec{p} \hspace{0.2in} \land \hspace{0.2in} \vec{k}\parallel -\vec{l} \hspace{0.2in} \land \hspace{0.2in} k \geq l\\
&A=2, B=4: &&\vec{k} \parallel -\vec{p} \hspace{0.2in} \land \hspace{0.2in} \vec{k}\parallel -\vec{l} \hspace{0.2in} \land \hspace{0.2in} p\geq k \geq l\\
&A=3, B=4: && \vec{k}=\vec{l}=\vec{p}=0 \\
&A=3, B=2: && \vec{k} \parallel \vec{l} \hspace{0.2in} \land \hspace{0.2in} \vec{k}\parallel -\vec{p} \hspace{0.2in} \land \hspace{0.2in} k \geq p\\
&A=4, B=2: &&\vec{k} \parallel -\vec{l} \hspace{0.2in} \land \hspace{0.2in} \vec{k}\parallel -\vec{p} \hspace{0.2in} \land \hspace{0.2in} l\geq k \geq p\\
&A=4, B=3: && \vec{k}=\vec{l}=\vec{p}=0\,.
\end{align}
If we plug these special cases into the dissipator, which contains a projection of $\vec{l}$ and also of $\vec{p}$ onto the plane perpendicular to $\vec{k}$, all extra terms containing $\vec{k} \parallel \pm\vec{l}$ and $\vec{k} \parallel \pm\vec{p}$ vanish. Then only the special solution $\vec{k}=\vec{l}=\vec{p}=0$ remains in all six cases (due to polar coordinates and thus also spherical coordinates being non-unique for zero radius, in that case still all directions are possible). However, as this is only one point regarding the radius integrations, it will vanish under the integral. Thus all the extra correction terms vanish\footnote{The same is the case for the result in \cite{Lagouvardos:2020laf}: There appear the $L_{(\sigma,\lambda)}^{++}(\vec{k},\vec{p})$ in the $J^1_{(\sigma,\lambda)}(\vec{k},\vec{p})$ with $\vec{k}\parallel\vec{p}$. To show this, one can use the definition $L_{(\sigma,\lambda)}^{++}(\vec{k},\vec{p}) = \sqrt{\Omega_p\Omega_{p+k}} (\epsilon_{-\sigma}(\vec{k}) \cdot \epsilon_{-\lambda}(\vec{p})) (\epsilon_{-\sigma}(\vec{k})\cdot \epsilon_\lambda(\vec{k}+\vec{p}))$ and that one can express the circular polarisation in terms of two linear polarisations $\epsilon_s(\vec{k}) = \frac{1}{\sqrt{2}} [\epsilon^1(\vec{k})+is\epsilon^2(\vec{k})]$ with $\epsilon^2(\vec{k}) := \frac{\hat{u}_0 \times \hat{k}}{|\sin \gamma_{u_0k}|}$ with an arbitrary unit vector $\hat{u}_0$ that is not (anti-)parallel to $\vec{k}$ and the angle $\gamma_{u_0k}$ between $\hat{u}_0$ and $\vec{k}$, as well as $\epsilon^1(\vec{k}) := \epsilon^2(\vec{k}) \times \hat{k}$, see also \cite{Cho:2021gvg}. As $\vec{p}\parallel\vec{k}$, we can see that $\epsilon_s(\vec{k}) = \epsilon_s(\vec{p}) = \epsilon_s(\vec{k}+\vec{p})$. Using also $\epsilon_{-s}(\vec{k}) = \epsilon^*_s(\vec{k})$ and $\epsilon^*_s(\vec{k}) \epsilon_t(\vec{k}) = \delta_{ts}$ which can be proven right away, we get that $[\epsilon_\sigma^*(\vec{k}) \cdot \epsilon_{-\lambda}(\vec{k})][\epsilon_\sigma^*(\vec{k})\cdot \epsilon_\lambda(\vec{k})]=0$, thus all additional terms containing $J^1$ vanish, as in our case, and only additional terms with $\vec{k}=\vec{p}=\vec{l}=0$ remain, which are of measure zero.} and we are left with the dissipator after the RWA in the form
\begin{align}\label{eq:apdisafrwa}
\mathcal{D}[\rho_S]= - \frac{\kappa}{2}\sum_{r\in\{+,-\}} \sum_{A=2}^4 \int_{\mathbb{R}^3} &\frac{d^3k\: d^3p\: d^3l}{(2\pi)^3}\frac{N(k)}{\Omega_k}\nonumber\\ &\cdot \Bigg\{ \left[J_r^A(\vec{k},\vec{p})^\dagger,\left[ J^A_r(\vec{k},\vec{l}),\rho_S(t)\right]\right]\; f(\Omega_k+\omega^A(\vec{k},\vec{l})) +h.c.\Bigg\}\Bigg|_{\omega^A(\vec{k},\vec{p})=\omega^A(\vec{k}, \vec{l})}\,.
\end{align}
In order to explicitly write the hermitian conjugate, we split $f(\omega) = f_{\delta}(\omega) + f_{PV}(\omega)$, as the two parts behave differently under complex conjugation (the first part is real, the second one purely imaginary) and compute them in the following two subsections.  

\subsection{Computation of the delta terms in the RWA}
The delta terms remains unaffected by the complex conjugation\footnote{As $\int_{\mathbb{R}} dx\; f(x) \delta^*(x) = \left[\int_{\mathbb{R}} dx\; f^*(x) \delta(x)\right]^*= [f^*(0)]^*=f(0)= \int_{\mathbb{R}} dx\; f(x) \delta(x)$ for a test function $f:\mathbb{R}\mapsto \mathbb{C}$, from which one can conclude that $\delta^*(x)=\delta(x)$.}, thus we get, using that the terms in the first line, which are independent of $N(k)$, vanished due to the Markov approximation:
\begin{align}
\mathcal{D}_\delta[\rho_S]= \kappa\sum_{r\in\{+,-\}} \sum_{A=2}^4 \int_{\mathbb{R}^3} &\frac{d^3k\: d^3p\: d^3l}{(2\pi)^3} \frac{f_\delta(\Omega_k+\omega^A(\vec{k},\vec{l}))}{\Omega_k} N(k)\nonumber\\
&\cdot \Bigg\{\left( J^A_r(\vec{k},\vec{p})^\dagger \rho J^A_r(\vec{k},\vec{l}) - \frac{1}{2} \left\{ \rho, J^A_r(\vec{k},\vec{l}) J^A_r(\vec{k},\vec{p})^\dagger\right\} \right)  \nonumber \\ &\hspace{0.2in}+  \left(J^A_r(\vec{k},\vec{l})\rho J^A_r(\vec{k},\vec{p})^\dagger - \frac{1}{2} \left\{ \rho, J^A_r(\vec{k},\vec{p})^\dagger J^A_r(\vec{k},\vec{l})\right\} \right) \Bigg\}\Bigg|_{\omega^A(\vec{k},\vec{p})=\omega^A(\vec{k}, \vec{l})}\,.
\end{align}
Using the following equalities:
\begin{align}
    J_r^2(\vec k,\vec p)^\dagger &= J_r^2(-\vec k,\vec k + \vec p)  & \omega^2(-\vec k,\vec p+\vec k) &= -\omega^2(\vec k,\vec p) \\
    J_r^3(\vec k,\vec p)^\dagger &= J_r^4(-\vec k,-\vec p) & \omega^3(-\vec k,-\vec p) &= -\omega^4(\vec k,\vec p) \\
    J_r^4(\vec k,\vec p)^\dagger &= J_r^3(-\vec k,- \vec p) & \omega^4(-\vec k,-\vec p) &= -\omega^3(\vec k,\vec p)\,, 
\end{align}
we can simplify the $\delta$-part of the dissipator and obtain
\begin{align}
\mathcal{D}_\delta[\rho_S]= \kappa\sum_{r\in\{+,-\}} \sum_{A=2}^4 \int_{\mathbb{R}^3} &\frac{d^3k\: d^3p\: d^3l}{(2\pi)^3} \frac{f_\delta(\Omega_k+\omega^A(\vec{k},\vec{l}))+f_\delta(\Omega_k-\omega^A(\vec{k},\vec{l}))}{\Omega_k} N(k)\nonumber\\
&\cdot \Bigg\{\left(J^A_r(\vec{k},\vec{l})\rho J^A_r(\vec{k},\vec{p})^\dagger - \frac{1}{2} \left\{ \rho, J^A_r(\vec{k},\vec{p})^\dagger J^A_r(\vec{k},\vec{l})\right\} \right) \Bigg\}\Bigg|_{\omega^A(\vec{k},\vec{p})=\omega^A(\vec{k}, \vec{l})}\,.
\end{align}
Additionally, all terms involving the extended projection, i.e. all terms where $A=3$ or $A=4$, vanished due to the Markov approximation, hence the $\delta$-part of the dissipator simplifies even further:
\begin{align}
\mathcal{D}_\delta[\rho_S]= \kappa\sum_{r\in\{+,-\}} \int_{\mathbb{R}^3} \frac{d^3k\: d^3p\: d^3l}{(2\pi)^3} &\frac{f_\delta(\Omega_k+\omega^2(\vec{k},\vec{l}))+f_\delta(\Omega_k-\omega^2(\vec{k},\vec{l}))}{\Omega_k} N(k)\nonumber\\
&\cdot \Bigg\{\left(J^2_r(\vec{k},\vec{l})\rho J^2_r(\vec{k},\vec{p})^\dagger - \frac{1}{2} \left\{ \rho, J^2_r(\vec{k},\vec{p})^\dagger J^2_r(\vec{k},\vec{l})\right\} \right) \Bigg\}\Bigg|_{\omega^2(\vec{k},\vec{p})=\omega^2(\vec{k}, \vec{l})}\,.
\end{align}
As shown above, $f_\delta \propto \delta(k)$, hence the RWA condition reads
\begin{equation}
    \omega^2(\vec k, \vec p) = \omega_p -\omega_{k+p} = 0 = \omega_l-\omega_{k+l}=\omega^2(\vec k, \vec l)  
\end{equation}
and is therefore automatically fulfilled, thus it can be dropped. One can furthermore evaluate the $f_\delta$ and obtains, using \eqref{eq:evalDelta2Mka}:
\begin{align}
    f_\delta(\Omega_k+\omega^2(\vec{k},\vec{l}))+f_\delta(\Omega_k-\omega^2(\vec{k},\vec{l})) = \pi\delta(k) \left[ \frac{1}{1+\frac{l \cos(\theta_l)}{\omega_l}} +\frac{1}{1-\frac{l \cos(\theta_l)}{\omega_l}}\right] =\delta(k) \frac{2\pi}{1-\frac{l^2 \cos^2(\theta_l)}{\omega_l^2}}\,,
\end{align}
where $\theta_l$ denotes the angle between $\vec k$ and $\vec l$ and $l=|\vec l|$, $k=|\vec k|$. This yields
\begin{align}
\mathcal{D}_\delta[\rho_S]= \kappa\sum_{r\in\{+,-\}} \int_{\mathbb{R}^3} \frac{d^3k\: d^3p\: d^3l}{(2\pi)^3} &\frac{\delta(k)}{\Omega_k} \frac{2\pi}{1-\frac{l^2 \cos^2(\theta_l)}{\omega_l^2}} N(k)\nonumber\\
&\cdot \Bigg\{\left(J^2_r(\vec{k},\vec{l})\rho J^2_r(\vec{k},\vec{p})^\dagger - \frac{1}{2} \left\{ \rho, J^2_r(\vec{k},\vec{p})^\dagger J^2_r(\vec{k},\vec{l})\right\} \right) \Bigg\}\,.
\end{align}
The rotating wave condition further implies
\begin{align}
    &\frac{\partial}{\partial |k|} \omega^2(\vec k,\vec p) =\frac{\partial}{\partial |k|} \omega^2(\vec k,\vec l) \nonumber\\
    &\iff \frac{k+p \cos(\theta_p)}{\omega_{k+p}} = \frac{k+l \cos(\theta_l)}{\omega_{k+l}}\,.
\end{align}
We therefore can use that
\begin{align}
    \delta(k) \frac{2\pi}{1-\frac{l^2 \cos^2(\theta_l)}{\omega_l^2}} = 2\pi \delta(k) \frac{1}{1-\frac{(k+l\cos(\theta_l))^2}{\omega_{k+l}^2}} =  2\pi \delta(k) \frac{1}{\sqrt{1-\frac{(k+l\cos(\theta_l))^2}{\omega_{k+l}^2}}} \frac{1}{\sqrt{1-\frac{(k+p\cos(\theta_p))^2}{\omega_{k+p}^2}}}
\end{align}
and, defining the Lindblad operators
\begin{equation}
    L_r(\vec k) := \int_{\mathbb{R}^3} d^3p \; \frac{1}{\sqrt{1-\frac{(k+p\cos(\theta_p))^2}{\omega_{k+p}^2}}} J_r^2(\vec k,\vec p)\,,
\end{equation}
we can recast the dissipator in Lindblad form:
\begin{equation}\label{eq:qplbfdis}
\mathcal{D}_\delta[\rho_S]= \kappa\sum_{r\in\{+,-\}} \int_{\mathbb{R}^3} \frac{d^3k}{(2\pi)^2}\delta(k) \frac{N(k)}{\Omega_k} \left(L_r(\vec{k})\rho L_r(\vec{k})^\dagger - \frac{1}{2} \left\{ \rho, L_r(\vec{k})^\dagger L_r(\vec{k})\right\} \right)\,.
\end{equation}

As the rotating wave approximation dropped the same terms as the single-particle projection and led to a condition on the frequencies that is already implemented in $\delta(k)$, which is present in every term of the $\delta$-part of the dissipator after the Markov approximation, the RWA does not change the form of the $\delta$-part of the dissipator compared to its form after the Markov approximation in \eqref{eq:delPart2Mark}.

\subsection{Computation of the Cauchy principal value terms in the RWA}
The contributions involving the Cauchy principal value, denoted as the PV-part of the dissipator \eqref{eq:apdisafrwa} in the following reads after renormalisation, which removes the terms independent of $N(k)$:
\begin{align}
&\mathcal{D}_{PV}[\rho_S]\nonumber\\ &= - \frac{\kappa}{2}\sum_{r\in\{+,-\}} \sum_{A=2}^4 \int_{\mathbb{R}^3} \frac{d^3k\: d^3p\: d^3l}{(2\pi)^3}\frac{N(k)}{\Omega_k}\nonumber\\ &\hspace{1.4in}\cdot \Bigg\{ \left[J_r^A(\vec{k},\vec{p})^\dagger,\left[ J^A_r(\vec{k},\vec{l}),\rho_S(t)\right]\right]\; f_{PV}(\Omega_k+\omega^A(\vec{k},\vec{l})) +h.c.\Bigg\}\Bigg|_{\omega^A(\vec{k},\vec{p})=\omega^A(\vec{k}, \vec{l})}\,.
\end{align}
As $f(\omega)$ is purely imaginary, it switches sign under the hermitian conjugation and we obtain:
\begin{align}
\mathcal{D}_{PV}[\rho_S]&=- \frac{\kappa}{2}\sum_{r\in\{+,-\}} \sum_{A=2}^4 \int_{\mathbb{R}^3} \frac{d^3k\: d^3p\: d^3l}{(2\pi)^3}\frac{N(k)}{\Omega_k} \nonumber \\ &\hspace{1.3in} \cdot\Bigg\{ \; \left[ \left[J_r^A(\vec{k},\vec{p})^\dagger, J^A_r(\vec{k},\vec{l})\right],\rho_S(t)\right]\; f_{PV}(\Omega_k+\omega^A(\vec{k},\vec{l})) \Bigg\}\Bigg|_{\omega^A(\vec{k},\vec{p})=\omega^A(\vec{k}, \vec{l})}\,.
\end{align}

We can then rewrite this part of the dissipator as
\begin{equation}
\mathcal{D}_{PV}[\rho_S]= -i\frac{\kappa}{2} \left[V_{LS},\rho\right]
\end{equation}
with
\begin{align}
V_{LS}&=- \sum_{r\in\{+,-\}} \sum_{A=2}^4 \int_{\mathbb{R}^3} \frac{d^3k\: d^3p\: d^3l}{(2\pi)^3}\frac{N(k)}{\Omega_k} PV\left( \frac{1}{\Omega_k+\omega^A(\vec{k},\vec{l})}\right)\left[ J_r^A(\vec{k},\vec{p})^\dagger, J^A_r(\vec{k},\vec{l})\right]\Bigg|_{\omega^A(\vec{k},\vec{p})=\omega^A(\vec{k}, \vec{l})}\,.
\end{align}
Note that the rotating wave approximation hence removed the imaginary terms in the fifth to seventh line of the Markovian master equation in \eqref{eq:meqnsmap}, while it did not change the other imaginary terms. These were vanishing when working with the extended projection, hence in that case we find $V_{LS}=0$. If the non-extended projection is used, only the terms for $A=2$ are left and there the RWA condition is already implemented in the one-particle projection, as the case $(1,1)$ includes $\delta(\vec p -\vec l)$, see table (I) and (II) at the beginning of section \ref{tab1}. Hence the RWA does not change anything in the remaining PV-terms. The final one-particle master equation then becomes
\begin{align}
\frac{\partial}{\partial t} \rho(\vec{u},\vec{v},t) = &-i \rho(\vec{u},\vec{v},t) \; (\omega_u-\omega_v) \nonumber\\
&+\frac{\kappa }{16\pi\beta} \Bigg\{ -\frac{10}{3} (u^2+v^2) +2 (\omega_u^2+\omega_v^2) -\frac{2}{\omega_u u} m^4 \,\text{arctanh} \left( \frac{u}{\omega_u}\right) - \frac{2}{\omega_v v} m^4 \,\text{arctanh} \left( \frac{v}{\omega_v}\right)\nonumber\\
    &\hspace{.7in} - \frac{\omega_u}{\omega_v} \left(v^2-3\frac{(\vec u \cdot \vec v)^2}{u^2}\right) \left[ \frac{2}{3} - \frac{m^2}{u^2} + \frac{m^4}{\omega_u u^3} \text{arctanh}\left(\frac{u}{\omega_u}\right)\right] \nonumber\\
    &\hspace{.7in}-\frac{\omega_v}{\omega_u} \left(u^2-3 \frac{(\vec u \cdot \vec v)^2}{v^2}\right) \left[ \frac{2}{3} -\frac{m^2}{v^2} + \frac{m^4}{\omega_v v^3} \text{arctanh}\left(\frac{v}{\omega_v}\right)\right]\Bigg\}\rho(\vec u,\vec v,t)\nonumber\\
    &-(1-\delta_P) \frac{i \kappa }{2(2\pi)^2} \lim_{\epsilon \rightarrow 0} \bigg[\frac{u^4}{\omega_u}\int_\epsilon^\infty dk \int_0^\pi d\theta\;  \sin^5(\theta) k N(k) \frac{1 - \frac{\omega_u}{\omega_{u-k}}}{k^2 - (\omega_{u-k} - \omega_u)^2}\nonumber\\
    &\hspace{1.5in} - \frac{v^4}{\omega_v}\int_\epsilon^\infty dk \int_0^\pi d\theta\;  \sin^5(\theta) k N(k) \frac{1 - \frac{\omega_v}{\omega_{v-k}}}{k^2 - (\omega_{v-k} - \omega_v)^2}\bigg]\rho(\vec u,\vec v,t)\,.
\end{align}

\end{appendices}

\bibliographystyle{h-physrev}
\bibliography{references.bib}

\begin{thebibliography}{10}

\bibitem{Fahn:2022zql}
M.~J. Fahn, K.~Giesel, and M.~Kobler,
\newblock Class. Quant. Grav. {\bf 40}, 094002 (2023), 2206.06397.

\bibitem{Breuer:2002pc}
H.-P. Breuer {\em et~al.},
\newblock {\em The theory of open quantum systems} (Oxford University Press on Demand, 2002).

\bibitem{Hornberger}
K.~Hornberger,
\newblock {\em Introduction to Decoherence Theory} (Springer Berlin Heidelberg, 2009), p. 221–276.

\bibitem{Weiss:2021uhm}
U.~Weiss,
\newblock {\em {Quantum Dissipative Systems}} (World Scientific, 2021).

\bibitem{Lindblad:1975ef}
G.~Lindblad,
\newblock Commun. Math. Phys. {\bf 48}, 119 (1976).

\bibitem{Gorini:1975nb}
V.~Gorini, A.~Kossakowski, and E.~C.~G. Sudarshan,
\newblock J. Math. Phys. {\bf 17}, 821 (1976).

\bibitem{Benatti:2000ph}
F.~Benatti and R.~Floreanini,
\newblock JHEP {\bf 02}, 032 (2000), hep-ph/0002221.

\bibitem{Blencowe:2012mp}
M.~P. Blencowe,
\newblock Phys. Rev. Lett. {\bf 111}, 021302 (2013), 1211.4751.

\bibitem{Anastopoulos:2013zya}
C.~Anastopoulos and B.~L. Hu,
\newblock Class. Quant. Grav. {\bf 30}, 165007 (2013), 1305.5231.

\bibitem{Guzzo:2014jbp}
M.~M. Guzzo, P.~C. de~Holanda, and R.~L.~N. Oliveira,
\newblock Nucl. Phys. B {\bf 908}, 408 (2016), 1408.0823.

\bibitem{Oniga:2015lro}
T.~Oniga and C.~H.~T. Wang,
\newblock Phys. Rev. D {\bf 93}, 044027 (2016), 1511.06678.

\bibitem{Lagouvardos:2020laf}
M.~Lagouvardos and C.~Anastopoulos,
\newblock Class. Quant. Grav. {\bf 38}, 115012 (2021), 2011.08270.

\bibitem{DEsposito:2023psn}
V.~D'Esposito and G.~Gubitosi,
\newblock (2023), 2306.14778.

\bibitem{Bassi:2017szd}
A.~Bassi, A.~Gro\ss{}ardt, and H.~Ulbricht,
\newblock Class. Quant. Grav. {\bf 34}, 193002 (2017), 1706.05677.

\bibitem{Anastopoulos:2021jdz}
C.~Anastopoulos and B.-L. Hu,
\newblock AVS Quantum Sci. {\bf 4}, 015602 (2022), 2111.02462.

\bibitem{fgke2024photon}
M.~J. Fahn, K.~Giesel, and R.~Kemper,
\newblock In preparation  (2025).

\bibitem{Lisi:2000zt}
E.~Lisi, A.~Marrone, and D.~Montanino,
\newblock Phys. Rev. Lett. {\bf 85}, 1166 (2000), hep-ph/0002053.

\bibitem{Sakharov:2009rn}
A.~Sakharov, N.~Mavromatos, A.~Meregaglia, A.~Rubbia, and S.~Sarkar,
\newblock J. Phys. Conf. Ser. {\bf 171}, 012038 (2009), 0903.4985.

\bibitem{Coelho:2017byq}
J.~a. A.~B. Coelho and W.~A. Mann,
\newblock Phys. Rev. D {\bf 96}, 093009 (2017), 1708.05495.

\bibitem{Carpio:2018gum}
J.~A. Carpio, E.~Massoni, and A.~M. Gago,
\newblock Phys. Rev. D {\bf 100}, 015035 (2019), 1811.07923.

\bibitem{Burrage:2018pyg}
C.~Burrage, C.~K\"ading, P.~Millington, and J.~Min\'a\v{r},
\newblock Phys. Rev. D {\bf 100}, 076003 (2019), 1812.08760.

\bibitem{weinberg2005quantum}
S.~Weinberg,
\newblock {\em The quantum theory of fields: Volume 1, foundations} (Cambridge university press, 2005).

\bibitem{tong}
D.~Tong,
\newblock Quantum field theory (lecture notes), 2007.

\bibitem{fleming2010rotating}
C.~Fleming, N.~Cummings, C.~Anastopoulos, and B.-L. Hu,
\newblock Journal of Physics A: Mathematical and Theoretical {\bf 43}, 405304 (2010).

\bibitem{Burgarth:2023ppw}
D.~Burgarth, P.~Facchi, R.~Hillier, and M.~Ligab\`o,
\newblock Quantum {\bf 8}, 1262 (2024), 2301.02269.

\bibitem{Wang:2023dkf}
P.~Wang, E.~Hiltunen, and J.~C. Schotland,
\newblock (2023), 2311.02670.

\bibitem{Domi:2024ypm}
A.~Domi {\em et~al.},
\newblock JCAP {\bf 11}, 006 (2024), 2403.03106.

\bibitem{Xu:2020lhc}
Q.~Xu and M.~P. Blencowe,
\newblock New J. Phys. {\bf 24}, 113048 (2022), 2005.02554.

\bibitem{Deser:1960zzc}
S.~Deser, R.~Arnowitt, and C.~W. Misner,
\newblock J. Math. Phys. {\bf 1}, 434 (1960).

\bibitem{Ashtekar:1986yd}
A.~Ashtekar,
\newblock Phys. Rev. Lett. {\bf 57}, 2244 (1986).

\bibitem{BarberoG:1994eia}
J.~F. Barbero~G.,
\newblock Phys. Rev. D {\bf 51}, 5507 (1995), gr-qc/9410014.

\bibitem{Immirzi:1996di}
G.~Immirzi,
\newblock Class. Quant. Grav. {\bf 14}, L177 (1997), gr-qc/9612030.

\bibitem{Ashtekar:1989ju}
A.~Ashtekar, J.~D. Romano, and R.~S. Tate,
\newblock Phys. Rev. D {\bf 40}, 2572 (1989).

\bibitem{rovelli_2004}
C.~Rovelli,
\newblock {\em Quantum Gravity}Cambridge Monographs on Mathematical Physics (Cambridge University Press, 2004).

\bibitem{Thiemann:2007pyv}
T.~Thiemann,
\newblock {\em {Modern Canonical Quantum General Relativity}}Cambridge Monographs on Mathematical Physics (Cambridge University Press, 2007).

\bibitem{Thiemann:1993zq}
T.~Thiemann,
\newblock Class. Quant. Grav. {\bf 12}, 181 (1995), gr-qc/9910008.

\bibitem{PoissonBook:2014}
E.~Poisson and C.~M. Will,
\newblock {\em {Gravity: Newtonian, Post-Newtonian, Relativistic}} (Cambridge University Press, 2014).

\bibitem{Rovelli:1990ph}
C.~Rovelli,
\newblock Class. Quant. Grav. {\bf 8}, 297 (1991).

\bibitem{Rovelli:1990pi}
C.~Rovelli,
\newblock Class. Quant. Grav. {\bf 8}, 317 (1991).

\bibitem{Rovelli:2001bz}
C.~Rovelli,
\newblock Phys. Rev. D {\bf 65}, 124013 (2002), gr-qc/0110035.

\bibitem{Vytheeswaran:1994np}
A.~S. Vytheeswaran,
\newblock Annals Phys. {\bf 236}, 297 (1994).

\bibitem{Dittrich:2004cb}
B.~Dittrich,
\newblock Gen. Rel. Grav. {\bf 39}, 1891 (2007), gr-qc/0411013.

\bibitem{Dittrich:2005kc}
B.~Dittrich,
\newblock Class. Quant. Grav. {\bf 23}, 6155 (2006), gr-qc/0507106.

\bibitem{Dittrich:2006ee}
B.~Dittrich and J.~Tambornino,
\newblock Class. Quant. Grav. {\bf 24}, 757 (2007), gr-qc/0610060.

\bibitem{Zwanzig:1960gvu}
R.~Zwanzig,
\newblock J. Chem. Phys. {\bf 33}, 1338 (1960).

\bibitem{Nakajima:1958pnl}
S.~Nakajima,
\newblock Prog. Theor. Phys. {\bf 20}, 948 (1958).

\bibitem{breuer2007non}
H.-P. Breuer,
\newblock Physical Review A—Atomic, Molecular, and Optical Physics {\bf 75}, 022103 (2007).

\bibitem{DAbbruzzo:2022hgp}
A.~D'Abbruzzo, V.~Cavina, and V.~Giovannetti,
\newblock SciPost Phys. {\bf 15}, 117 (2023), 2211.04400.

\bibitem{Arimitsu:1985ez}
T.~Arimitsu and H.~Umezawa,
\newblock Prog. Theor. Phys. {\bf 74}, 429 (1985).

\bibitem{Takahashi:1996zn}
Y.~Takahashi and H.~Umezawa,
\newblock Int. J. Mod. Phys. B {\bf 10}, 1755 (1996).

\bibitem{Khanna:2009zz}
F.~C. Khanna, A.~P.~C. Malbouisson, J.~M.~C. Malbouisson, and A.~R. Santana,
\newblock {\em {Thermal quantum field theory - Algebraic aspects and applications}} (, 2009).

\bibitem{Benatti:2001fa}
F.~Benatti and R.~Floreanini,
\newblock Phys. Rev. D {\bf 64}, 085015 (2001), hep-ph/0105303.

\bibitem{Donoghue:1994dn}
J.~F. Donoghue,
\newblock Phys. Rev. D {\bf 50}, 3874 (1994), gr-qc/9405057.

\bibitem{Arteaga:2003we}
D.~Arteaga, R.~Parentani, and E.~Verdaguer,
\newblock Phys. Rev. D {\bf 70}, 044019 (2004), gr-qc/0311065.

\bibitem{Hatfield:2019sox}
B.~Hatfield,
\newblock {\em {Quantum Field Theory Of Point Particles And Strings}} (CRC Press, 2019).

\bibitem{Nathan:2020}
F.~Nathan and M.~S. Rudner,
\newblock Physical Review B {\bf 102} (2020).

\bibitem{Jaynes:1963zz}
E.~T. Jaynes and F.~W. Cummings,
\newblock IEEE Proc. {\bf 51}, 89 (1963).

\bibitem{shore1993jaynes}
B.~W. Shore and P.~L. Knight,
\newblock Journal of Modern Optics {\bf 40}, 1195 (1993).

\bibitem{larson2021jaynes}
J.~Larson and T.~Mavrogordatos,
\newblock {\em The Jaynes--Cummings model and its descendants: modern research directions} (IoP Publishing, 2021).

\bibitem{BalieiroGomes:2018gtd}
G.~Balieiro~Gomes, D.~V. Forero, M.~M. Guzzo, P.~C. De~Holanda, and R.~L.~N. Oliveira,
\newblock Phys. Rev. D {\bf 100}, 055023 (2019), 1805.09818.

\bibitem{Boriero:2017tkh}
D.~Boriero, D.~J. Schwarz, and H.~Velten,
\newblock Universe {\bf 5}, 203 (2019), 1704.06139.

\bibitem{Pfenniger:2006rd}
D.~Pfenniger and V.~M.~G. Observatory,
\newblock Astron. Astrophys. {\bf 456}, 45 (2006), astro-ph/0605354.

\bibitem{Bernardini:2012uc}
A.~E. Bernardini,
\newblock EPL {\bf 103}, 30005 (2013), 1204.1504.

\bibitem{Milburn:1991zkb}
G.~J. Milburn,
\newblock Phys. Rev. A {\bf 44}, 5401 (1991).

\bibitem{Milburn:2003zj}
G.~J. Milburn,
\newblock New J. Phys. {\bf 8}, 96 (2006), gr-qc/0308021.

\bibitem{Diosi:2004iq}
L.~Diosi,
\newblock Braz. J. Phys. {\bf 35}, 260 (2005), quant-ph/0412154.

\bibitem{Breuer:2008rh}
H.~P. Breuer, E.~Goklu, and C.~Lammerzahl,
\newblock Class. Quant. Grav. {\bf 26}, 105012 (2009), 0812.0420.

\bibitem{Adkins:1982zk}
G.~S. Adkins,
\newblock Phys. Rev. D {\bf 27}, 1814 (1983).

\bibitem{Kaplanek:2022xrr}
G.~Kaplanek and E.~Tjoa,
\newblock Phys. Rev. A {\bf 107}, 012208 (2023), 2207.13750.

\bibitem{Colas:2022hlq}
T.~Colas, J.~Grain, and V.~Vennin,
\newblock Eur. Phys. J. C {\bf 82}, 1085 (2022), 2209.01929.

\bibitem{davidovic2020completely}
D.~Davidovi{\'c},
\newblock Quantum {\bf 4}, 326 (2020).

\bibitem{Giesel:2022pzh}
K.~Giesel and M.~Kobler,
\newblock Mathematics {\bf 10}, 4248 (2022), 2207.08749.

\bibitem{MORGAN2006311}
D.~Morgan, E.~Winstanley, J.~Brunner, and L.~F. Thompson,
\newblock Astroparticle Physics {\bf 25}, 311 (2006).

\bibitem{coloma2018decoherence}
P.~Coloma, J.~Lopez-Pavon, I.~Martinez-Soler, and H.~Nunokawa,
\newblock The European Physical Journal C {\bf 78}, 614 (2018).

\bibitem{icecube2024search}
Nature Physics {\bf 20}, 913 (2024).

\bibitem{aiello2025search}
S.~Aiello {\em et~al.},
\newblock Journal of Cosmology and Astroparticle Physics {\bf 2025}, 039 (2025).

\bibitem{Cho:2021gvg}
H.-T. Cho and B.-L. Hu,
\newblock Phys. Rev. D {\bf 105}, 086004 (2022), 2112.08174.

\end{thebibliography}

\end{document}